\documentclass[%
    aip,
    amsmath,amssymb,
    tightenlines,
    floatfix,
    reprint
]{revtex4-1}

\usepackage[american]{babel}
\usepackage[utf8x]{inputenc}

\usepackage{newtxtext,newtxmath}
\usepackage{microtype}

\usepackage{bbm}
\usepackage{mathtools}
\usepackage{physics}
\usepackage{dsfont}
\usepackage{braket}
\usepackage{xparse}
\usepackage{stackengine}

\usepackage[cal=cm,scr=dutchcal]{mathalpha}

\usepackage{graphicx}
\usepackage[svgnames]{xcolor}
\usepackage{siunitx}

\usepackage{ragged2e}
\usepackage{array}
\usepackage{tabularx}
\usepackage{booktabs}
\usepackage[normalem]{ulem}

\definecolor{cset-aps-blueberry}{RGB}{28,128,158}
\definecolor{cset-aps-blue}{RGB}{46,44,184}
\definecolor{cset-aps-turquoise}{RGB}{0,67,88}
\definecolor{cset-aps-limegreen}{RGB}{190,219,67}
\definecolor{cset-aps-green}{RGB}{31,138,112}
\definecolor{cset-aps-yellow}{RGB}{255,225,25}
\definecolor{cset-aps-orange}{RGB}{253,116,0}
\definecolor{cset-aps-red}{RGB}{219,0,43}

\definecolor{cset-aps-kobalt-medium}{RGB}{62,54,222}
\definecolor{cset-aps-kobalt-dark}{RGB}{28,24,150}

\definecolor{cset-aps-my-label-red}{RGB}{202,0,17}
\definecolor{cset-aps-my-label-blue}{RGB}{53,71,140}
\definecolor{cset-aps-my-label-gray}{RGB}{145,145,145}

\usepackage{hyperref}
\hypersetup{%
    colorlinks=true,
    linkcolor={cset-aps-red},
    linkbordercolor={cset-aps-red},
    filecolor={cset-aps-orange},
    filebordercolor={cset-aps-orange},
    citecolor={cset-aps-kobalt-medium},
    citebordercolor={cset-aps-kobalt-medium},
    urlcolor={cset-aps-kobalt-dark},
    urlbordercolor={cset-aps-kobalt-dark},
    menucolor={cset-aps-limegreen},
    menubordercolor={cset-aps-limegreen},
    breaklinks=true,
    pdfborderstyle={/S/U/W 2},
    pdfpagemode=UseOutlines,
    pdfstartpage={1},
}
\usepackage{enumitem}

\newcommand{\ee}{\text{e}}
\newcommand{\ii}{\text{i}}
\renewcommand{\vec}[1]{\boldsymbol{\mathbf{#1}}}

\newcommand{\orcid}[1]{\href{https://orcid.org/#1}{\includegraphics[width=7pt]{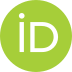}}}

\allowdisplaybreaks

\AtBeginDocument{%

}

\begin{document}

\preprint{AIP/123-QED}

\title[Finite Pulse-Time Effects in Long-Baseline Quantum Clock Interferometry]{Finite Pulse-Time Effects in Long-Baseline Quantum Clock Interferometry}
\author{Gregor Janson~\orcid{0009-0002-7509-4550}}
\email{gregor.janson@uni-ulm.de}
\author{Alexander Friedrich~\orcid{0000-0003-0588-1989}}
\author{Richard Lopp}
\address{Institut f{\"u}r Quantenphysik and Center for Integrated Quantum
    Science and Technology (IQST), Universit{\"a}t Ulm, Albert-Einstein-Allee 11, D-89069 Ulm, Germany}

\begin{abstract}
\noindent Quantum-clock interferometry has been suggested as a quantum probe to test the universality of free fall (UFF) and the universality of gravitational redshift (UGR). In typical experimental schemes it seems advantageous to employ Doppler-free E1-M1 transitions which have so far been investigated in quantum gases at rest. Here, we consider the fully quantized atomic degrees of freedom and study the interplay of the quantum center-of-mass (COM) -- that can become delocalized -- together with the internal clock transitions.  In particular, we derive a model for finite-time E1-M1 transitions with atomic intern-extern coupling and arbitrary position-dependent laser intensities. We further provide generalizations to the ideal expressions for perturbed recoilless clock pulses. Finally, we show at the example of a Gaussian laser beam that the proposed quantum-clock interferometers are stable against perturbations from varying optical fields for a sufficiently small quantum delocalization of the atomic COM.
\end{abstract}

\maketitle

\section{Introduction}
\label{sec.Intr}
Light-pulse atom interferometry (LPAI) has demonstrated its versatility in a myriad of applications: Starting from measuring gravitational acceleration~\cite{Kasevich1991, Peters1998, Peters1999} and rotation~\cite{Lan2012} to field applications~\cite{Barrett2016,Wu2019,Templier2022,Stray2022} and mobile gravimetry~\cite{Wu2019}, the measurement of Newton's gravitational constant~\cite{Rosi2014} as well as the so-far most accurate determination of the fine structure constant~\cite{Parker2018,Morel2020b}. In the last decade, there have been proposals for mid-band gravitational wave detection~\cite{Dimopoulos2008b,Dimopoulos2009,Graham2016}, complementary to LIGO/VIRGO and LISA, and recently construction has started on first prototypes which might be sensitive to ultra-light dark matter signals~\cite{Geraci2016,Arvanitaki2018,Badurina2023a} and serve as testbeds for gravitational wave antennas~\cite{Abe2021,Bertoldi2021,Badurina2021} based on atom interferometry.

These advancements have paved the way to perform tests on the fundamental physical principles underlying today's best physical theories with high precision atomic sensors~\cite{Safronova2018, Dimopoulos2007, Dimopoulos2008, Asenbaum2020}.
On the other hand, ever-increasing precision goals require an upscaling of the interferometers' spacetime areas. For that reason, several very-large baseline projects are currently being planned globally, hoping to reach the kilometer scale: AION-km in the UK~\cite{Badurina2020}, MAGIS-km in the USA~\cite{Abe2021}, MIGA/ELGAR in Europe~\cite{Canuel2018,Canuel2020}, and ZAIGA in China~\cite{Zhan2019}.   

A side beneficiary of these endeavors will be new long baseline tests of the Einstein equivalence principle, encapsulated in its three pillars~\cite{DiPumpo2023a} consisting of local Lorentz invariance, the universality of free fall (UFF) and local position invariance which in turn contains the universality of gravitational redshift (UGR) and universality of clock rates (UCR). Together these principles form the backbone of general relativity~\cite{Will2014,DiCasola2015}. 
All aspects of the equivalence principle have proven to be extremely resilient to experimental challenges over an extremely large regime ranging from the microscopic to the cosmic scale~\cite{Vessot1980, Chou2010, Herrmann2018, Delva2019, Takamoto2020, Bothwell2022, Brewer2019, Oelker2019, Madjarov2019, Touboul2019, Schlippert2014,Archibald2018,Zheng2023}.

UFF in particular has been tested via LPAI by comparison of the free fall rates of different atomic isotopes and species~\cite{Schlippert2014,Asenbaum2020,Rosi2017} as well as for different internal states~\cite{Zhou2021Aug} of the same atomic species. However, LPAI using quantum clocks as initial states have been shown~\cite{Loriani2019} to be insensitive to UGR violations in a linear gravitational field without additional internal transitions \emph{during}~\cite{Giulini2012,Roura2020,Ufrecht20202,DiPumpo2021} the interferometer. Recently, two different LPAI schemes were introduced for UGR and UFF tests by Roura~\cite{Roura2020} and Ufrecht et al.~\cite{Ufrecht2020} In contrast to the original proposals by Zych et al.~\cite{Zych2011} and Sinha et al.~\cite{Sinha2011} to detect general relativistic time dilation by the interference of quantum clocks in a gravitational field, both proposals are predicated on the essential step of initializing the atomic clock inside the interferometer in order to unequivocally isolate such a signal. Without this crucial step one is stuck with the no-go result~\cite{Loriani2019}. Therefore the scheme of Roura~\cite{Roura2020} needs a superposition of internal states to gain UGR sensitivity (and being insensitive to UFF violations); the alternative approach of Ufrecht et al.~\cite{Ufrecht2020} does not require superpositions of internal states (as seen by the laboratory frame). In turn it becomes sensitive to both, UGR and UFF violations. Similarly, other proposals~\cite{DiPumpo2021,DiPumpo2023a} can test different aspects of local position invariance like UCR.

Ideally one would like to initialize an atomic clock inside the interferometer without disturbing the center-of-mass (COM) motion of the atomic test masses which serve as inertial reference. Hence, recoilless internal transitions are strongly beneficial or might even be necessary since they can ease the experimental constraints and the implementation significantly. In this study, we examine recoilless transitions implemented via two-photon E1-M1 couplings, i.e. two-photon transitions consisting of one electric dipole (E1) and one magnetic dipole (M1) transition. This type of two-photon process has previously been investigated for Doppler-free two-photon spectroscopy~\cite{Grynberg1983,Rahaman2023} and for the application in optical atomic vapor clocks~\cite{Alden2014,AldenDiss2014} without COM motion. In contrast to these previous studies, we will consider the full quantum nature of all atomic degrees of freedom -- internal and COM. Due to the quantized nature of the COM degrees of freedom one would \textit{a priori} expect that the LPAI phase shift suffers from the corresponding delocalizing light-matter interaction~\cite{Lopp2021}. In particular, we find after incorporating COM motion that additional branch-dependent phases as well as momentum kicks and thus branches appear when considering realistic spatial laser profiles. The effect of phase shifts due to finite-time pulses with Doppler-shifted detunings for different classical Rabi frequencies has been investigated by Gillot et al.~\cite{Gillot2016} for Mach-Zehnder-type atom interferometers. Here, in contrast, the effects arise due to position-dependent Rabi frequencies in addition to the finite pulse time and structured laser beams when considering the quantized atomic COM. However, we show that the protocols of Roura~\cite{Roura2020} and Ufrecht et al.~\cite{Ufrecht2020} are resilient to leading order effects in the induced COM spread when compared to the interferometer size. Nonetheless, our results can serve as a guide when such or similar corrections need to be accounted for or modeled in future high precision experiments.

\subsection*{Overview \& Structure}
Our article is structured as follows:
In Sec.~\ref{sec: Interferometer Albert und Christian} we will recapitulate the two interferometer schemes presented by Roura~\cite{Roura2020} and Ufrecht et al.~\cite{Ufrecht2020}, and put them into the context of the dynamical mass energy of composite particles.
In Sec.~\ref{sec.Idealized model for E1M1 transitions}, we will introduce an idealized model for E1-M1 transitions using plane waves for the electromagnetic field, as an intermediate step but often serving as the foundation in the literature~\cite{Grynberg1983,Alden2014,AldenDiss2014,Weiss1994}, and that takes into account the quantized atomic COM and finite pulse times.
The internal structure of the atom will be described by a three-level system that can be reduced to an effective two-level system using adiabatic elimination.
To achieve the absorption of two counter-propagating photons a specific polarization scheme is needed~\cite{Alden2014,AldenDiss2014}. We will verify explicitly that standard Rabi oscillations are recovered to lowest order, canceling any quantum COM delocalization effects.
In Sec.~\ref{sec.Finite Pulse-Time effects} we will extend these results for E1-M1 transitions by taking into account position-dependent laser intensities. The generalized $\pi$- and $\pi/2$-pulse operators will be obtained in Sec.~\ref{sec: Fundamental Gaussian Laser Beam} for the experimentally relevant case of a Gaussian laser beam to lowest order.
In Sec.~\ref{sec. Effects in interferometry} we will come back to the two interferometer schemes~\cite{Roura2020,Ufrecht2020} and analyze the implications due to the finite pulse times and position-dependent Rabi frequencies, in particular their impact on the phase and visibility of the interferometers.
We conclude with a summary, discussion and contextualization of our results in Sec.~\ref{sec.Conclusion}.

\section{UGR and UFF tests with Quantum Clock Interferometry}
\label{sec: Interferometer Albert und Christian}
\begin{figure*}[!ht]
    \centering
    \includegraphics[width=\textwidth]{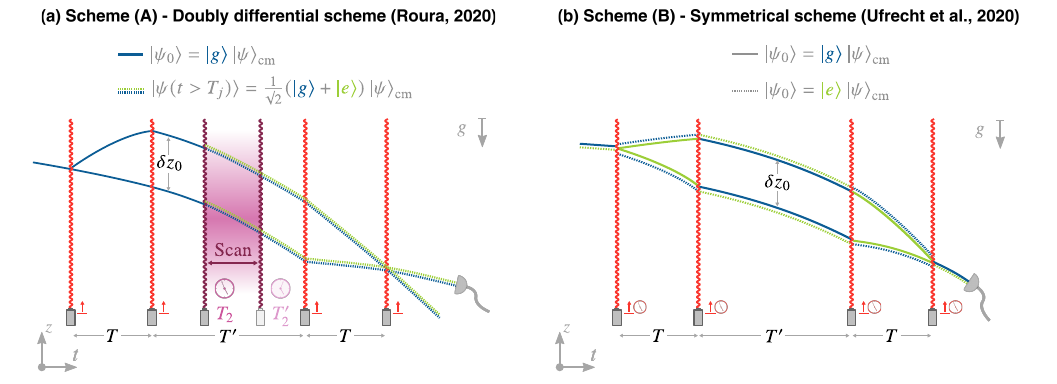}
\caption{Two LPAI proposals utilizing atomic clocks to detect UGR and UFF violations.
Panel {\sffamily\bfseries (a)} Scheme {\sffamily\bfseries (A)} is doubly differential: The atom enters the interferometer sequence in the ground state $\ket g$ and is split up onto two branches, e.g. via a Bragg pulse. At time $T_2$ a recoilless E1-M1 pulse drives the atom into a superposition of excited (denoted by $\ket e$) and ground state, corresponding to the initialization of an atomic clock. The branches are recombined afterwards and one can measure the intensities in the ground and excited state channels. The experiment is then repeated with a different initialization time $T_2'=T_2+\tau$. The COM wave function is denoted by $\ket \psi_{\text{cm}}$.
Panel {\sffamily\bfseries (b)} Scheme {\sffamily\bfseries (B)} on the other hand is symmetrical: The atom enters the interferometer sequence in the ground/excited state and is split up onto two branches, e.g. via a double Bragg diffraction pulse. In the middle segment, in between the times $t=T$ and $t=T+T'$, the internal state is changed to the excited/ground state via a recoilless E1-M1 transition. After recombining the two branches the detection is performed. A single run of the experiment consists of two runs of the interferometer sequence with different initial internal states.}
\label{fig1:ugr_schemes}
\end{figure*}
Here, we will briefly review the two interferometer schemes~\cite{Roura2020,Ufrecht2020} employing quantum clocks to test UGR and UFF. In the following we will denote the scheme proposed by Roura~\cite{Roura2020} as scheme (A) and the one proposed by Ufrecht et al.~\cite{Ufrecht2020} as scheme (B). 
Before starting this discussion we will introduce the relevant aspects of the dynamical mass energy (or mass defect) of atoms which is the underlying connection to test the UGR and UFF in an interferometer with quantum clocks. We note that our introduction only serves as a sketch of the ingredients necessary for incorporating a description of dynamical mass energy perturbatively into atoms.

While Einstein's mass energy equivalence \mbox{$E=Mc^2$} has been known for more than 100 years now, its impact on quantum interference due to the possibility of obtaining which-path information for composite particles with time-evolving internal structure has only recently been highlighted in the works of Zych et al.~\cite{Zych2011} and Sinha et al.~\cite{Sinha2011} How dynamical mass energy manifests in a Mach-Zehnder atom interferometer was first sketched by Giulini~\cite{Giulini2012} in the context of the redshift debate~\cite{Mueller2010,Mueller2010b,Wolf2010,Wolf2011,Wolf2012,Schleich2013}.
Based on these initial considerations, significant progress has been made. For a review of these initial discussions and proposed experiments beyond the ones discussed here~\cite{Roura2020,Ufrecht2020} see e.g. the works of Pikovski et al.~\cite{Pikovski2017} or Di Pumpo et al.~\cite{DiPumpo2021} and references therein. 

However, to the authors' knowledge, dynamical mass energy itself was already discussed in the works of Sebastian~\cite{Sebastian1981,Sebastian1986} on semi-relativistic models for composite systems interacting with a radiation field.
There the author indicates that the appearance of these terms (including dynamical mass energy) is intimately linked to relativistic corrections to the COM coordinates first derived by Osborn et al.~\cite{Osborn1968,Close1970} and Krajcik et al.~\cite{Krajcik1974} over 50 years ago.
The last few years have seen significant efforts and discussions devoted to providing first principles derivations from atomic physics of dynamical mass energy. Specifically, we refer to the works of Sonnleitner et al.~\cite{Sonnleitner2018} and Schwartz et al.~\cite{Schwartz2019,Schwartz2020b} for systems with quantized COM motion, respectively without and with gravity. A field theoretical derivation has recently been performed by A{\ss}mann et al.~\cite{Assmann2023} Moreover Perche et al.~\cite{Perche2021,Perche2022} contains a discussion under which conditions and by which guiding principles effective models for composite systems can be constructed in curved spacetime. Extensions examining the coupling of Dirac particles to gravitational backgrounds have recently also been discussed~\cite{Ito2021,Perche2021,Alibabaei2023} yielding overall sensible but in the details slightly differing results in the weak-field limit. A general review discussing the issues and problems regarding such couplings of quantum matter to gravity is available in Guilini et al.~\cite{Giulini2022} 

\subsection{A Simple Model for the Dynamical Mass Energy of Atoms}
In the non-relativistic limit, a first-quantized Hamiltonian description of a particle of mass $M$ moving in a weak gravitational field is prescribed~\cite{Anastopoulos2018} by the sum of the kinetic COM  energy and its gravitational potential energy $U(\vec{R})$
\begin{align}
    \label{eq:defect0}
    \hat{H}(\hat{\vec{R}},\hat{\vec{P}};M)= Mc^2+MU(\hat{\vec{R}})+\frac{\hat{\vec{P}}^2}{2M}-\frac{\hat{\vec{P}} U(\hat{\vec{R}}) \hat{\vec{P}}}{2M},
\end{align}
 with COM position $\hat{\Vec{R}}$ and momentum $\hat{\vec{P}}$ where we have neglected any terms contributing at orders higher than $1/M$. Practically, the gravitational potential can often  be approximated as \mbox{$U(\hat{\Vec{R}})=U(\vec{R}_0)+\vec{g}^\intercal(\hat{\vec{R}}-\vec{R}_0)+ (\hat{\vec{R}}-\vec{R}_0)^\intercal \Gamma (\hat{\vec{R}}-\vec{R}_0)/2$} up to the gravity gradient contribution where we adopted a symmetric Weyl ordering for the operators $\hat{\vec{R}}$ and $\hat{\vec{P}}$ and have expanded around the point $\Vec{R}_0$ in whose vicinity the system is localized.
 Note, that the mass $M$ of the particle corresponds to its rest mass here. Hence, only in going beyond this non-relativistic model, the dynamical nature of mass energy can become relevant. 
 
 When considering the atom to be comprised of individual particles sub-leading relativistic corrections will appear~\cite{Sebastian1981,Sebastian1986,Pikovski2017,Schwartz2019,Schwartz2020b,MartinezLahuerta2022b,Wood2022} and change the Hamiltonian. The most impactful change resulting from this is the insight that the total atomic mass is no longer just the sum of the rest masses of its constituent particles but also contains a contribution from the internal Hamiltonian of the atom, as one would naively expect from mass energy equivalence. We incorporate this in our simple model by performing the replacement~\cite{Schwartz2020b}
\begin{align}
    M \mapsto \hat{M}=M + \frac{\hat{H}_{\mathrm{A}}}{c^2}.\label{eq:defect1}
\end{align}
Note, that we have introduced the abbreviation $\hat{M}$ for the quantity $\hat{M}=M+\hat{H}_A/c^2$ which behaves akin to a mass operator. In this we are guided by the fact that the eigenvalue equation \mbox{$\hat{M}\ket{n}=M_n\ket{n}$} leads to the eigenvalues $\mathcal{E}_j$ of the internal Hamiltionian $\hat{H}_A$. These are determined by the eigenvalue equation $\hat{H}_A\ket{n}=\mathcal{E}_n \ket{n}$ and directly connected to the eigenvalues of the mass operator
\mbox{$M_n=M+\mathcal{E}_j/c^2$}, that is they are scaled by the square of the speed of light and shifted by the rest mass. Since there is a one-to-one mapping between the eigenvalues and eigenstates of the the mass operator and the internal Hamiltionian no additional complexity of the system in terms of additional Hilbert spaces or new dynamics is gained, and to this order this looks like a simple reformulation in terms of different quantities.

Furthermore, if the smallest and largest eigenstate of the internal Hamiltonian $\hat{H}_A$ are separated by an energy  \mbox{$\Delta \mathcal{E}\ll Mc^2$}, then the intern-extern coupling can be treated perturbatively. This is a useful approximation e.g. in the case of optical clock transitions in ytterbium or strontium where the (relevant part of the) internal Hamiltonian has a spectral range in the optical regime and we can thus estimate\cite{Loriani2019} \mbox{$\Vert \hat{H}_A\Vert/(Mc^2)~\simeq ~10^{-11}$.} Thus, often we can assume the perturbative identification~\cite{Loriani2019} \mbox{$M^{-1}\big(1-\hat{H}_A/(Mc^2)\big)\simeq M^{-1}\big(1+\hat{H}_A/(Mc^2)\big)^{-1}$} via the geometric series. Consequently, we can also replace \mbox{$M^{-1}\mapsto\hat{M}^{-1}$} in the terms in Eq.~\eqref{eq:defect0} describing the potential and kinetic energy. 
The overall Hamiltonian $\hat{H}^{(\text{MD})}$, including the mass defect, accordingly takes the form
\begin{align}\label{eq:defect2}
  \begin{split}
  \hat{H}^{(\text{MD})}=\hat{H}(\hat{\vec{R}},\hat{\vec{P}};\hat{M})=
  \hat{M}c^2+\hat{M}U(\hat{\vec{R}})
   +\frac{\hat{\vec{P}}^2}{2\hat{M}}-\frac{\hat{\vec{P}} U(\hat{\vec{R}}) \hat{\vec{P}}}{2\hat{M}}.
  \end{split}
\end{align}
All but the first term in Eq.~\eqref{eq:defect2} induce a coupling of the internal atomic energies to the kinetic and potential energy of the COM.

Alternatively, using the energy eigenstates $\ket{n}$ and the mass operator eigenvalues $M_n=M+\mathcal{E}_n/c^2$, the Hamiltonian $\hat{H}^{(\text{MD})}$ can be rewritten as
\begin{align}
    \label{eq:defect3}
    \hat{H}^{(\text{MD})} =\; \hat{H}^{(\text{ME})}= \sum_n \hat{H}^{\text{(ME)}}_{n}\dyad{n}{n}
\end{align}
with $\hat{H}^{\text{(ME)}}_{n}=\hat{H}(\hat{\vec{R}},\hat{\vec{P}};M_n)$, which is equivalent to a collection of single particles, characterized by the state-dependent masses $M_n$. In this form the Hamiltonian directly embodies the equivalence of inertial and gravitational mass~\cite{Anastopoulos2018}. While we have omitted the coupling to external (electromagnetic) fields in all our considerations for simplicity, they are in principle instrumental to actually prepare and manipulate the atomic wave packet in experiments. These may be accounted for in an interaction Hamiltonian $\hat{H}_\text{int}$ added to Eq.~\eqref{eq:defect2} or Eq.~\eqref{eq:defect3}, since the mass eigenstates are identical to the internal energy eigenstates except for an energy shift. The details of this interaction Hamiltonian can be quite complicated~\cite{Sonnleitner2018,Schwartz2020b,Assmann2023} when all corrections from the mass defect are included. However, to leading order it consists of the standard electric or magnetic dipole transitions described by
\begin{align}\label{eq:dipole-interaction}
    \hat{H}_\text{int}=-\hat{\vec{d}}\cdot\vec{E}(t,\hat{\vec{R}})+\hat{\vec{\mu}}\cdot\vec{B}(t,\hat{\vec{R}}),
\end{align}
where  $\hat{\vec{d}}$ and $\hat{\vec{\mu}}$ are the atomic electric and magnetic dipole operators, respectively.  Note that we have neglected here terms to leading order in the electric charge and Bohr radius that are further suppressed by the atomic mass $M$, such as the R\"ontgen term~\cite{Lopp2021}. In conclusion we arrive at the total model Hamiltonian (excluding higher order contributions for the electromagnetic field coupling)
\begin{align}
 \hat{\mathcal{H}} = \hat{H}^\text{(MD)} +\hat{H}_\text{int}
 = \hat{H}^\text{(ME)} +\hat{H}_\text{int}
\end{align}
for an atom with internal structure interacting with an external electromagnetic field.

While our introduction here can only serve as a sketch, motivated by mass energy equivalence, it turns out that the derivation of the intern-extern coupling can be made fairly rigorous~\cite{Anastopoulos2018,Loriani2019,Wood2021,Schwartz2019,Schwartz2020b,MartinezLahuerta2022b,Assmann2023}, however with serious gains in the theoretical complexity of the model depending on the setting as well as the starting point. Nevertheless, the basic premises and leading order results do not change significantly.

\subsection{Phase Shift in a Light-Pulse Atom Interferometer}
There are multiple methods available to calculate the phase shift in a LPAI. In simple cases, with quadratic Hamiltonians and for instantaneous beam splitter pulses, one can often rely on path-integral methods~\cite{Storey1994}. However path-integrals become quite unwieldy in case of non-quadratic systems as there are no or only few standard methods available for their solution~\cite{Kleinert2009}. In these more involved cases, e.g. with multiple internal states and complicated external potentials involved, the Hamiltonian approach~\cite{Schleich2013,Kleinert2015,Ufrecht2021} offers a more versatile toolbox. Moreover, phase-space methods~\cite{Giese2014,Roura2014,Zimmermann2021} are also available and sometimes helpful for interpretation.

However, in all cases the interference signal in an exit port of a two-path interferometer arises from the superposition of two branches characterized by the evolutions $\hat{U}_{1}$ and $\hat{U}_{2}$ and is determined by the expectation value~\cite{Schleich2013,DiPumpo2023a} 
\begin{align}\label{eq:two-path-signal}
    I_{\phi_{\text{exit}}} = 
    \bra{\psi_0} \hat{U}_\text{tot}^\dagger \hat{\Pi}_\text{exit}^{\phantom{x}} \hat{U}_\text{tot}^{}\ket{\psi_0},
\end{align}
with the overall evolution given by $\hat{\Pi}_\text{exit}\hat{U}_\text{tot}=\hat{U}_1+\hat{U}_2$. Here $\hat{U}_\text{tot}$ is the total time evolution, $\hat{\Pi}_{\text{exit}}$ is a projection operator with the property $\hat{\Pi}_{\text{exit}}^2=\hat{\Pi}_{\text{exit}}$  characteristic to the detection process occurring in the exit port and $\ket{\psi_0}$ is the initial state at the start of the interferometer. Note that here the individual evolutions $\hat{U}_1$ and $\hat{U}_2$ need not be unitary by themselves. In fact, even in a Mach-Zehnder interferometer they are not. This is due to the fact that only half of the atoms participates in each branch of the interferometer. Furthermore, the individual nature of the beam splitters creating the interferometer decides the balance  between the interferometer branches. On the other hand the total evolution $\hat{U}_\text{tot}$ usually is unitary, unless e.g. atom losses occur or not all paths are included in the modelling of the interferometer and thus $\hat{U}_\text{tot}$ becomes an open system evolution.

After expanding the sum over the individual branches in the exit port signal, defined in Eq.~\eqref{eq:two-path-signal}, it takes the form
\begin{align}\label{eq:interferometer-2path-signal}
    \begin{split}
    I_{\phi_{\text{exit}}} &= 
    \bra{\psi_0} \hat{U}_1^\dagger \hat{U}_1^{\phantom{\dagger}}\ket{\psi_0}+
    \bra{\psi_0} \hat{U}_2^\dagger \hat{U}_2^{\phantom{\dagger}}\ket{\psi_0} \\
    & \quad \quad \quad +\bra{\psi_0} \hat{\mathcal{O}}_{21}\ket{\psi_0}
    + \text{c.c.}
    \end{split}
\end{align}
where we introduced the amplitude 
\begin{align}\label{eq:overlap-expectation}
\langle\hat{\mathcal{O}}_{21}\rangle = \bra{\psi_0} \hat{U}_2^\dagger \hat{U}_1^{\phantom{\dagger}} \ket{\psi_0} =\mathcal{V}_{21} \exp(\ii \Delta\phi_{21})
\end{align}
of the so-called overlap operator~\cite{Schleich2013} ~\mbox{$\hat{\mathcal{O}}_{21}=\hat{U}_2^\dagger \hat{U}_1^{\phantom{\dagger}}$ }  between the branches. The absolute value of this amplitude is the visibility \mbox{$\mathcal{V}_{21}=|\bra{\psi_0} \hat{U}_2^\dagger\hat{U}_1^{\phantom{\dagger}} \ket{\psi_0}|$} of the interference signal, while the argument \mbox{$\Delta \phi_{21} =\text{arg}\bra{\psi_0} \hat{U}_2^\dagger \hat{U}_1^{\phantom{\dagger}} \ket{\psi_0}$} is the interferometer phase~\cite{Schleich2013,Ufrecht2020,DiPumpo2023a}.

In general, the situation in a realistic LPAI can be a bit more complex and the overall signal $I_{\phi_{\text{exit}}}$ detected in an exit port results from the pair-wise interference of all paths through the interferometer contributing to the exit port. Practically, additional and often undesired paths can originate e.g. from imperfect diffraction processes~\cite{Jenewein2022,Kirsten-Siemss2023} or perturbing potentials acting during the interferometer.

However, any interfering pair of paths contributing to the signal amplitude of the exit port in such a multi-path LPAI has a contribution of the form of the expectation value of an overlap
\begin{align}
    I_{\phi_{\text{exit}}}^{(lm)}=\bra{\psi_0} \hat{U}^\dagger_{\ell}\hat{U}_m^{\phantom{x}}\ket{\psi_0} + \text{c.c.} = \mathcal{V}_{\ell m} \exp(\ii \Delta \phi_{\ell m}) + \text{c.c.}
\end{align}
Here we have introduced the relative path visibility $\mathcal{V}_{\ell m}$ and relative phase between paths $\Delta \phi_{\ell m}$ which generalizes the same quantities from the two-path case. Summation over the signal amplitude contributions $I_{\phi_{\text{exit}}}^{(lm)}$ with respect to the indices $\ell$ and $m$ directly leads to the overall exit port signal
\begin{align}
    I_{\phi_{\text{exit}}} =   \sum_{\ell\geq 1} \mathcal{V}_{\ell\ell} + \sum_{\substack{\ell,m\geq 1\\ \ell\neq m}} \mathcal{V}_{\ell m} \exp(\ii \Delta \phi_{\ell m}).
\end{align}
When we also note that the relative phase between paths obeys the relation \mbox{$\Delta \phi_{\ell m}=-\Delta \phi_{m \ell}$} we arrive at the expression
\begin{align}\label{eq:multipathsignal}
I_{\phi_{\text{exit}}} = \sum_{\ell} \mathcal{V}_{\ell\ell} +\sum_{\substack{\ell,m \geq 1\\ \ell\neq m}} \mathcal{V}_{\ell m} \cos\Delta \phi_{\ell m}
\end{align}
for the exit port signal. This expression is a superposition of the cosines of the \emph{relative path phases} weighted by the \emph{relative path visibilities}. In an (open) two-path interferometer the sums terminate after two terms, and is thus identical to Eq.~\eqref{eq:interferometer-2path-signal}.

\subsection{Interferometer Phase, (Classical) Action and Proper time}
Usually, the interferometer phase in a LPAI is linked to the (classical) action by appealing to the relativistic action of a massive particle in a gravitational background~\cite{Storey1994,Schleich2013,Loriani2019,DiPumpo2021} and a subsequent non-relativistic expansion. The resulting expression is then quantized and introduced as governing action $\mathcal{S}$ of an appropriate path integral for the particle. Afterwards one identifies the quantum mechanical phase~\cite{Loriani2019} acquired along the trajectory via
\begin{align}\label{eq:phase-and-proper-time}
    \phi=-\omega_{C} \tau=-\frac{1}{\hbar}\int\!\!\!\dd t ~ \mathcal{L}(\vec{R},\dot{\vec{R}},t)+S_\text{em}/\hbar,
\end{align}
where $\omega_C=M c^2/\hbar$ is the Compton frequency and $\mathcal{L}$ is the classical Lagrangian $\mathcal{L}(\vec{R},\dot{\vec{R}})$ corresponding to the Hamiltonian,~Eq.~\eqref{eq:defect0}, of the particle. Here $S_\text{em}$ is the action corresponding to the Lagrangian for the electromagnetic interaction,~Eq.~\eqref{eq:dipole-interaction}, needed for manipulation of the atom. Fundamentally, this interpretation originates from a semi-classical approximation for the Feynman path-integral~\cite{Storey1994,Kleinert2009} being a valid approximation. This is due to the fact that only in the semi-classical limit the dominant contributions to the path-integral come from the classical trajectories, resulting from solving the Euler-Lagrange equations for the (classical) Lagrangian~\cite{Kleinert2009}. Ultimately, this is what makes the identification between proper time and the action in~Eq.~\eqref{eq:phase-and-proper-time} possible also for quantum particles but only in the semi-classical limit.

\subsection{UGR Sensitive Scheme (A)}
The interferometer scheme (A)~\cite{Roura2020}, shown in Fig.~\ref{fig1:ugr_schemes}(a), initializes an atomic clock by a recoilless $\pi/2$-pulse so that the atoms that enter the interferometer in the ground state are in a 50:50 superpostion of excited and ground state atoms after the clock initialization. Due to the atoms having a different mass $M_{g,e}$ in their respective internal ground and excited states, the Compton frequency $\omega_{g,e}$ becomes state-dependent. 
One can measure the frequency in the ground and excited state exit port between the two branches via the differential phase shift $\Delta\phi_{g,e}$ and separate out the gravitational redshift by a double-differential measurement, i.e. calculating the phase difference $\Delta\phi_-$ between the excited and ground state exit port and performing two runs of the experiment with different initialization times $T_2$ of the atomic clock:
\begin{align}
\Delta\phi_{-}(T_2)-\Delta\phi_-(T_2+\tau) = -\frac{\Delta M}{\bar{M}}g k_p \delta T \tau,
\end{align}
where
\begin{align}
\Delta\phi_- (T_2) = \Delta\phi_g(T_2) - \Delta\phi_e(T_2),
\end{align}
$\bar{M}=(M_e+M_g)/2$ is the mean mass, $g$ is the gravitational acceleration, $k_p$ is the wave number of the laser that drives the atoms onto the two branches, $\delta T$ is the separation time, and $\Delta M=M_e - M_g$ is the mass difference due to the mass defect. Since the rest mass $M$ and the mean mass $\bar{M}$ are equivalent to our order of approximation, i.e. to order $\mathcal{O}(c^{-2})$~\cite{Loriani2019}, we may identify the mean mass as  $M$.

\subsection{UGR and UFF Sensitive Scheme (B)}
The interferometer scheme (B)~\cite{Ufrecht2020} (cf. Fig.~\ref{fig1:ugr_schemes}(b)) is sensitive to both, the UGR and UFF. In contrast to scheme (A) it does not require a superpostion of internal states. The sensitivity arises from the specific space-time geometry of the interferometer and a change of internal states so that the atoms are in the same state at equal times (in the laboratory frame). The total phase
\begin{align}
\Delta \Phi = \Delta \Phi_M - \frac{\Delta M c^2}{2\hbar} \sum_n \lambda_\pm \Delta\tau_n
\end{align}
consists of two contributions: the contribution $ \Delta \Phi_M$ is independent of the mass defect $\Delta M$ and is obtained via the reference Hamiltonian $\hat H_M$ at the mean mass $M$. This part of the total phase can be used for tests of UFF~\cite{Ufrecht2020}. The proper time differences $\Delta\tau_n$ in each segment $n$ of the interferometer enter the phase proportional to the mass defect $\Delta M$ such that it can be associated with the ticking rate of an atomic clock~\cite{DiPumpo2021}. The $\lambda_\pm$ indicate the internal state for each segment: $\lambda_-=-1$ for the ground state and $\lambda_+=+1$ for the excited state. Since the sum $\Delta\tau=\Delta\tau_1+\Delta\tau_2+\Delta\tau_3$ of the proper-time differences vanishes in this geometry, the proper-time difference in the middle segment can be written as $\Delta\tau_2=-(\Delta\tau_1+\Delta\tau_3)$. Changing the internal state in the middle segment (associated with $\Delta\tau_2$), the total phase becomes
\begin{align}
\Delta\Phi = \Delta\Phi_M\pm\frac{\Delta M c^2}{\hbar}\Delta\tau_2,
\end{align}
depending on the choice of the initial internal state. Again, performing two runs of the experiment with different initial internal states one can separate the UFF and the UGR effects by adding or subtracting the phases, respectively:
\begin{subequations}
    \begin{align}
        \Delta\phi_+&=2\Delta\phi_M,\\
        \Delta\phi_-&=2\frac{\Delta M c^2}{\hbar}\Delta\tau_2.
    \end{align}
\end{subequations}

\subsection{Common Challenges}
Both interferometer schemes presented above require the manipulation of the internal states during the interferometer sequence. While this manipulation can be achieved by (technically challenging) optical Double-Raman diffraction~\cite{Sarkar2022} in scheme (B), i.e. kicking the atoms and changing the internal states simultaneously, scheme (A) requires recoilless internal transitions. There are several reasons why one would like to avoid Double-Raman diffraction: first of all, to drive this kind of transitions one needs quite long laser pulses leading to finite pulse-time effects. Secondly, the single-photon detuning cannot be chosen arbitrarily large if one still wants to have significant Rabi frequencies. This constraint for the detunings leads to problems with spontaneous emision. Furthermore, Double-Raman diffraction requires a high stability for the difference of the two laser frequencies during the pulse. Replacing the Double-Raman diffraction by a momentum-transfer pulse, e.g. Double-Bragg diffraction, and a state-changing pulse could alleviate these issues. These recoilless transitions can be achieved by E1-M1 transitions where the atom absorbs two counter-propagating photons with equal frequency $\omega$ so that the total momentum kick caused by the two-photon transition vanishes. Such E1-M1 transitions were only investigated without (quantized) COM motion in the context of optical clocks~\cite{Alden2014, AldenDiss2014,ZanonWillette2023}. However, in atom interferometry the COM motion plays a crucial role. Hence, its influence also needs to be included when modeling the pulses to account for possible corrections. This is the task of the following sections.

\section{Idealized model for E1-M1 transitions}
\label{sec.Idealized model for E1M1 transitions}
In this section we will derive an effective model for the finite-time E1-M1 transition processes during the LPAI schemes discussed in the previous section for an atomic cloud in a gravitational potential, see Fig.~\ref{fig:three-level atom}. The cloud will be modelled as a fully first-quantized atom, including its quantized COM motion. In particular, this can be applied to general initial atomic wavepackets.  We will assume, for now,  that the electromagnetic fields of the laser beam are classical plane waves. We will explicitly show that to leading order standard Rabi oscillations are recovered, since the COM dependence drops out for fields of constant intensity in space. The extension to realistic position-dependent laser intensities as well as finite pulse-time effects will be treated in Sec.~\ref{sec.Finite Pulse-Time effects}. 

\subsection{Model}
For an arbitrary three-level atom of mass $M$ in a gravitational field along $-Z$ and via the dipole approximation the Hamiltonian reads
\begin{align}\label{eq: final approximated Hamiltonian}
\hat{\tilde{H}}= \frac{\hat{\vec{P}}^2}{2M}+\sum_n \mathcal{E}_n \dyad{n}{n}-\hat{\vec{d}}\cdot\vec{E}(t,\hat{\vec{R}})+\hat{\vec{\mu}}\cdot\vec{B}(t,\hat{\vec{R}})+Mg\hat{Z},
\end{align}
where $\mathcal{E}_n$ are the atomic internal energies.
For simplicity, we will assume the electric and magnetic fields, $\vec{E}$ and $\vec{B}$, to be plane waves with frequency $\omega$ for now.
Note that we have neglected the mass defect, cf. Eq.~\eqref{eq:defect1}, during the interaction with the laser since the pulse time $t$ is much smaller than the characteristic interferometer time $T$.
The internal-state dependent mass energy enters the phase via $\Delta M c^2\cdot t/\hbar$ and $\Delta M c^2\cdot T/\hbar$, respectively. The effects of the mass defect during the laser pulse compared to the effects during the rest of the interferometer sequence is therefore negligible. In particular, the $\mathcal{O}(c^{-2})$ correction of Eq.~\eqref{eq:defect2} is subdominant with respect to the dipolar interaction terms.
Moreover, for the same reason we have only retained the linear potential contribution from the gravitational potential energy and do not consider the higher order contributions due to gravity gradients and kinetic-energy to position couplings from Eq.~\eqref{eq: final Hamiltonian}. If necessary they could be included perturbatively, similar to the optical potentials in~Sec.~\ref{sec.Finite Pulse-Time effects}. 
\begin{figure}[h]
        \centering
        \includegraphics{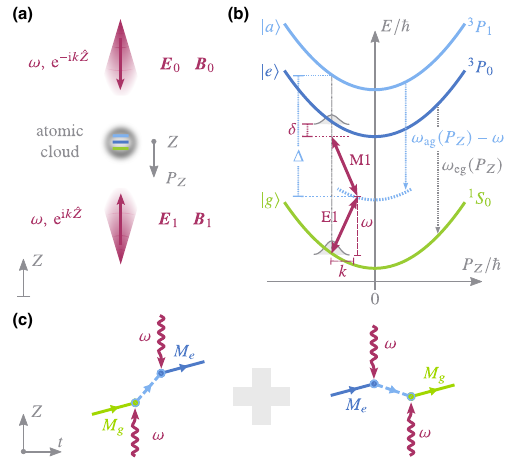}
        \caption{Panel {\sffamily\bfseries (a)} Incoming and reflected electromagnetic waves with frequency $\omega$ and amplitudes $\vec{E}_i$ and $\vec{B}_i$ driving E1-M1 transition in a three-level atom with quantized COM motion (black arrow).
        Panel {\sffamily\bfseries (b)} Three-level atom with ground state $\ket{g}$, excited state $\ket{e}$ and ancilla state $\ket{a}$ modelled by the states $^1S_0$, $^3P_0$ and $^3P_1$, respectively. Counter-propagating fields drive E1-M1 transitions, i.e. an E1 transition between $\ket{g}$ and $\ket{a}$ with single-photon detuning $\Delta$ and subsequent M1 transition between $\ket{a}$ and $\ket{e}$ leading to the overall detuning $\delta$. The ancilla state $\ket{a}$ lies then virtually between the energy levels of $\ket{g}$ and $\ket{e}$.
        Panel {\sffamily\bfseries (c)} Left: Two-photon excitation by absorbing two counter-propagating photons with momentum $\pm\hbar\vec{k}_L$. The first absorption leads to a transition from $\ket{g}$ to $\ket{a}$ and a momentum kick $\hbar\vec{k}_L$, the second absorption is a transition from $\ket{a}$ to $\ket{e}$ with momentum kick $-\hbar\vec{k}_L$. Right: Two-photon decay by stimulated emission of two photons in opposite directions. The first emission leads to a transition from $\ket{e}$ to $\ket{a}$ and a momentum kick $-\hbar\vec{k}_L$ and the second emission is a transition from $\ket{a}$ to $\ket{g}$ with momentum kick $\hbar\vec{k}_L$. Both two-photon processes have a vanishing netto momentum kick.}
        \label{fig:three-level atom}
\end{figure}

In order to describe the laser beam in a retro-reflective geometry, we consider two counter-propagating electromagnetic plane waves.
To obtain a recoilless two-photon transition one has to ensure that the atom absorbs two counter-propagating photons. This can be achieved by choice of a certain polarization scheme as we will discuss later on in Sec.~\ref{sec:doppler-free}; for now we will keep the polarization arbitrary. Then the fields can be written as:
\begin{subequations}
\begin{align}
    \vec{E}(\hat Z)&=\sum_{j=0}^1 \mathrm{i}\,\vec{E}_j\mathrm{e}^{(-1)^j\mathrm{i}\,k_L\hat{Z}}\mathrm{e}^{-\mathrm{i}\,\omega t}+\text{h.c.},\\
    \vec{B}(\hat Z)&=\sum_{j=0}^1 \mathrm{i}\,\vec{B}_j\mathrm{e}^{(-1)^j\mathrm{i}\,k_L\hat{Z}}\mathrm{e}^{-\mathrm{i}\,\omega t}+\text{h.c.}
\end{align}
\end{subequations}
We particularize to E1 transitions only between the ground state $\ket{g}$ and the ancilla state $\ket{a}$ and M1 transitions only between the ancilla state $\ket{a}$ and the excited state $\ket{e}$, i.e. the matrix elements $\vec{d}_{ea}=\bra{e}\hat{\vec{d}}\ket{a}$ and $\vec{\mu}_{ag}=\bra{a}\hat{\vec{\mu}}\ket{g}$ vanish. This can be ensured by considering the selection rules for electric and magnetic dipole transitions that are discussed subsequently. Thus, in the internal atomic eigenenergy basis, the electric and magnetic dipole moment operators reduce, respectively, to
\begin{align}\label{eq: electric dipole moment interaction picture}
\hat{\vec{d}}(t)=\vec{d}_{ag}\dyad{a}{g} + \text{h.c.},~
\hat{\vec{\mu}}(t)=\vec{\mu}_{ae}\dyad{a}{e}+\text{h.c.}
\end{align}
We further define  $\hbar\omega_{ij}=\hbar(\omega_i-\omega_j)$ as the energy spacings between the internal atomic states $\ket{i}$ and $\ket{j}$ (\mbox{$\{i,j\}\in \{a,e,g\}$}). Then, we can introduce the single-photon detuning \mbox{$\Delta=\omega_{ag}-\omega$} for the E1 transition between the ground state $\ket{g}$ and the ancilla state $\ket{a}$ and the overall detuning of the two-photon process, i.e. $\delta=\omega_{eg}-2\omega$, as shown in Fig.~\ref{fig:three-level atom}. The time dependence of the Hamiltonian with respect to the atomic frequencies can be simplified via the unitary transformation
\begin{align}\label{eq: unitary transformation detunings}
\hat{U}=\mathrm{e}^{\mathrm{i}(\omega_a+\Delta) t}\dyad{a}{a}+\mathrm{e}^{\mathrm{i}(\omega_e+\delta) t}\dyad{e}{e}+\mathrm{e}^{\mathrm{i}\omega_g t}\dyad{g}{g},
\end{align}
leading to the interaction Hamiltonian in the (modified) internal atomic interaction picture
\begin{align}
	\begin{split}
\hat{H}_{\mathrm{rot}}=&\hat{U}^\dagger \hat{H}\hat{U}+\mathrm{i}\hbar\left(\frac{\mathrm{d}}{\mathrm{d} t}\hat{U}^\dagger\right)\hat{U}\\
=& \sum_{j=0}^1\left\{\mathrm{e}^{\mathrm{i}\omega t} \vec{d}_{ag}\cdot\bigg[-\mathrm{i}\vec{E}_j\mathrm{e}^{(-1)^j\mathrm{i}k_L\hat{Z}}\mathrm{e}^{-\mathrm{i}\omega t}+\mathrm{h.c.}\bigg]\dyad{a}{g}\right.\\
&\!\quad\quad +\mathrm{e}^{-\mathrm{i}\omega t} \vec{\mu}_{ae}\cdot\bigg[\mathrm{i}\vec{B}_j\mathrm{e}^{(-1)^j\mathrm{i}k_L\hat{Z}}\mathrm{e}^{-\mathrm{i}\omega t}+\mathrm{h.c.}\bigg]\dyad{a}{e}\\
&\!\quad\quad+ \mathrm{h.c.}\biggr\}+\frac{\hat{\vec{P}}^2}{2M}+\hbar\Delta\dyad{a}{a}+\hbar\delta\dyad{e}{e}+Mg\hat{Z}.
	\end{split}
\end{align}
Performing a displacement transformation
\begin{align}\label{eq: Displacement gravitation}
    \hat{D}(t)=\exp\left(-\frac{\ii}{\hbar}(Z_\text{cl}(t)\hat{P}_z-P_\text{cl}(t)\hat{Z})\right)
\end{align}
corresponding to $\hat{Z}\rightarrow\hat{Z}+Z_\text{cl}(t)$ and \mbox{$\hat{P}_z\rightarrow\hat{P}_z+P_\text{cl}(t)$}, with $Z_\text{cl}(t)=-\frac{1}{2}gt^2$ and $P_\text{cl}(t)=-Mgt$ being the solutions of the classical equation of motion in the gravitational potential, yields the Hamiltonian
\begin{align}
\hat{H}_{\mathrm{rot}}^\prime
=& \sum_{j=0}^1\left\{\mathrm{e}^{\mathrm{i}\omega t} \vec{d}_{ag}\cdot\bigg[-\mathrm{i}\vec{E}_j\mathrm{e}^{(-1)^j\mathrm{i}k_L\hat{Z}(t)}\mathrm{e}^{-\mathrm{i}\omega t}+\mathrm{h.c.}\bigg]\dyad{a}{g}\right.\nonumber\\
&\!\quad\quad +\mathrm{e}^{-\mathrm{i}\omega t} \vec{\mu}_{ae}\cdot\bigg[\mathrm{i}\vec{B}_j\mathrm{e}^{(-1)^j\mathrm{i}k_L\hat{Z}(t)}\mathrm{e}^{-\mathrm{i}\omega t}+\mathrm{h.c.}\bigg]\dyad{a}{e}\nonumber\\
&\!\quad\quad+ \mathrm{h.c.}\biggr\}+\frac{\hat{\vec{P}}^2}{2M}+\hbar\Delta\dyad{a}{a}+\hbar\delta\dyad{e}{e},
\end{align}
where a time-dependent energy shift acting on the identities of the Hilbert spaces is omitted. The quadratic time-dependency of the phase of the electromagnetic fields~\cite{Marzlin1996} via $k_L\hat{Z}(t)=k_L\hat{Z}-k_Lgt^2/2$ will be compensated through chirping in the following.
\subsection{Adiabatic Elimination}
Next, we wish to reduce the atomic three-level system to an effective two-level system by adiabatic elimination of the ancilla state $\ket{a}$. 
The idea behind it is that if the detuning $\Delta$ is large compared to the coupling frequencies, i.e. the single-photon Rabi frequencies, and the overall detuning $\delta$, the ancilla state gets populated by the electric dipole transition and depopulated by the magnetic dipole transition so fast that the ancilla state is only virtually populated, i.e. the probability of finding the atom in the state $\ket{a}$ is vanishingly small. To see this, we are forcing the atomic three-level system into a form where the ancilla state is separate from the other two by writing the Schrödinger equation as
\begin{align}\label{eq: Schrödinger equation three-level system}
\begin{split}
\mathrm{i}\frac{\mathrm{d}}{\mathrm{d} t}\begin{pmatrix}\ket{\psi_a}\\ \ket{\psi_e}\\ \ket{\psi_g}\end{pmatrix}=\mathrm{i}\frac{\mathrm{d}}{\mathrm{d} t}\begin{pmatrix}\ket{\psi_a}\\ \ket{\vec{\psi}}\end{pmatrix}
=\begin{pmatrix}
\Delta(\hat{\vec P})			&	\vec{\Omega}^\dagger(\hat Z)\\
\vec{\Omega}(\hat Z)	&	\delta(\hat{\vec P})
\end{pmatrix}
\begin{pmatrix}\ket{\psi_a}\\ \ket{\vec{\psi}}\end{pmatrix},\end{split}
\end{align}
where we have collected the excited and ground state into the vector $\ket{\vec{\psi}}$ and defined the detuning operators
\begin{align}
	\Delta	(\hat{\vec P}) = \frac{\hat{\vec{P}}^2}{2M\hbar}+\Delta\;\;\text{and}\;\;
	\delta(\hat{\vec P}) = \left(\begin{array}{cc}
	\frac{\hat{\vec{P}}^2}{2M\hbar}+\delta	&	0\\
	0										&	\frac{\hat{\vec{P}}^2}{2M\hbar}
	\end{array}\right).
\end{align}
Furthermore we defined the transition operator between the ancilla state and the two-level system as
\begin{align}\label{eq: Omega Vektor}
\begin{split}
\vec{\Omega}&(\hat{Z}) = \frac{\mathrm{i}}{\hbar}\left(\begin{array}{c}
\left(\vec{\mu}^*_{ae}\cdot\vec{B}_0\right)\mathrm{e}^{\mathrm{i}k_L\hat{Z}}+\left(\vec{\mu}^*_{ae}\cdot\vec{B}_1\right)\mathrm{e}^{-\mathrm{i}k_L\hat{Z}}\\
\left(\vec{d}^*_{ag}\cdot\vec{E}^*_0\right)\mathrm{e}^{-\mathrm{i}k_L\hat{Z}}+
\left(\vec{d}^*_{ag}\cdot\vec{E}^*_1\right)\mathrm{e}^{\mathrm{i}k_L\hat{Z}}
	\end{array}\right)\\
&-\frac{\mathrm{i}}{\hbar}\left(\begin{array}{c}
\left(\vec{\mu}^*_{ae}\cdot\vec{B}^*_0\right)\mathrm{e}^{-\mathrm{i}k_L\hat{Z}}\mathrm{e}^{2\mathrm{i}\omega t}+\left(\vec{\mu}^*_{ae}\cdot\vec{B}^*_1\right)\mathrm{e}^{\mathrm{i}k_L\hat{Z}}\mathrm{e}^{2\mathrm{i}\omega t}\\
\left(\vec{d}^*_{ag}\cdot\vec{E}_0\right)\mathrm{e}^{\mathrm{i}k_L\hat{Z}}\mathrm{e}^{-2\mathrm{i}\omega t}+\left(\vec{d}^*_{ag}\cdot\vec{E}_1\right)\mathrm{e}^{-\mathrm{i}k_L\hat{Z}}\mathrm{e}^{-2\mathrm{i}\omega t}
\end{array}\right).
\end{split}
\end{align}
The population of the ancilla state $\ket{\psi_a}$ can be expressed in terms of the two-level system $\ket{\vec{\psi}}$ by defining the quasi-projector $\hat{\Pi}$ which projects the two-level system onto the ancilla state via
\begin{align}
    \ket{\psi_a}=\hat{\Pi}\ket{\vec{\psi}}.
\end{align}
Next, we shall derive an explicit expression for $\hat\Pi$. For simplicity, let us assume that this projector does not depend on time, i.e. $\hat{\Pi}\neq \hat{\Pi}(t)$. Note that corrections due to the time dependence will not be present to our order of expansion, see Bott et al.~\cite{Bott2023} for a treatment including weakly time-dependent detunings. We then obtain from Eq.~\eqref{eq: Schrödinger equation three-level system} the differential equation
\begin{align}\label{eq: derivative ancilla ket}
\begin{split}
\mathrm{i}\frac{\mathrm{d}}{\mathrm{d} t}\hat{\Pi}\ket{\vec{\psi}}\approx&\mathrm{i}\hat{\Pi}\frac{\partial}{\partial t}\ket{\vec{\psi}}\\
=&\Delta(\hat{\vec{P}})\ket{\psi_a}+\vec{\Omega}^\dagger(\hat{Z})\ket{\vec{\psi}}=\left(\Delta(\hat{\vec{P}})\hat{\Pi}+\vec{\Omega}^\dagger(\hat{Z})\right)\ket{\vec{\psi}}
\end{split}
\end{align}
for the ancilla state and
\begin{align}\label{eq: derivative vector ket}
\mathrm{i}\frac{\mathrm{d}}{\mathrm{d} t}\ket{\vec{\psi}}=\delta(\hat{\vec{P}})\ket{\vec{\psi}}+\vec{\Omega}(\hat{Z})\ket{\psi_a}=\left(\delta(\hat{\vec{P}})+\vec{\Omega}(\hat{Z})\hat{\Pi}\right)\ket{\vec{\psi}}
\end{align}
for the two-level system. Comparing these two equations -- where the latter has to be multiplied by $\hat{\Pi}$, i.e. projecting the two-level system onto the ancilla state -- leads to the so-called Bloch equation~\cite{Bloch1946,Sanz2015} given by
\begin{align}\label{eq: Bloch equation}
\begin{split}
&\Delta(\hat{\vec{P}})\hat{\Pi}+\vec{\Omega}^\dagger(\hat{Z}) = \hat{\Pi}\delta(\hat{\vec{P}})+\hat{\Pi}\vec{\Omega}(\hat{Z})\hat{\Pi}\\
\Leftrightarrow ~&\hat{\Pi}=\Delta^{-1}(\hat{\vec{P}})\left(-\vec{\Omega}^\dagger(\hat{Z})+\hat{\Pi}\delta(\hat{\vec{P}})+\hat{\Pi}\vec{\Omega}(\hat{Z})\hat{\Pi}\right).
\end{split}
\end{align}
Assuming that the single-photon detuning $\Delta$ is much larger than the coupling frequencies and the overall detuning $\delta$ we can thus define adiabaticity parameters
\begin{align}\label{eq: Adiabaticity parameters}
	\epsilon_\Omega=\frac{\Vert\vec{\Omega}(\hat{Z})\Vert}{\Vert\Delta(\hat{\vec{P}})\Vert}\ll 1\;\;\text{and}\;\;
	\epsilon_\delta=\frac{\Vert\delta(\hat{\vec{P}})\Vert}{\Vert\Delta(\hat{\vec{P}})\Vert} \ll 1,
\end{align}
where
\begin{align}\label{eq: Definition norm}
    \left\Vert\hat{A}\right\Vert=\expval{\hat{A}}{\Psi}/\braket{\Psi}
\end{align}
is the norm of an operator $\hat{A}$ conditioned on the state $\ket\Psi$ of our wave packet.
Solving the Bloch equation, Eq.~\eqref{eq: Bloch equation}, analytically is in most cases intractable and exact solutions are in general not known~\cite{Sanz2015}. Hence, we use a perturbative ansatz
\begin{align}\label{eq: perturbative ansatz}
\hat{\Pi}=\sum_{k=0}\hat{\Pi}_k\;\text{with}\; \hat{\Pi}_k\sim\Delta^{-(k+1)}(\hat{\vec{P}}),
\end{align}
where we expand in powers of the inverse operator-valued detuning $\Delta(\hat{\vec{P}})$, which can be approximated by \mbox{$\Delta^{-1}(\hat{\vec{P}})=\Delta^{-1}\left(1+\hat{\vec{P}}^2/(2M\hbar\Delta)\right)^{-1}\approx\Delta^{-1}$} for sufficiently non-relativistic COM motion, i.e. $\Vert\hat{\vec{P}}^2/(2M\hbar)\Vert\ll\abs{\Delta}$.
The $\hat{\Pi}_k$ can then be determined recursively by
\begin{align}
\begin{split}
\hat{\Pi}_{k+1}=&\Delta^{-1}(\hat{\vec{P}})\hat{\Pi}_k\delta(\hat{\vec{P}})+\Delta^{-1}(\hat{\vec{P}})\sum_{j=0}^{k-1}\hat{\Pi}_{k-j-1}\vec{\Omega}(\hat{Z})\hat{\Pi}_j,\\
\hat{\Pi}_0=&-\Delta^{-1}\vec{\Omega}^\dagger(\hat{Z}),
\end{split}
\end{align}
where $\hat{\Pi}_0$ follows directly from Eq.~\eqref{eq: Bloch equation} since it has to solve the equation to the order $\mathcal{O}(\Delta^{-1}(\hat{\vec{P}}))$. For large detunings $\Delta$ we can truncate this expansion after the first order, i.e., only keeping the lowest order term $\hat{\Pi}_0$. Thus, the slowly evolving dynamics of Eq.~\eqref{eq: derivative vector ket} become an effective two-level transition:
\begin{align}\label{eq: adiabatic system real case}
\mathrm{i}\frac{\mathrm{d}}{\mathrm{d} t}\ket{\vec{\psi}}=\left(\hat{\delta}-\vec{\Omega}(\hat{Z})\Delta^{-1}(\hat{\vec{P}})\vec{\Omega}^\dagger(\hat{Z})\right)\ket{\vec{\psi}}.
\end{align}
Finally, Eq.~\eqref{eq: Omega Vektor} will be inserted into Eq.~\eqref{eq: adiabatic system real case}. The internal states' dynamics contain then position-dependent terms that correspond to two-photon transitions where the atom absorbs two photons from the same direction. These terms lead to unwanted momentum kicks.
Here, the rotating-wave-approximation (RWA) can be applied by neglecting all terms involving $\mathrm{e}^{\pm\mathrm{i}\omega t}$ since the (rapidly) rotating terms average out during the pulse. Note, it is important however that the adiabatic elimination is carried out before the RWA~\cite{Fewell2005}, otherwise important terms of the form $\mathrm{i}\vec{d}_{ag}\cdot\hat{\vec{E}}_i^*$ and $\mathrm{i}\vec{\mu}_{ae}\cdot\hat{\vec{B}}_i$ will be lost. Later, we will see that in a retro-reflective geometry and for the $\sigma^+$- $\sigma^-$ polarization scheme these terms lead to a doubling of the AC Stark shift.

\subsection{Doppler-Free Two-Photon Transitions}\label{sec:doppler-free}
In order to obtain a Doppler-free interaction without momentum kicks, one has to eliminate the position-dependent terms which can be done formally by setting the Rabi frequencies \mbox{$\Omega_{B0}=\Omega_{E1}=0$} (or vice versa). This means, recalling Fig.~\ref{fig:three-level atom} and the field configuration shown there, that the two-photon transition is driven by counter-propagating photons. Practically, this can be done by using a certain polarization scheme suppressing the unwanted single-photon transitions~\cite{Grynberg1983,Alden2014, AldenDiss2014}. To find the right polarization configuration, one has to apply the selection rules of single-photon dipole transitions.  The selection rules for two-photon transitions can then be obtained by interpolating the sequential single-photon transitions.
\subsubsection{Selection Rules and Polarization Scheme}
\label{subsubsec: Selection rules}
In the following, we make use of the well-known dipole selection rules~\cite{Englert2014, GalindoPascualQM1, GalindoPascualQM2, ShoreBook, GardinerZoller1, GardinerZoller2, DalibardCohen-Tannoudji1989}:
\paragraph{\textbf{Electric Dipole Transitions}}
E1 transitions can only take place between two internal states with different parity and the change of angular momentum has to be $\Delta L = \pm 1$.
\paragraph{\textbf{Magnetic Dipole Transitions}}
M1 transitions can only take place between two internal states with the same parity. Therefore, the change of angular momentum has to be $\Delta L = 0$.

However, in both cases (E1 and M1 transitions) the total angular momentum $J=L+S$ has to change via $\Delta J = 0,\pm1$ while transitions from $J=0$ to $J'=0$ are forbidden. Furthermore, conservation of angular momentum leads us to selection rules for the magnetic quantum number $\mathcal{M}$, which changes depending on the polarization of the light: linearly polarized light does not change the magnetic quantum number, i.e. $\Delta\mathcal{M}=0$, while positive (negative) circularly polarized light changes the magnetic quantum number via $\Delta\mathcal{M}=+1$ ($\Delta\mathcal{M}=-1$).
Note that the distinction whether it is positive circular ($\vec{\sigma}^+$) or negative circular ($\vec{\sigma}^-$) depends on the propagation direction and the quantization axis.
Coming back to our setup displayed in Fig.~\ref{fig:three-level atom}, the selection rules for the change of angular momentum $\Delta L$ are fulfilled since between $\ket{g}=^1\!\!S_0$ and $\ket{a}=^3\!\!P_1$ (the E1 transition) we have $\Delta L = 1$ and between $\ket{a}=^3\!\!P_1$ and $\ket{e}=^3\!\!P_0$ (the M1 transition) we have $\Delta L = 0$.

To suppress unwanted transitions, i.e. ensuring that the atom absorbs two counter-propagating photons, we use now a \mbox{$\sigma^+$ - $\sigma^-$} scheme, where the two counter-propagating laser beams have positive (negative) circular polarization, respectively. The above selection rules together with this polarization scheme require $\vec{d}_{ag}\cdot\vec{E}_0\neq 0$ while \mbox{$\vec{d}_{ag}\cdot\vec{E}_1=0$}, and $\vec{\mu}_{ae}\cdot\vec{B}^*_0=0$ while $\vec{\mu}_{ae}\cdot\vec{B}^*_1\neq 0$, given the electric field satisfies $\vec{E}_0\propto\vec{\sigma}^+$ and $\vec{E}_1\propto \vec{\sigma}^-$. Note that if the electric field has positive circular polarization, the corresponding magnetic field has negative circular polarization and vice versa.

In experiments one would typically use a retro-reflective geometry. The circular polarization of the laser beam can then be rotated by a quarter-wave plate. The laser beam traverses the quarter-wave plate twice resulting in an effective half-wave plate. However, the intensity of the two counter-propagating laser beams stays the same, i.e., $\vert \vec{E}_0\vert = \vert \vec{E}_1\vert$ and $\vert \vec{B}_0\vert = \vert \vec{B}_1\vert$. We can then define the single-photon Rabi frequencies
\begin{align}\label{eq: Rabi frequencies}
	\frac{\hbar\Omega_{Ei}}{2}:=-\mathrm{i}\vec{d}_{ag}\cdot\vec{E}_i\;\;\;\text{and}\;\;\;
	\frac{\hbar\Omega_{Bi}}{2}:=-\mathrm{i}\vec{\mu}_{ae}\cdot\vec{B}^*_i,
\end{align}
which describe the corresponding dipole transitions. With the above considerations and using the $\sigma^+$ - $\sigma^-$ polarization scheme, Eq.~\eqref{eq: adiabatic system real case} reduces to (having applied the RWA)
\begin{align}\label{eq: two-level system not position-dependent}
\mathrm{i}\frac{\mathrm{d}}{\mathrm{d} t}\ket{\vec{\psi}}=\left(\begin{array}{cc}
\frac{\hat{\vec{P}}^2}{2M\hbar}+\delta-\frac{1}{2\Delta}\vert\Omega_{B1}\vert^2 & \frac{\Omega}{2} \\
\frac{\Omega^*}{2} & \frac{\hat{\vec{P}}^2}{2M\hbar}-\frac{1}{2\Delta}\vert\Omega_{E0}\vert^2 \\
\end{array} \right)\ket{\vec{\psi}},
\end{align}
where we have defined the two-photon Rabi frequency
\begin{align}
\label{eq: two-photon Rabi frequency}
    \Omega = -\frac{\Omega^*_{B1}\Omega^{\phantom{*}}_{E0}}{2\Delta}.
\end{align}
Since electromagnetic fields have to satisfy Maxwell's equations, the M1 couplings are suppressed by a factor of the inverse of the speed of light $c^{-1}$. Thus, they are much weaker than E1 transitions at typical laser intensities, and the two-photon Rabi frequency, Eq.~\eqref{eq: two-photon Rabi frequency}, is quite small when comparing to the Rabi frequency associated with two E1 transitions. Accordingly, one needs pretty long or relatively intense laser pulses to achieve $\pi$- or $\pi/2$-pulses. That is why finite pulse-time effects become important for E1-M1 transitions. Since the transition between $\ket{g}$ and $\ket{e}$ is forbidden for single-photon transitions, however, we can still neglect spontaneous emission.

Evidently, Eq.~\eqref{eq: two-level system not position-dependent} has no longer any dependence on the atomic COM position. Consequently, there is no effective momentum kick caused by the two-photon transition on the atom. In momentum space, with
\begin{align}
\psi_{e,g}(\vec{P}) = \braket{\vec{P}|\psi_{e,g}},
\end{align}
the dynamics of the effective two-level system is described by
\begin{align}
\mathrm{i}\frac{\partial}{\partial t} \left(\begin{array}{c}\psi_e(\vec{P})\\\psi_g(\vec{P})\end{array}\right) = \frac{1}{2}\left(\begin{array}{cc}
\bar{\gamma}+\gamma	& \Omega\\
\Omega^*		& \bar{\gamma}-\gamma
\end{array}\right) \left(\begin{array}{c}\psi_e(\vec{P})\\\psi_g(\vec{P})
\end{array}\right),
\end{align}
where
\begin{subequations}
	\begin{align}
	\bar{\gamma} 	&= \frac{\vec{P}^2}{M\hbar}+\delta-\frac{|\Omega_{E0}|^2+|\Omega_{B1}|^2}{2\Delta}=
   \frac{\vec{P}^2}{M\hbar}+\delta -\omega_{\mathrm{AC}}^{(+)},\\
	\gamma			&= \delta + \frac{|\Omega_{E0}|^2-|\Omega_{B1}|^2}{2\Delta}=\delta+ \omega_{\mathrm{AC}}^{(-)}
	\end{align}
\end{subequations}
are the mean detuning $\bar{\gamma}$ and relative detuning $\gamma$ as well as \mbox{$\omega_{\mathrm{AC}}^{(+)} = \left(|\Omega_{E0}|^2+|\Omega_{B1}|^2\right)/(2\Delta)$} the mean AC Stark shift and \mbox{$\omega_{\mathrm{AC}}^{(-)} = \left(|\Omega_{E0}|^2-|\Omega_{B1}|^2\right)/(2\Delta)$} the differential AC Stark shift. Note that the relative detuning $\gamma$ does not depend on the COM momentum but on the overall detuning $\delta$ and the AC Stark shift. Thus, the overall detuning can be set in such a way that it compensates the AC Stark shift $\omega_{\mathrm{AC}}^{(-)}$. After going into another interaction picture with respect to the mean detuning $\bar{\gamma}$, the new time evolution operator can be easily obtained by calculating the corresponding matrix exponential such that
\begin{align}
\hat{\tilde{U}}(t) = \cos\frac{\Omega_{\mathrm{eff}}t}{2}\mathds{1} - \frac{\ii}{\Omega_{\mathrm{eff}}} \sin\frac{\Omega_\mathrm{eff}t}{2} \left(\begin{array}{cc}
\gamma	& \Omega\\
\Omega^*		& -\gamma
\end{array}\right),
\end{align}
where we have defined the effective two-photon Rabi frequency \mbox{$\Omega_{\mathrm{eff}} = \sqrt{|\Omega|^2+\gamma^2}$} which depends on the relative detuning $\gamma$. Since the transformations leading to this result are unitary transformations on the diagonal of the Hamiltonian, the transformed states are physically equivalent to the old ones.

\begin{figure}
	\centering
	\includegraphics{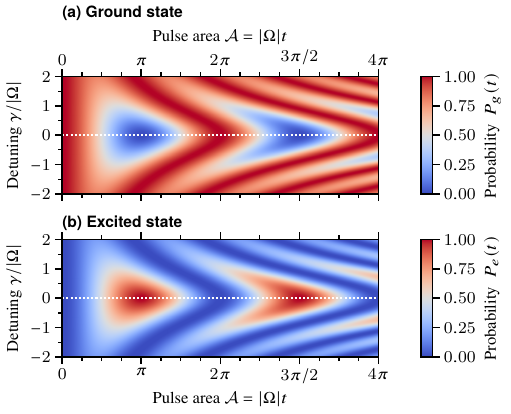}
	\caption{%
    Time evolution of the probability (density) $P_j$ with $j=g,e$ of finding an atom in the ground state {\sffamily\bfseries (a)~} and excited state {\sffamily\bfseries (b)~} during the laser pulse (plane waves) driving the ideal E1-M1 transitions. Initially the atom is assumed to be prepared in the ground state and the probability density is plotted as a function of relative detuning $\gamma$ in units of the Rabi frequency $|\Omega|=\Omega_\text{eff}\vert_{\gamma=0}$ at vanishing detuning. Increasing the relative detuning $\gamma$ increases the frequency of the Rabi oscillations, changes the amplitudes and thus shifts them to a different location in time at fixed relative detuning $\gamma$.}
\label{fig: Rabi oscillation plot}
\end{figure}

Depending on the initial state we observe the well-known Rabi oscillations between the ground and excited state. For instance if the atom is initially in the ground state, the probability to find the atom in the excited or in the ground state at time $t$ is given, respectively, by
\begin{subequations}\label{eq: Rabi oscillations}
	\begin{align}
	P_e(t) &= \left(\frac{|\Omega|}{\Omega_{\mathrm{eff}}}\right)^2
    \sin^2{\frac{\Omega_{\mathrm{eff}}t}{2}},\\
	P_g(t) &= \cos^2{\frac{\Omega_{\mathrm{eff}}t}{2}} 
    + \left(\frac{\gamma}{\Omega_{\mathrm{eff}}}\right)^2\sin^2{\frac{\Omega_{\mathrm{eff}}t}{2}}.
	\end{align}
\end{subequations}
In Fig.~\ref{fig: Rabi oscillation plot} we plot the ground and excited state probabilities for different values of the relative detuning $\gamma$. The highest amplitude is achieved for a vanishing relative detuning, i.e. when the detuning $\delta$ compensates the AC Stark shift. Increasing the relative detuning $\gamma$ leads to a decreasing amplitude and an increasing effective Rabi frequency $\Omega_{\mathrm{eff}}$. For $\gamma>|\Omega|$, it is no longer possible to achieve a 50:50 superposition of excited and ground state. Therefore, we have explicitly shown that taking into account finite pulse-time effects of beams with position-independent intensities -- even with  delocalizable atomic COM clouds --  reproduces Rabi oscillations, connecting to  known results in general contexts, see Weiss et al.~\cite{Weiss1994} where finite-time Raman transitions for COM momentum eigenstates were considered. In the following we will investigate the generalization to arbitrarily structured laser beams.

\section{Finite Pulse-Time effects of Arbitrary Beams}
\label{sec.Finite Pulse-Time effects}
In the idealized scenario of Sec.~\ref{sec.Idealized model for E1M1 transitions} we considered plane waves for the electromagnetic fields. However, a realistic laser beam has a position-dependent intensity, e.g. a Gaussian beam profile. At the same time, the Rabi frequencies from Eq.~\eqref{eq: Rabi frequencies} depend on the amplitudes of the electric and magnetic fields, thereby also  on the intensity. As a consequence, and due to the operator-valued nature of the atomic COM, the atoms might experience small, possibly state-dependent potentials due to the position dependency of the laser intensity while falling during a laser pulse. In particular, small perturbations in the already small magnetic field amplitude might have large effects. In order to find the effective time-evolution operator $\hat{U}_\mathrm{eff}$ for the atomic wave packet in weakly position-dependent pulses we replace
\begin{subequations}\label{eq: Rabi operators position-dependent}
	\begin{align}
	-\frac{\Omega^*_{B1} \Omega_{E0}}{2\Delta}&\rightarrow-\frac{\Omega^*_{B1}(\hat{\vec{R}}) \Omega_{E0}(\hat{\vec{R}})}{2\Delta} =: \Omega(\hat{\vec{R}})\mathrm{e}^{\mathrm{i}\Phi(\hat{\vec{R}})}\label{eq: Operator Omega + Phi},\\
	\frac{\vert\Omega_{E0}\vert^2}{2\Delta}&\rightarrow\frac{\vert\Omega_{E0}(\hat{\vec{R}})\vert^2}{2\Delta} =: \omega_{\mathrm{AC},0}(\hat{\vec{R}})\label{eq: Operator Lambda_0},\\
	\frac{\vert\Omega_{B1}\vert^2}{2\Delta}&\rightarrow\frac{\vert\Omega_{B1}(\hat{\vec{R}})\vert^2}{2\Delta} =: \omega_{\mathrm{AC},1}(\hat{\vec{R}}).\label{eq: Operator Lambda_1}
	\end{align}
\end{subequations}
Adding the atomic rest energy, the Hamiltonian describing the effective two-level atom via Eq.~\eqref{eq: adiabatic system real case} becomes
\begin{align}
\begin{split}
\hat{H} =&\left(\frac{\hat{\vec{P}}^2}{2M} + Mc^2\right)\!\mathds{1} + \hbar\left(\!\begin{array}{cc}
\delta-\omega_{\mathrm{AC},1}(\hat{\vec{R}})& \frac{\Omega(\hat{\vec{R}})}{2}\mathrm{e}^{\mathrm{i}\Phi(\hat{\vec{R}})} \\
\frac{\Omega(\hat{\vec{R}})}{2}\mathrm{e}^{-\mathrm{i}\Phi(\hat{\vec{R}})} & -\omega_{\mathrm{AC},0}(\hat{\vec{R}}) 
\end{array} \!\!\right)\!.
\end{split}
\end{align}
The particular effects of the position dependency of the laser intensity can be separated by unitary transformations that cancel specific operator-valued terms in the Hamiltonian. First of all, let us cancel out the phase in the off-diagonal part in the Hamiltonian. This can be achieved by the unitary displacement transformation $\ket{\vec{\psi}}\rightarrow\hat{U}_1^\dagger\ket{\vec{\psi}}$, where
\begin{align}\label{eq: unitary displacement transformation}
\hat{U}_1^\dagger = \left(\begin{array}{cc} \hat{\mathcal{D}}^\dagger & 0\\
0 & 1\end{array}\right)
\;\;\text{and}\;\;
\hat{\mathcal{D}}^\dagger = \exp\left(-\frac{\mathrm{i}}{\hbar}\left[\vec{\eta}\hat{\vec{R}}-\vec{\xi}\hat{\vec{P}}+\alpha\right]\right)
\end{align}
is the displacement operator. Assuming the wave packet to be, without loss of generality, initially centered around $\vec{R}=0$, the phase in Eq.~\eqref{eq: Operator Omega + Phi} can thus be expanded around the origin in the COM position via
\begin{align}\label{eq: phase operator expansion}
\Phi(\hat{\vec{R}}) = \Phi(0) + \hat{\vec{R}}\cdot \nabla\Phi|_{\vec{R}=0} + \varphi(\hat{\vec{R}}),
\end{align}
where $\varphi(\hat{\vec{R}})=\mathcal{O}(\hat{\vec{R}}^2)$, provided that the spatial extension of the wave packet is small enough compared to characteristic scales of the laser beam, e.g. the beam waist and the Rayleigh length for a Gaussian laser beam, and the laser pulse time is sufficiently small  as the atom is falling, i.e. moving away from the initial position. This phase can then easily be eliminated by choosing the specific transformation parameters
\begin{align}\label{eq: displacement transformation parameters}
	\alpha 		= \hbar \Phi(0),\;\;
	\vec{\xi} 	= 0\;\; \text{and}\;\;
	\vec{\eta} 	= \hbar\nabla\Phi|_{\vec{R}=0}.
\end{align}
Thus, the unitary transformation, Eq.~\eqref{eq: unitary displacement transformation}, corresponds to a small momentum kick
\begin{align}\label{eq: Momentum kick wave vector}
\hbar\vec{k} = \hbar\nabla\Phi|_{\vec{R}=0}.
\end{align}
Consequently, one can identify the recoil frequency $\omega_k$, the Doppler detuning $\nu(\hat{\vec{P}})$ and the (position-dependent) mean AC Stark shift $\omega_{\mathrm{AC}}^{(+)}(\hat{\vec{R}})$ (expanded around the origin) via the definitions
\begin{align}\label{eq: recoil frequency, Doppler detuning and mean AC Stark shift}
\begin{split}
	\omega_k &= \frac{\hbar\vec{k}^2}{2M},\;\;
	\nu(\hat{\vec{P}}) = \frac{\vec{k}\cdot\hat{\vec{P}}}{M},\\
	\text{and}\;\;\omega_{\mathrm{AC}}^{(+)}(\hat{\vec{R}}) &= \omega_{\mathrm{AC},0}(\hat{\vec{R}})+\omega_{\mathrm{AC},1}(\hat{\vec{R}})\\ 
    &= \omega_{\mathrm{AC},0}^{(+)} + \hat{\vec{R}}\cdot\nabla\omega_{\mathrm{AC}}^{(+)}|_{\vec{R}=0} + \mathcal{S}(\hat{\vec{R}}),
\end{split}
\end{align}
where $\mathcal{S}(\hat{\vec{R}})=\mathcal{O}(\hat{\vec{R}}^2)$ is the second (and higher) order part of the expansion of the mean AC Stark shift $\omega_{\mathrm{AC}}^{(+)}(\hat{\vec{R}})$. The transformed Hamiltonian reads then
\begin{align}\label{eq: first transformed Hamiltonian}
\hat{H}' = \left(\begin{array}{cc}
\hat{\bar{H}}+\frac{\Delta(\hat{\vec{R}})}{2}-\frac{\hbar}{2}\mathcal{S}(\hat{\vec{R}})	& \hat{H}_{\mathrm{off}}\\
\hat{H}_{\mathrm{off}}^\dagger					& \hat{\bar{H}}-\frac{\Delta(\hat{\vec{R}})}{2}-\frac{\hbar}{2}\mathcal{S}(\hat{\vec{R}})
\end{array}\right),
\end{align}
where
\begin{align}\label{eq: mean hamiltonian}
\begin{split}
\hat{\bar{H}} =& \frac{\hat{\vec{P}}^2}{2M} + Mc^2 + \frac{\hbar}{2}\Big[\hat{\nu}+\omega_k + \delta\\
& -\left(\omega_{\mathrm{AC},0}^{(+)} + \nabla\omega_{\mathrm{AC}}^{(+)}|_{\vec{R}=0}\hat{\vec{R}}\right) + \dot{\vec{k}}\hat{\vec{R}} +  \dot{\Phi}(0)\Big]
\end{split}
\end{align}
is the mean Hamiltonian,
\begin{align}\label{eq: Detuning operator}
\Delta(\hat{\vec{R}}) = \hbar\left(\hat{\nu} + \omega_k + \delta + \omega_{\mathrm{AC}}^{(-)}(\hat{\vec{R}}) + \dot{\vec{k}}\hat{\vec{R}}+\dot{\Phi}(0)\right)
\end{align}
is the detuning operator and
\begin{align}
\hat{H}_{\mathrm{off}} = \hbar\,\frac{\Omega(\hat{\vec{R}})}{2}\ee^{\mathrm{i}\varphi(\hat{\vec{R}})}
\end{align}
is the off-diagonal part of the Hamiltonian; we denoted partial derivatives in time with a dot. Let us now transform into the interaction picture where we cancel out the dynamics of the mean Hamiltonian $\hat{\bar{H}}$, i.e. a unitary transformation with
\begin{align}
\hat{\bar{U}} = \mathcal{T}\exp\left(-\frac{\mathrm{i}}{\hbar}\int_{0}^{t}\hat{\bar{H}}\;\mathrm{d}t'\right),
\end{align}
where $\mathcal{T}$ is the time ordering operation. Since the Hamiltonian $\hat{H}'$, Eq.~\eqref{eq: first transformed Hamiltonian}, is a function of the COM momentum $\hat{\vec{P}}$ and the COM position $\hat{\vec{R}}$, the remaining operators of the transformed Hamiltonian are evaluated on the Heisenberg trajectories generated by $\hat{\bar{H}}$ via the Heisenberg equations of motion:
\begin{align}\label{eq: Heisenberg equation}
\frac{\mathrm{d}\hat{\vec{R}}_H}{\mathrm{d}t} = \frac{\mathrm{i}}{\hbar}[\hat{\bar{H}},\hat{\vec{R}}_H]\;\;\text{and}\;\;\frac{\mathrm{d}\hat{\vec{P}}_H}{\mathrm{d}t} = \frac{\mathrm{i}}{\hbar}[\hat{\bar{H}},\hat{\vec{P}}_H].
\end{align}
We denote the Heisenberg picture with a subscript $H$ and obtain the new Hamiltonian
\begin{align}
\hat{H}''= \left(\begin{array}{cc}
\frac{\Delta_H(\hat{\vec{R}},t)}{2}-\frac{\hbar}{2} \mathcal{S}_H(\hat{\vec{R}},t)	&	\hat{H}_{\mathrm{off},H}\\
\hat{H}^\dagger_{\mathrm{off},H}						&	-\frac{\Delta_H(\hat{\vec{R}},t)}{2}-\frac{\hbar}{2} \mathcal{S}_H(\hat{\vec{R}},t)
\end{array}\right).
\end{align}
Note that terms of quadratic or higher order in COM position in the mean AC Stark shift $\omega_{\mathrm{AC}}^{(+)}(\hat{\vec{R}})$ are encapsulated by $\mathcal{S}(\hat{\vec{R}})$ so that the mean Hamiltonian $\hat{\bar{H}}$ of Eq.~\eqref{eq: mean hamiltonian} is at most linear in atomic position $\hat{\vec{R}}$. This will facilitate later on the back-transformation to the full unitary time evolution.
From the calculations in Sec.~\ref{sec.Idealized model for E1M1 transitions} we know that in the idealized case without position dependency of the laser amplitude the internal dynamics of the atom is described by Rabi oscillations, see Eq.~\eqref{eq: Rabi oscillations}. These internal transitions can be canceled out of our Hamiltonian by the unitary transformation
\begin{align}
\hat{U}_\Omega = \exp\left(-\mathrm{i}\frac{\Omega(0)}{2} t \hat{\sigma}_x\right),
\end{align}
with  $\hat\sigma_x$ being the Pauli x operator.
After this final unitary transformation we are left with the Hamiltonian
\begin{align}
&\hat{H}_3(t)= \hat{H}_0(t)\,\mathds{1} + \sum_{j=\{x,y,z\}}\hat{H}_j(t)\hat{\sigma}_j\nonumber\\
&= -\frac{\hbar}{2} \mathcal{S}_H(\hat{\vec{R}},t) \mathds{1} + \frac{\hbar}{2}\left[\Omega_H(\hat{\vec{R}},t)\cos\left(\varphi_H(\hat{\vec{R}},t)\right)-\Omega(0)\right]\hat{\sigma}_x\nonumber\\
&\quad+\frac{1}{2} \Big[\Delta_H(\hat{\vec{R}},t)\sin\left(\Omega(0) t\right)\nonumber\\
&\left.\qquad-\hbar\Omega_H(\hat{\vec{R}},t)\sin\left(\varphi_H(\hat{\vec{R}},t)\right)\cos\left(\Omega(0) t\right)\right]\hat{\sigma}_y \nonumber\\
&\quad+\frac{1}{2} \Big[\Delta_H(\hat{\vec{R}},t)\cos\left(\Omega(0) t\right)\nonumber\\
&\left.\qquad+\hbar\Omega_H(\hat{\vec{R}},t)\sin\left(\varphi_H(\hat{\vec{R}},t)\right)\sin\left(\Omega(0) t\right)\right]\hat{\sigma}_z, \label{eq: final Hamiltonian}
\end{align}
where  $\hat\sigma_y$ and  $\hat\sigma_z$
are the remaining Pauli operators. This Hamiltonian can be treated perturbatively to find the effective time evolution operator $\hat{U}_3(t)$ which allows us to provide the full evolution of the system after transforming back to the original picture, i.e.
\begin{align}
    \hat{U}(t) = \hat{D}(t)\hat{U}_1\hat{\bar{U}}(t)\hat{U}_\Omega(t)\hat{U}_3(t)\hat{U}_1^\dagger\hat{D}^\dagger(0),
\end{align}
where we have used $\hat{\bar{U}}^\dagger(0)=\hat{U}_\Omega^\dagger(0)=\mathds{1}$.

\subsection{Example: Fundamental Gaussian Laser Beam}
\label{sec: Fundamental Gaussian Laser Beam}
\begin{figure}
\centering
\includegraphics{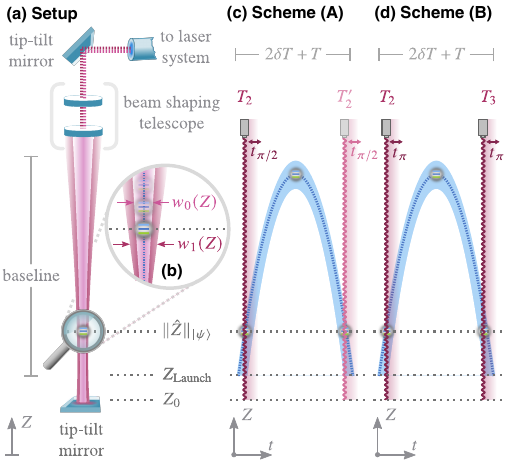}
\caption{%
    Panel {\sffamily\bfseries (a)} Schematic optical setup for long-baseline clock atom interferometry with a double tip-tilt mirror system for rotation and gravity gradient compensation~\cite{Loriani2020,Ufrecht2021}, modeled after the setup described in Glick et al.~\cite{Glick2023} with two in-vacuum tip-tilt mirror systems as well as an in-vacuum beam shaping telescope. Inside the baseline the (simplified) optical field is modeled as a superposition of two Gaussian beams with different $Z$-dependent beam waists $w_i(Z)$ due to imperfectly retro-reflected beams (not to scale). Outside the interferometery region we only indicate the propagating optical beams by dashed lines along the optical axis. Inset {\sffamily\bfseries (d)} Magnified situation close to the location of the atoms at the beginning of the E1-M1 clock pulse when they pass the height $\Vert\hat{Z}\Vert_{\ket{\psi}}$ after being launched at position $Z_{\text{Launch}}$ above the mirror placed at $Z_0$. Panel {\sffamily\bfseries (b/c)} Two possible configurations for the schemes (A) and (B) in a setup which maximize the overall experimental duration and, thus, the signal of interest via fountain geometries. Explicit interferometer paths are not shown, only the central trajectories (dashed light blue) and their envelope. For scheme (A) the clock initialization pulse is placed alternatingly at the beginning or end of the central element of the interferometer sequence in subsequent runs. For scheme (B) the clock pulses are performed in each run at the beginning and end of the central element of the interferometer sequence.
}
\label{fig:beamshape-plot}
\end{figure}

We now continue with the simplest example for a position-dependent laser intensity profile: the Gaussian laser beam. In our case we assume that the superposition of incoming and the retro-reflected beams have slightly different beam waist sizes $w_{0,i}$. Accordingly, they also have slightly different Rayleigh lengths $Z_{R,i}$ while the distance of the atoms to the co-located beam waists is $Z_0$. In Fig.~\ref{fig:beamshape-plot} we show in {\sffamily\bfseries (a)} an optical setup for this case as well as in {\sffamily\bfseries (c/d)} the corresponding interferometer sequences with the optimal placement of the E1-M1 clock transitions in them. In the following we neglect beam distortion effects, assuming here a negligible influence of random phase and intensity noise. As shown by Bade et al.~\cite{Bade2018}, there are, however, correlations of phase and intensity noise which may prevent the averaging out even for many iterations of the interferometric experiments, and as such a proper inclusion of the distortion effects is left as necessary future work.
Introducing the radial position operator
\begin{align}
\hat{\varrho} = \sqrt{\hat{X}^2 + \hat{Y}^2}
\end{align}
and assuming the atoms to be falling along the optical axis of the laser beam, i.e. the $Z$-axis, we can write the Gaussian~\cite{Meschede2015} electromagnetic field in cylindrical coordinates. Thus~\footnote{Note, that the plane wave factor in $Z$-direction is already considered in the equations of the previous section.}
\begin{subequations}\label{eq: Gaussian beam}
	\begin{align}
\vec{E}_0(\hat{\vec{R}}) &=  \frac{\vec{E}_0 w_{0,0}}{w_0(\hat{\tilde{Z}})} \exp\left[-\frac{\hat{\varrho}^2}{w_0^2(\hat{\tilde{Z}})}-\mathrm{i}k_L\frac{\hat{\varrho}^2}{2R_0(\hat{\tilde{Z}})}+\mathrm{i}\zeta_0(\hat{\tilde{Z}})\right],\\
\vec{B}_1(\hat{\vec{R}}) &= \frac{\vec{B}_1  w_{0,1}}{w_1(\hat{\tilde{Z}})} \exp\left[-\frac{\hat{\varrho}^2}{w_1^2(\hat{\tilde{Z}})}+\mathrm{i}k_L\frac{\hat{\varrho}^2}{2R_1(\hat{\tilde{Z}})}-\mathrm{i}\zeta_1(\hat{\tilde{Z}})\right],
	\end{align}
\end{subequations}
where
\begin{align}
    \hat{\tilde{Z}} &= \hat{Z} - Z_0
\end{align}
describes the distance of the atom to the beam waist,
\begin{align}
\begin{split}
w_i^{-1}(\hat{Z}) &= \left[w_{0,i} \sqrt{1+\left(\frac{\hat{Z}}{Z_{R,i}}\right)^2}\right]^{-1}
\end{split}
\end{align}
is the inverse of the spot size parameter,
\begin{align}
\begin{split}
R_i^{-1}(\hat{Z}) &= \hat{Z}^{-1}\left[1+\left(\frac{Z_{R,i}}{\hat{Z}}\right)^{2}\right]^{-1}
\end{split}
\end{align}
is the inverse of the radius of curvature and
\begin{align}
\zeta_i(\hat{Z}) = \arctan\frac{\hat{Z}}{Z_{R,i}}
\end{align}
is the Gouy phase.
Furthermore, we can describe the corresponding Rayleigh lengths and beam waists by its mean and differential values via
\begin{align}
    Z_{R,i} &= Z_R \pm \frac{\delta Z_R}{2},\\
    w_{0,i} &= w_0 \pm \frac{\delta w_0}{2}.
\end{align}
In Table~\ref{table: Order of magnitudes} we provide some order of magnitude values for the possible realistic length scales.
Scaling all $Z$-components by $Z_R$ and all radial components by $w_0$, i.e. the introduction of dimensionless operators \mbox{$\hat{\tilde{\mathcal{Z}}}=\hat{\tilde{Z}}/Z_R$} and $\hat{\rho}=\hat{\varrho}/w_0$ as well as, $\mathcal{Z}_0=Z_0/Z_R$, $\delta \mathcal{Z}_R = \delta Z_R/Z_R$ and $\delta \mathscr{w}_0 = \delta w_0/w_0$, we can use the definitions of the single-photon Rabi frequencies, Eq.~\eqref{eq: Rabi frequencies}, and determine  the operators that are present in the final Hamiltonian, Eq.~\eqref{eq: final Hamiltonian}, via their definitions, Eqs.~\eqref{eq: Rabi operators position-dependent},~\eqref{eq: recoil frequency, Doppler detuning and mean AC Stark shift} and~\eqref{eq: Detuning operator}, expanded to the second order in these scaled parameters in terms of the Heisenberg trajectories of $\hat{\bar{H}}$:
\begin{subequations}\label{eq: Operators up to second order in space}
	\begin{align}
	\Omega_H(\hat{\vec{R}},t) &\approx \Omega_0\left[1-\hat{\tilde{\mathcal{Z}}}_H(t)^2-2\hat{\rho}^2_H(t)\right],\\
	\label{eq: Rabi Frequency Phase}
    \Phi_H(\hat{\vec{R}},t) &\approx -\delta\mathcal{Z}_R\mathcal{Z}_0 + \delta\mathcal{Z}_R\hat{\tilde{\mathcal{Z}}}_H(t)\\
	\begin{split}
	\Delta_H(\hat{\vec{R}},t) &\approx \hbar \left[\nu(\hat{\vec{P}}_H(t))+\omega_k+\delta\right.\\
 &\left. \quad+\omega_{\mathrm{AC},0}^{(-)}\left(1-\hat{\tilde{\mathcal{Z}}}_H(t)^2-2\;\rho^2_H(t)\right)\right],
	\end{split}\\
	\mathcal{S}_H(\hat{\vec{R}},t) &\approx -\omega_{\mathrm{AC},0}^{(+)}\left(\hat{\mathcal{Z}}^2_H(t)+2\;\rho^2_H(t)\right),
	\end{align}
\end{subequations}
where we used $k_L = 2\pi/\lambda$ and $Z_R = \pi w_0^2/\lambda$ and defined \mbox{$\Omega_0 = \vert\Omega_E\vert\,\vert\Omega_B\vert/(2\Delta)$} as well as \mbox{$\omega_{\mathrm{AC},0}^{(\pm)} = (|\Omega_E|^2\pm|\Omega_B|^2)/(2\Delta)$.} Note that we have defined the position-independent Rabi frequencies $\Omega_E$ and $\Omega_B$ in total analogy to Eq.~\eqref{eq: Rabi frequencies}. Further note that the extension to, e.g., Gillot et al.~\cite{Gillot2016} becomes quite clear in our example: quantization of the atomic COM and structured beam shapes not only change, to leading order, the detuning but also the Rabi frequency (in this case the phase of the two-photon Rabi frequency, cf. Eq.~\eqref{eq: Rabi Frequency Phase}). Recalling Eqs.~\eqref{eq: Rabi frequencies},~\eqref{eq: Rabi operators position-dependent} and~\eqref{eq: Momentum kick wave vector}, the effective kick due to the Gaussian beam shape is then given by
\begin{align}\label{eq: k-vector}
\hbar\vec{k} = \hbar\nabla\Phi|_{\vec{R}=0} = \hbar\frac{\delta Z_R}{Z_R^2}\vec{e}_Z.
\end{align}
\begin{table}
	\caption{Order of magnitudes of the relevant physical quantities of the Rabi problem with COM used for the approximations in this study.}\label{table: Order of magnitudes}
	\centering
	\renewcommand{\arraystretch}{1.3}
	\begin{tabular}{ccl}
		\toprule
		 {\sffamily\bfseries Symbol}	    &    {\sffamily\bfseries Description}      &   {\sffamily\bfseries Order of magnitude}\\ 
        \midrule
		$w_0$		                     &	Beam waist	                            &   $\sim10^{-2}~\mathrm{m}$~\cite{Glick2023}\\ 
		$Z_R$		                     &	Rayleigh length                         &   $\sim10^{3}~\mathrm{m}$\\
        $\delta Z_0$                     &  Distance of interferometer arms         &   $\sim10^{-1}~\mathrm{m}$~\cite{Abend2023}\\
        $\delta w_0$                     &  Change of beam waist                    &   $\sim10^{-4}~\mathrm{m}$\\
        $\delta Z_R$                     &  Change of Rayleigh length               &   $\sim10~\mathrm{m}$\\
		\bottomrule
	\end{tabular}
	\renewcommand{\arraystretch}{1}
\end{table}
Assuming now that the atoms are near the beam waist (on the optical axis), i.e. $\Vert\hat{\rho}\Vert_{\ket{\psi}} \ll \Vert\hat{\tilde{\mathcal{Z}}}\Vert_{\ket{\psi}} \ll \delta\mathcal{Z}_R$, see Table~\ref{table: Order of magnitudes} and Fig.~\ref{fig:beamshape-plot}, the dominant effect of the beam profile originates from the change of the Rayleigh length $\delta\mathcal{Z}_R$. We can thus neglect all other effects so that the Hamiltonian $\hat{H}_3$ of Eq.~\eqref{eq: final Hamiltonian} reads
\begin{align}\label{eq: final Hamiltonian Gaussian beam}
    \hat{H}_3 = \frac{\hbar}{2}\nu(\hat{\vec{P}}_H(t))\left(\sin(\Omega(0)t)\hat{\sigma}_y+\cos(\Omega(0)t)\hat{\sigma}_z\right).
\end{align}
Furthermore, we chose the overall detuning $\delta=-\omega_k-\omega_\mathrm{AC}^{(-)}(0)$ to compensate the differential AC Stark shift $\omega_\mathrm{AC}^{(-)}(0)$ and the recoil frequency $\omega_k$. The Heisenberg trajectories for the mean Hamiltonian Eq.~\eqref{eq: mean hamiltonian} can be calculated via the Heisenberg equations of motion, Eq.~\eqref{eq: Heisenberg equation},
which leads to
	\begin{align}
	\hat{\vec{R}}_H(t) &= \left(\frac{\hat{\vec{P}}}{M}+\frac{\hbar\vec{k}}{2M}\right)t+\hat{\vec{R}},\quad
	\hat{\vec{P}}_H(t) =  \hat{\vec{P}}. \label{eq: Heisenberg trajectories}
	\end{align}
Note, that the Heisenberg momentum operator $\vec{P}_H(t)$ is constant in time, so that we can neglect the subscript $H$ in Eq.~\eqref{eq: final Hamiltonian Gaussian beam}. Since we do not go beyond the first order of $\delta\mathcal{Z}_R$, the Hamiltonian of Eq.~\eqref{eq: final Hamiltonian Gaussian beam} is (quasi-)commuting at different times. Calculating the time-evolution operator,
\begin{align}\label{eq: Dyson series}
\hat{U}_3(t) = \mathcal{T}\exp\left(-\frac{\mathrm{i}}{\hbar}\int_{0}^{t}\hat{H}_3(t')\;\mathrm{d}t'\right),
\end{align}
the time-ordering operation can thus be ignored. Instead, we can directly compute the integral in the exponent. Introducing further the dimensionless time $\tau=\Omega(0) t$, we obtain the time-evolution operator
\begin{align}
\begin{split}
&\hat{U}_3(\tau) = \exp\left\{-\frac{\mathrm{i}\nu(\hat{\vec{P}})}{2\Omega(0)}\left[\left(1-\cos(\tau)\right)\hat{\sigma}_y+\sin(\tau)\hat{\sigma}_z\right]\right\}\\
&=\cos\left[\frac{\nu(\hat{\vec{P}})}{\Omega(0)}\sin\left(\frac{\tau}{2}\right)\right]\mathds{1}-\mathrm{i}\sin\left[\frac{\nu(\hat{\vec{P}})}{\Omega(0)}\sin\left(\frac{\tau}{2}\right)\right]\\
&\quad\times\left(\sin\left(\frac{\tau}{2}\right)\hat{\sigma}_y+\cos\left(\frac{\tau}{2}\right)\hat{\sigma}_z\right).
\end{split}
\end{align}
To consider now all finite pulse-time effects for a Gaussian laser beam, the unitary transformations done in this section as well as the displacement transformation, Eq.~\eqref{eq: Displacement gravitation}, have to be reversed. Doing this subsequently and using \mbox{$\hat{\bar{U}}^\dagger(0)=\hat{U}_\Omega^\dagger(0)=\mathds{1}$}, we end up with
\begin{align}
\hat{U}(\tau)=\hat{D}(\tau)\hat{U}_1\hat{\bar{U}}(\tau)\hat{U}_\Omega(\tau)\hat{U}_3(\tau)\hat{U}_1^\dagger\hat{D}^\dagger(0)
\end{align}
being the total time evolution in the initial picture, cf. Eq.~\eqref{eq: adiabatic system real case}. Inserting $\tau_\pi=\pi$ ($\tau_{\pi/2}=\pi/2$) for the duration of a $\pi$-pulse ($\pi/2$-pulse) we obtain the generalized $\pi$-pulse and $\pi/2$-pulse operators:
\begin{align}\label{eq: Pi and Pi-half pulse operators}
	\hat{U}_{\pi} = \left(\begin{array}{cc}
	\hat{U}_{\pi,ee}	&	\hat{U}_{\pi,ge}\\
	\hat{U}_{\pi,eg}	&	\hat{U}_{\pi,gg}
	\end{array}\right)\;\;\text{and}\;\;
	\hat{U}_{\frac{\pi}{2}} = \left(\begin{array}{cc}
	\hat{U}_{\frac{\pi}{2},ee}	&	\hat{U}_{\frac{\pi}{2},ge}\\
	\hat{U}_{\frac{\pi}{2},eg}	&	\hat{U}_{\frac{\pi}{2},gg}
	\end{array}\right)
\end{align}
with
\begin{subequations}
\label{eq: Pi pulse operator}
	\begin{align}
	\hat{U}_{\pi,ee} &=- \mathrm{i}\hat{D}(\tau)\hat{\mathcal{D}}\hat{\bar{U}}\left(\pi\right)\sin\left(\frac{\nu(\hat{\vec{P}})}{\Omega(0)}\right)\hat{\mathcal{D}}^\dagger\hat{D}^\dagger(0),\\
	\hat{U}_{\pi,ge} &= -\mathrm{i}\hat{D}(\tau)\hat{\mathcal{D}}\hat{\bar{U}}\left(\pi\right)\cos\left(\frac{\nu(\hat{\vec{P}})}{\Omega(0)}\right)\hat{D}^\dagger(0),\\
	\hat{U}_{\pi,eg} &= -\mathrm{i}\hat{D}(\tau)\hat{\bar{U}}\left(\pi\right)\cos\left(\frac{\nu(\hat{\vec{P}})}{\Omega(0)}\right)\hat{\mathcal{D}}^\dagger\hat{D}^\dagger(0),\\
	\hat{U}_{\pi,gg} &= +\mathrm{i}\hat{D}(\tau)\hat{\bar{U}}\left(\pi\right)\sin\left(\frac{\nu(\hat{\vec{P}})}{\Omega(0)}\right)\hat{D}^\dagger(0),
	\end{align}
\end{subequations}
and
\begin{subequations}
\label{eq: Pi/2 pulse operator}
	\begin{align}
 \begin{split}
	\hat{U}_{\frac{\pi}{2},ee} &= \frac{1}{\sqrt{2}}\hat{D}(\tau)\hat{\mathcal{D}}\hat{\bar{U}}\left(\frac{\pi}{2}\right)\bigg[\cos\left(\frac{\nu(\hat{\vec{P}})}{\sqrt{2}\Omega(0)}\right)\\
 &\quad-\sqrt{2}\ii\sin\left(\frac{\nu(\hat{\vec{P}})}{\sqrt{2}\Omega(0)}\right)\bigg]\hat{\mathcal{D}}^\dagger\hat{D}^\dagger(0),
 \end{split}\\
	\hat{U}_{\frac{\pi}{2},ge} &= -\frac{\ii\hat{D}(\tau)\hat{\mathcal{D}}\hat{\bar{U}}\left(\frac{\pi}{2}\right)\cos\left(\frac{\nu(\hat{\vec{P}})}{\sqrt{2}\Omega(0)}\right)\hat{D}^\dagger(0)}{\sqrt{2}},\\
	\hat{U}_{\frac{\pi}{2},eg} &= -\frac{\ii\hat{D}(\tau)\hat{\bar{U}}\left(\frac{\pi}{2}\right)\cos\left(\frac{\nu(\hat{\vec{P}})}{\sqrt{2}\Omega(0)}\right)\hat{\mathcal{D}}^\dagger\hat{D}^\dagger(0)}{\sqrt{2}},\\
 \begin{split}
	\hat{U}_{\frac{\pi}{2},gg} &= \frac{1}{\sqrt{2}}\hat{D}(\tau)\hat{\bar{U}}\left(\frac{\pi}{2}\right)\bigg[\cos\left(\frac{\nu(\hat{\vec{P}})}{\sqrt{2}\Omega(0)}\right)\\
 &\quad+\sqrt{2}\ii\sin\left(\frac{\nu(\hat{\vec{P}})}{\sqrt{2}\Omega(0)}\right)\bigg]\hat{D}^\dagger(0).
 \end{split}
	\end{align}
\end{subequations}
Recall the displacement operators $\hat{D}(t)$, Eq.~\eqref{eq: Displacement gravitation}, and $\hat{\mathcal{D}}$, Eq.~\eqref{eq: unitary displacement transformation}, and the time evolution operator $\hat{\bar{U}}$ corresponding to the mean Hamiltonian $\hat{\bar{H}}$, Eq.~\eqref{eq: mean hamiltonian}. In the limit $Z_R\to \infty$ (accordingly also $w_0\to \infty$ due to the relation $Z_R=\pi w_0^2/\lambda$, where $\lambda$ is the wavelength of the laser beam, and \mbox{$\vec{k}=\delta Z_R\vec{e}_Z/Z_R^2\to 0$}) they reduce to the well-known ideal $\pi$ and $\frac{\pi}{2}$-pulse operators, respectively,
	\begin{align}
	\hat{U}_{\pi,\mathrm{ideal}} &=\left(\begin{array}{cc}
	0			&	-\mathrm{i}\\
	-\mathrm{i}	&	0
	\end{array}\right), \quad
	\hat{U}_{\frac{\pi}{2},\mathrm{ideal}} =\frac{1}{\sqrt{2}}\left(\begin{array}{cc}
	1			&	-\mathrm{i}\\
	-\mathrm{i}	&	1
	\end{array}\right), \label{eq: Pi and Pi/2 pulse operator ideal}
	\end{align}
i.e. the plane wave solution of the previous section where the COM momentum $\hat{\vec{P}}$ has no effect on the E1-M1 transitions.
\subsubsection{Discussion of the Generalized $\pi$ and $\pi/2$-Pulse Operators}
Comparing the generalized $\pi$ and $\pi/2$-pulse operators, Eqs.~\eqref{eq: Pi pulse operator} and ~\eqref{eq: Pi/2 pulse operator}, with the ideal operators, Eq.~\eqref{eq: Pi and Pi/2 pulse operator ideal}, one can observe some additional effects due to the finite pulse time and the position dependency of the intensity of a Gaussian laser beam:
\paragraph{Additional momentum kicks and branch-dependent phase}
The displacement operators $\hat{\mathcal{D}}$ and $\hat{\mathcal{D}}^\dagger$ correspond to a transfer of momentum \mbox{$\pm\hbar\vec{k}=\pm\hbar \delta Z_R\vec{e}_Z/Z_R^2$} and to an imprinting of a branch-dependent phase $\Phi(Z_0)=\mp\delta Z_R Z_0/Z_R^2$, i.e. the transitions from ground to excited state and vice versa are accompanied by small additional momentum kicks and phase shifts. Note, that although there are displacement operators present in the terms $\hat{U}_{\pi,ee}$ and $\hat{U}_{\frac{\pi}{2},ee}$, i.e. the atoms remaining in the excited state during the laser pulse, the momentum of the atom is identical before and after the pulse. There is only a momentum shift occuring during the interaction with the laser.
\paragraph{Action of $\hat{\bar{U}}(\tau)$}
The time-evolution operator 
\begin{align}\label{eq: Uquer}
\hat{\bar{U}}(\tau) = \exp\!\left[\frac{-\mathrm{i}\tau}{\hbar\Omega(0)}\left(\frac{\hat{\vec{P}}^2}{2M}+\frac{\hbar\nu(\hat{\vec{P}})}{2}+Mc^2-\frac{\hbar\vert\Omega_E\vert^2}{2\Delta}\right)\!\right]
\end{align}
associated with the mean Hamiltonian, Eq.~\eqref{eq: mean hamiltonian}, corresponds to a displacement $-\hbar\tau\delta Z_R/(MZ_R^2\Omega(0))$ in $Z$-direction (recall Eqs.~\eqref{eq: recoil frequency, Doppler detuning and mean AC Stark shift} and~\eqref{eq: k-vector})
and a laser phase. Recall that we chose the overall detuning $\delta=-\omega_k-\omega_\mathrm{AC}^{(-)}(0)$ so that the recoil frequency $\omega_k$ and the part of the mean AC Stark shift $\omega_{\mathrm{AC}}^{(+)}(0)$ corresponding to the M1 transition in zeroth order is compensated in the mean Hamiltonian. Furthermore, the first order of the mean AC Stark shift vanishes for the $\text{TEM}_{00}$ laser mode.
\paragraph{Additional branches}
The $\pi$ and $\pi/2$-pulses further contain operators of the form $\sin(\xi\hat{P}_z/\hbar)$ and $\cos(\xi\hat{P}_z/\hbar)$, i.e. a splitting of the branches in opposite directions. Moreover, we see that the $\pi$ and $\pi/2$-pulse operators do not transform all the atoms to the appropriate internal state (in contrast to the ideal case). Both, the $\pi$- and $\pi/2$-pulses lead therefore to a splitting into four branches.

\begin{table}
	\caption{Electric (E1) and magnetic (M1) dipole matrix elements, wavelength for the excitation scheme $\lambda$ and E1-M1 two-photon Rabi frequencies for typical atoms used in atom interferometry. The laser intensity is $6\times 10^6~\mathrm{W}/\mathrm{m^2}$.\cite{Alden2014,AldenDiss2014}}\label{table: Rabi frequencies}
	\centering
	\renewcommand{\arraystretch}{1.3}
	\begin{tabular}{lllllc}
		\toprule
		 {\sffamily\bfseries Atom}    &   {\sffamily\bfseries E1/$ea_0$} &   {\sffamily\bfseries M1/$\mu_B$}  &   {\sffamily\bfseries $\boldsymbol{\lambda}\;(\mathrm{nm})$}  &   {\sffamily\bfseries $\boldsymbol{\Omega}~(\mathrm{Hz})$} & {$t_{\pi/2}=\pi/(2\Omega) \;(\mathrm{ms})$} \\ 
        \midrule
		Yb    &   $0.54$~\cite{Beloy2012}  &   $\sqrt{2}$~\cite{Beloy2012}    &   $1157$~\cite{Hong2005}         &   $150$~\cite{Alden2014,AldenDiss2014} &  $10.47$ \\ 
		Sr    &   $0.15$~\cite{Porsev2001} &   $\sqrt{2}$~\cite{Curtis2001}   &   $1397$~\cite{Ludlow2006}        &   $52.8$~\cite{Alden2014,AldenDiss2014} & $29.75$ \\
		\bottomrule
	\end{tabular}
	\renewcommand{\arraystretch}{1}
\end{table}
However, the order of magnitude of the displacements in space during the pulse time, i.e. $\mathcal{O}(\hbar\delta Z_R/(Z_R^2 M \Omega(0)))$, is much smaller than the displacement due to the momentum kick which is of the order $\mathcal{O}(\hbar\delta Z_R T/(M Z_R^2))$, since $\Omega(0) T \gg 1$ (cf. Table~\ref{table: Rabi frequencies}), and where $T$ is a characteristic time of the interferometer sequence, e.g. $T=T_4-T_2$ in Fig.~\ref{fig: Albert Interferometer modified} or $T=T_3-T_2$ in Fig.~\ref{fig: Christian Interferometer}. Therefore we will neglect branch splitting from now on and continue with the $\pi$- and $\pi/2$-pulse operators given by
\begin{subequations}
\label{eq: Pi and Pi half pulses without splitting}
	\begin{align}
\hat{U}_\pi &= -\ii\hat{D}(\tau)\left(\begin{array}{cc}
0	&	\hat{\mathcal{D}}\hat{\bar{U}}'\left(\pi\right)\\
\hat{\bar{U}}'\left(\pi\right)\hat{\mathcal{D}}^\dagger	&	0
\end{array}\right)\hat{D}^\dagger(0),\\
\hat{U}_\frac{\pi}{2} &=\frac{1}{\sqrt{2}} \hat{D}(\tau)\left(\begin{array}{cc}
\hat{\bar{U}}'\left(\frac{\pi}{2}\right)	&	-\mathrm{i}\hat{\mathcal{D}}\hat{\bar{U}}'\left(\frac{\pi}{2}\right)\\
-\mathrm{i}\hat{\bar{U}}'\left(\frac{\pi}{2}\right)\hat{\mathcal{D}}^\dagger	&	\hat{\bar{U}}'\left(\frac{\pi}{2}\right)
\end{array}\right)\hat{D}^\dagger(0),
	\end{align}
\end{subequations}
where
\begin{align}\label{eq: Uquer ohne displacement}
\hat{\bar{U}}'(\tau) = \exp\left[-\frac{\mathrm{i}\tau}{\hbar\Omega(0)}\left(Mc^2+\frac{\hat{\vec{P}}^2}{2M}-\hbar\frac{\vert\Omega_E\vert^2}{2\Delta}\right)\right]
\end{align}
is the time-evolution due to the mean Hamiltonian Eq.~\eqref{eq: mean hamiltonian} without the term inducing spatial translations since
\begin{align}
\left\Vert\frac{\delta Z_R\hat{P}_z}{MZ_R^2\Omega(0)}\right\Vert\ll 1.
\end{align}
\section{Effects on the interferometer phase from additional momentum kicks}
\label{sec. Effects in interferometry}
In this section we investigate the main finite pulse-time effects for E1-M1 transitions, namely the falling of the atom during the laser pulse and the additional momentum kicks, for UGR and UFF tests using the interferometer schemes (A)~\cite{Roura2020} and (B)~\cite{Ufrecht2020} (recall Sec.~\ref{sec: Interferometer Albert und Christian}). We assume ideal momentum kick operators 
\begin{align}
\hat{\mathcal{D}}_p=\frac{1}{\sqrt{2}}\ee^{\ii k_p \hat{Z}}
\end{align}
for the (magic) Bragg pulses (which should not be confused with the recoils due to the E1-M1 pulses), i.e. momentum transfer pulses that do not change the internal state, as well as for the $\pi$ and $\pi/2$-pulses, see Eq.~\eqref{eq: Pi and Pi half pulses without splitting} in Sec.~\ref{sec.Finite Pulse-Time effects}. The evolution of the atom in the gravitational field from time $T_k$ to $T_i$ in between laser pulses in its ground/excited state can be described via~\cite{Loriani2019, Roura2020, Ufrecht2020}
\begin{align}
\hat{U}(T_i,T_k) = \sum_{\mathclap{n=g,e}}\ee^{-\frac{\mathrm{i}}{\hbar}\hat{H}^{\text{(ME)}}_{n}\left(T_i-T_k\right)}\dyad{n}{n}= \sum_{\mathclap{n=g,e}} \hat{U}_n(T_i,T_k)\dyad{n}{n}.
\end{align}
Since this Hamiltonian is diagonal in the internal states, calculating the evolution during the free fall of the atoms is particularly easy, and reduces to finding the time evolution operators $\hat{U}_n(T_i,T_k)$ corresponding to the Hamiltonian \mbox{$\hat{H}_n^{(\text{ME})}=H(\hat{\vec{R}},\hat{\vec{P}};M_n)$}. On the other hand, one could also use the mass defect representation of the Hamiltonian $\hat{H}^{(\text{MD})}$ and calculate the evolution via \mbox{$\hat{U}(T_i,T_k)=\ee^{-\frac{\ii}{\hbar}\hat{H}^{(\text{MD})}(T_i-T_k)}$} which is however much more complicated than rewriting the previous result.

We assume further that the initial COM state $\ket{\psi_0}$ of the atom corresponds to a $L^2$-normalized Gaussian wave packet 
\begin{align}\label{eq: Gaussian wave packet}
\psi_0(\vec{P}) 
=\frac{\exp\left({-\frac{1}{4}(\vec{P}-\vec{P}_0)^T \vec{\sigma}^{-1}(\vec{P}-\vec{P}_0)}\right)}{(2 \pi)^{3/4} \operatorname{det}^{1/4}\vec{\sigma}},
\end{align}
with covariance matrix  $\vec\sigma=\text{diag}((\Delta p_x)^2,(\Delta p_y)^2,(\Delta p_z)^2)$ and mean momentum $\vec{P}_0$. We can thus describe the full initial atomic state by a product state
\begin{align}
\ket{\Psi_0} = \ket{\phi_\mathrm{i}}\otimes\ket{\psi_0},
\end{align}
where $\ket{\phi_\mathrm{i}}$ is the initial internal state of the atom. Using the exit port projection operator
\begin{align}\label{eq: exit port projection operator}
    \hat{\Pi}_\text{exit}^{} = \dyad{\phi_\mathrm{f}}\otimes \hat{\Pi}_\text{exit}^{\text{(COM)}},
\end{align}
the measured intensity in the respective exit port (excited or ground state) is then described by
\begin{align}
    I_{\phi_\mathrm{f}} = \matrixel{\psi_0}{\left[\matrixel{\phi_\mathrm{f}}{\hat{\Pi}_\text{exit}^{(\text{COM})}\hat{U}_\mathrm{tot}}{\phi_\mathrm{i}}\right]^\dagger\matrixel{\phi_\mathrm{f}}{\hat{\Pi}_\text{exit}^{(\text{COM})}\hat{U}_\mathrm{tot}}{\phi_\mathrm{i}}}{\psi_0},
\end{align}
where
\begin{align}
    \matrixel{\phi_\mathrm{f}}{\hat\Pi_\text{exit}^{(\text{COM})}\hat{U}_\mathrm{tot}}{\phi_\mathrm{i}}=\hat{U}_{l,\phi_\mathrm{i}\phi_\mathrm{f}}+\hat{U}_{u,\phi_\mathrm{i}\phi_\mathrm{f}}
\end{align}
is the sum of the evolution operators along the lower and the upper branches leading to the exit port of the interferometer characterized by the internal state $\ket{\phi_\mathrm{f}}$ and the projector $\hat{\Pi}_\text{exit}^{(\text{COM})}$ on the COM degrees of freedom corresponding to the exit port.

\subsection{UGR Tests Using Superpositions of Internal States}
In the interferometer scheme (A)~\cite{Roura2020} the atoms entering the interferometer in the ground state are divided into two branches, and in the middle segment a Doppler-free $\pi/2$-pulse is applied simultaneously on both branches to get a 50:50 superposition of excited and ground state atoms, i.e. the initialization of an atomic clock.
\begin{figure}[h]
	\centering
	\includegraphics{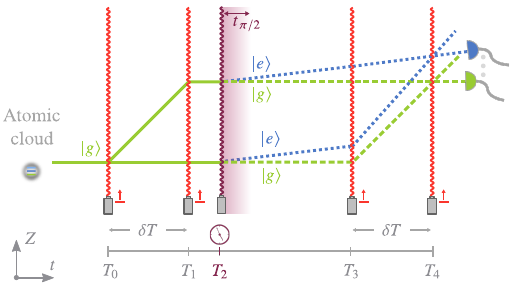}
	\caption{Interferometer scheme (A) of Roura~\cite{Roura2020} in the freely falling frame modified by finite pulse-time effects during the E1-M1 $\frac{\pi}{2}$-pulse. Initially the atomic wave packet is prepared in the ground state $\ket{g}$ (green) before being diffracted by (magic) Bragg pulses at times $T_0,~T_1,~T_3$ and $T_4$.
    During each of the Bragg diffraction light pulses (red) the momentum of the atoms changes only for parts of the atoms. 
    At all but the first Bragg pulse some of the atoms are not diffracted along the branches of the interferometer and hence do not propagate into the exit port. We do not illustrate these additional paths for clarity. At time $T_2$ an E1-M1 pulse of duration $t_{\frac{\pi}{2}}$ at time $T_2$ (violet pulse and shading) initialized a delocalized supersposition of ground (green) and excited state (blue) on both branches. Additional momentum kicks $\hbar\vec{k}$ originating from the optical potentials acting during the transition at time $T_2$ lead to an additional spatial delocalization of the clock states along the interferometer branches (illustrated unrealistically large). At the end of the interferometer this kick-induced delocalization along the branches between the clock states leads to slightly displaced detection locations for the ground (green) and excited state (blue) atoms.}
\label{fig: Albert Interferometer modified}
\end{figure}
However, considering finite pulse-time effects and the fact that the E1-M1 transitions are not Doppler-free anymore originating from the position dependency of the Rabi frequency, the modified trajectories of this interferometer are shown in Fig.~\ref{fig: Albert Interferometer modified}. Nevertheless, we can still measure the intensity in the ground state and the excited state exit ports. Describing the evolution along the lower and upper trajectories, respectively, by
\begin{subequations}\label{eq: Branches Albert gg}
	\begin{align}
	\hat{U}_{l,gg} &= \frac{1}{2}\hat{\mathcal{D}}_p^\dagger\hat{U}_g(T_4,T_3)\hat{\mathcal{D}}_p\hat{U}_g(T_3,T_2+t_{\frac{\pi}{2}})\hat{U}_{\frac{\pi}{2},gg}\hat{U}_g(T_2,T_0),\\
	\hat{U}_{u,gg} &=\frac{1}{2} \hat{U}_g(T_4,T_2+t_{\frac{\pi}{2}})\hat{U}_{\frac{\pi}{2},gg}\hat{U}_g(T_2,T_1)\hat{\mathcal{D}}_p^\dagger\hat{U}_g(T_1,T_0)\hat{\mathcal{D}}_p,
	\end{align}
\end{subequations}
which leads to
\begin{align}
\hat{U}_{l,gg}^\dagger\hat{U}_{l,gg} = \hat{U}_{u,gg}^\dagger\hat{U}_{u,gg} = \frac{1}{32},
\end{align}
the intensity in the ground state exit port is given by
\begin{align}
I_g &= \frac{1}{16} + \matrixel{\psi_0}{\left(\hat{U}_{u,gg}^\dagger\hat{U}_{l,gg}+\mathrm{h.c.}\right)}{\psi_0}\nonumber\\
&= \frac{1}{32}\left[2+\left(\mathcal{V}_g\mathrm{e}^{\mathrm{i}\Delta\phi_g}+\mathrm{c.c.}\right)\right],
\end{align}
where we defined the visibility $\mathcal{V}_g$ and the phase difference $\Delta\phi_g$ in the ground state exit port. Analogously we obtain for the excited state exit port the lower and upper branch evolution operators, 
\begin{subequations}
	\begin{align}
	\hat{U}_{l,ge} &= \frac{1}{2}\hat{\mathcal{D}}_p^\dagger\hat{U}_e(T_4,T_3)\hat{\mathcal{D}}_p\hat{U}_e(T_3,T_2+t_{\frac{\pi}{2}})\hat{U}_{\frac{\pi}{2},ge}\hat{U}_g(T_2,T_0),\\
	\hat{U}_{u,ge} &=\frac{1}{2} \hat{U}_e(T_4,T_2+t_{\frac{\pi}{2}})\hat{U}_{\frac{\pi}{2},ge}\hat{U}_g(T_2,T_1)\hat{\mathcal{D}}_p^\dagger\hat{U}_g(T_1,T_0)\hat{\mathcal{D}}_p,
	\end{align}
\end{subequations}
leading to the intensity in the excited state exit port
\begin{align}
I_e &=  \frac{1}{16} + \matrixel{\psi_0}{\left(\hat{U}_{u,ge}^\dagger\hat{U}_{l,ge}+\mathrm{h.c.}\right)}{\psi_0}\nonumber\\
&= \frac{1}{32}\left[2+\left(\mathcal{V}_e\mathrm{e}^{\mathrm{i}\Delta\phi_e}+\mathrm{h.c.}\right)\right]
\end{align}
with the visibility $\mathcal{V}_e$ and the phase difference $\Delta\phi_e$ in the excited state exit port. Inserting the Gaussian wave packet from Eq.~\eqref{eq: Gaussian wave packet} we can calculate the visibility and the phase
	\begin{align}
	\mathcal{V}_g &= 1, \quad
	\Delta\phi_g = \frac{1}{2}g k_p \delta T\left(T_4+T_3-\delta T\right)+\frac{\epsilon}{2}g k_p t_{\frac{\pi}{2}}\delta T
	\end{align}
in the ground state exit port and ($\delta Z_0(T_i)$ being the branch separation at time $T_i$)
\begin{align}
    \begin{split}
	\mathcal{V}_e &= \exp\left(-\frac{k_p^2\epsilon^2\Delta p_z^2\delta T^2}{2M^2}\right) = 1 + \mathcal{O}\left(\epsilon^2\right),\\
	\Delta\phi_e &= \frac{1}{2}g k_p \delta T \left(T_4+T_3-\delta T\right)-\frac{\hbar k k_p \delta T}{M}+\frac{\delta Z_R \delta Z_0(T_2)}{Z_R^2}\\
	&\quad\quad+\epsilon\left(\frac{\hbar(k+k_p)k_p\delta T}{2M}-\frac{1}{2}gk_p\delta T(t_{\frac{\pi}{2}}+2T_2)\right)
	\end{split}
\end{align}
for the excited state exit port, and where we expanded everything up to the first order in $\epsilon=\Delta M / M$. Accordingly, the visibility is one to our order in the approximations for both channels. The differential phase is then given by
\begin{align}\label{eq: Differential phase Albert}
\begin{split}
\Delta\phi_-(T_2) =&\Delta\phi_g - \Delta\phi_e \\
=& \frac{\hbar k k_p \delta T}{M}-\frac{\delta Z_R \delta Z_0(T_2)}{Z_R^2}\\
&-\epsilon\left(\frac{\hbar(k+k_p)k_p\delta T}{2M}-gk_p\delta T(t_{\frac{\pi}{2}}+T_2)\right),
\end{split}
\end{align}
where we still observe finite pulse-time effects, i.e. the terms containing the additional momentum kick $\hbar k$ and the $\pi/2$ pulse time $t_\frac{\pi}{2}$. However, the double differential phase, i.e. the difference of the differential phase, Eq.~\eqref{eq: Differential phase Albert}, with different initialization times of the atomic clock $T_2$, is given by
\begin{align}\label{eq: Double-differential phase Albert}
\Delta\phi_{-}(T_2)-\Delta\phi_-(T_2+\tau) = -\epsilon g k_p \delta T \tau.
\end{align}
Note that the branch separations at the clock initialization are the same, i.e. $\delta Z_0(T_2)=\delta Z_0(T_2+\tau)$.
Therefore, the finite pulse-time effects cancel each other between different runs of the experiment to the leading order in the expansion parameters.

\subsection{UGR and UFF Tests Without Superposition}
\begin{figure}
	\centering
	\includegraphics{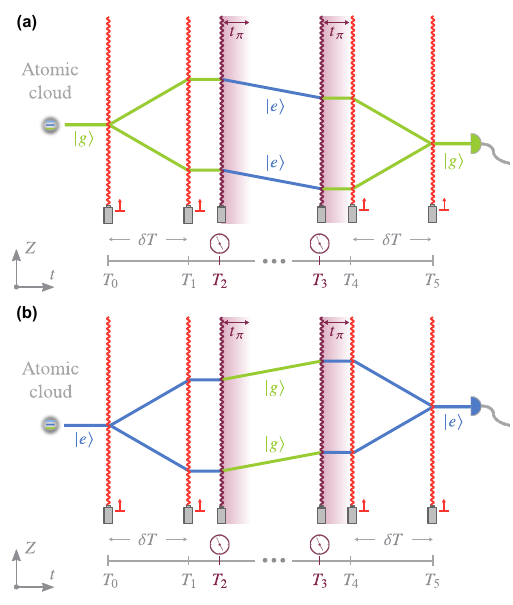}
	\caption{Interferometer scheme (B) of Ufrecht et al.~\cite{Ufrecht2020} in the freely falling frame for different initial states of the atoms. Panel {\sffamily\bfseries (a)} showcases the interferometer sequence with the atoms initially and at detection in the ground state $\ket{g}$. Panel {\sffamily\bfseries (b)} shows the corresponding sequence with atoms initially and at detection in the excited state $\ket{e}$. The E1-M1 $\pi$-pulses of duration $t_\pi$ are indicated by the violet lasers and subsequent violet shading after the beginning of the pulses at times $T_2$ and $T_3$. We exaggerated the separation between the momentum-kick pulses and the E1-M1 pulses as well as the effect of the modified momentum kicks $\pm\hbar\vec{k}$ due to the additional optical potentials acting during the finite-duration E1-M1 transitions starting at times $T_2$ and $T_3$ for better visualization.}
\label{fig: Christian Interferometer}
\end{figure}
The interferometer scheme (B)~\cite{Ufrecht2020} does not require superpositions of different internal states but a change of the internal state in the middle segment, i.e. a recoilless $\pi$-pulse. Again, we need two runs of the experiment: one with the initial ground state and one with the initial excited state. We found in Sec.~\ref{sec: Fundamental Gaussian Laser Beam} that the E1-M1 transitions are not perfectly recoilless for realistic beam shapes and finite pulse times. Likewise, the interferometer scheme in this section is modified by additional momentum kicks during the $\pi$-pulses, see Fig.~\ref{fig: Christian Interferometer}.

We can calculate the final intensity analogously to the previous section and set $T_3-T_2 = T_4-T_1=T$, so that we obtain the visibilities and phases
	\begin{align}
	\mathcal{V}_g &= 1, \quad
	\Delta\phi_g = 2gk_p\delta T\big(\delta T + T+\epsilon T\big)
	\end{align}
for an initial ground state and
	\begin{align}
	\mathcal{V}_e &= 1,  \quad
	\Delta\phi_e = 2gk_p\delta T\big(\delta T + T-\epsilon T\big)
	\end{align}
for an initial excited state. For this interferometer scheme we find that the final pulse-time effects cancel each other already in the non-differential phase due to the symmetry of the interferometer. The results~\cite{Ufrecht2020} can thus be reproduced immediately, i.e. the differential signals read
\begin{subequations}
	\begin{align}
	\Delta\phi_+ &= \Delta\phi_g + \Delta\phi_e = 4g k_p (\delta T + T)\delta T,\\
	\Delta\phi_- &= \Delta\phi_g - \Delta\phi_e = 4\epsilon g k_p T \delta T,
	\end{align}
\end{subequations}
separating the effects of UFF and UGR.

\section{Conclusion}
\label{sec.Conclusion}
Very-large baseline atom interferometry is built upon exploiting the beneficial scaling of the interferometer signal in terms of the enclosed space-time area. However, due to the resulting long free-fall and interaction times, imperfections and perturbations  act over much longer timescales and even small, accumulated effects can lead to a loss of visibility in the interference pattern in such devices. In case of (local) magnetic and gravitational field gradients~\cite{Wodey2020,Lezeik2023} these effects have to be studied in detail for upcoming large baseline setups in addition to the already available results for e.g. gravity gradients or rotations~\cite{Roura2014,Lan2012,Dickerson2013,Roura2017,Overstreet2018,Ufrecht2021}. In this spirit, our paper may be regarded as an extension of such studies to the case of E1-M1 transitions in quantum clock interferometry, performed under realistic conditions arising in large baseline atomic fountains.

Following the recent proposals of Roura~\cite{Roura2020} and Ufrecht et al.~\cite{Ufrecht2020} for LPAI schemes that are sensitive to UGR and UFF tests, we have investigated recoil-less clock transitions mediated by \mbox{E1-M1} processes. For this purpose, we took into account the fully quantized atomic degrees of freedom, including the quantized -- and possibly delocalizing -- COM motion as well as the intern-extern coupling/mass defect. As a simple starting point we considered electromagnetic plane waves in a $\sigma^+$- $\sigma^-$ polarization scheme such that the atom absorbs two counter-propagating field excitations. However, since the M1 transitions are much weaker than E1 transitions in typical atoms one needs quite large laser intensities and still long pulse times in an interferometer scheme containing E1-M1 transitions. Accordingly, finite pulse-time effects need to be included for a precise modeling.
After an adiabatic elimination procedure we found that the internal atomic dynamics yield standard Rabi oscillations, where the COM dependency drops out due to the fields having a constant intensity in space.

Further, we also took into account that a realistic laser beam has a spatially dependent intensity, i.e. the Rabi frequencies become position dependent. Here we considered the case of an atom falling through a general laser beam with arbitrary spatial coupling and finite pulse times, and provided an expression for the dynamics of such an atom. In the exemplary case of a Gaussian laser beam, we derived $\pi$ and $\pi/2$-pulse operators that generalize the standard expressions. In particular, our results show as the dominant effect additional momentum kicks $\pm\delta Z_R\hbar/Z_R^2\vec{e}_Z$ and an imprinting of a branch-dependent phase $\Phi(Z_0)=\mp\delta Z_R Z_0/Z_R^2$ and a splitting of the branches to the next order. As one would expect, these expressions reduce then to the idealized operators when we neglect the laser profile and assume a strictly localized atomic COM wave function.

It is important to emphasize that the laser inhomogeneities result in corrections to the idealized plane-wave model already for very localized atomic wave packets. In the case of atomic clouds that are suffering from strong dispersion, e.g. for long interferometer times or strong atom-atom interactions, the assumption of large scale separation between atomic and laser extension no longer holds, and the efficiency of optical pulses  is expected to be reduced even further; cf. for instance the efficiency of a Bragg beam splitter with two E1 transitions.~\cite{Neumann2021}

Finally, we returned to the initial question of the paper and applied our results for the leading order finite pulse-time effects, i.e. the additional momentum kicks, to the proposed UGR and UFF tests of Roura, scheme (A)~\cite{Roura2020}, and Ufrecht et al., scheme (B)~\cite{Ufrecht2020}. In both cases we derived the interferometer phases of the modified schemes and found that the additional momentum kicks are canceling each other out to leading order in the respective signal of interest. In the interferometer scheme (B)~\cite{Ufrecht2020} these effects cancel out already in the non-differential phase due to symmetry, while in the interferometer scheme (A)~\cite{Roura2020} they are only canceled out in the double-differential phase.

\section*{Acknowledgements}
We are grateful to W. P. Schleich for his stimulating input and continuing support.
We are thankful to C. Ufrecht, E. Giese, T. Aßmann, F. Di Pumpo and S. Böhringer as well as the QUANTUS and \mbox{INTENTAS} teams for fruitful and interesting discussions.
A.F. is grateful to the Carl Zeiss Foundation (Carl-Zeiss-Stiftung) and IQST for funding in terms of the project MuMo-RmQM.
The QUANTUS and INTENTAS projects are supported by the German Space Agency at the German Aerospace Center (Deutsche Raumfahrtagentur im Deutschen Zentrum f\"ur Luft- und Raumfahrt, DLR) with funds provided by the Federal Ministry for Economic Affairs and Climate Action (Bundesministerium f\"ur Wirtschaft und Klimaschutz, BMWK) due to an enactment of the German Bundestag under Grant Nos. 50WM2250D-2250E (QUANTUS+), as well as 50WM2177-2178 (INTENTAS).

\section*{Author Declarations}
\subsection*{Conflict of Interest Statement}
The authors have no conflicts to disclose.

\subsection*{Author Contributions }
\noindent{\sffamily\small\textbf{Gregor Janson}} Conceptualization (support); Formal analysis (lead); Validation (equal); Investigation (lead); Methodology (equal); Visualization (equal); Writing - original draft (lead); Writing – review and editing (equal).
{\sffamily\small\textbf{Alexander Friedrich}}
Conceptualization (lead); Formal analysis (support); Validation (equal); Investigation (support); Methodology (equal); Visualization (equal); Writing – review and editing (equal); Supervision (equal).
{\sffamily\small\textbf{Richard Lopp}} Conceptualization (support); Formal analysis (support); Validation (equal); Investigation (support); Methodology (equal); Writing - original draft (Support); Writing – review and editing (equal); Supervision (equal).

\section*{Data Availability}

The data that support the findings of this study are available
within the article.

\bibliography{main.bib}

\begin{thebibliography}{126}%
\makeatletter
\providecommand \@ifxundefined [1]{%
 \@ifx{#1\undefined}
}%
\providecommand \@ifnum [1]{%
 \ifnum #1\expandafter \@firstoftwo
 \else \expandafter \@secondoftwo
 \fi
}%
\providecommand \@ifx [1]{%
 \ifx #1\expandafter \@firstoftwo
 \else \expandafter \@secondoftwo
 \fi
}%
\providecommand \natexlab [1]{#1}%
\providecommand \enquote  [1]{``#1''}%
\providecommand \bibnamefont  [1]{#1}%
\providecommand \bibfnamefont [1]{#1}%
\providecommand \citenamefont [1]{#1}%
\providecommand \href@noop [0]{\@secondoftwo}%
\providecommand \href [0]{\begingroup \@sanitize@url \@href}%
\providecommand \@href[1]{\@@startlink{#1}\@@href}%
\providecommand \@@href[1]{\endgroup#1\@@endlink}%
\providecommand \@sanitize@url [0]{\catcode `\\12\catcode `\$12\catcode
  `\&12\catcode `\#12\catcode `\^12\catcode `\_12\catcode `\%12\relax}%
\providecommand \@@startlink[1]{}%
\providecommand \@@endlink[0]{}%
\providecommand \url  [0]{\begingroup\@sanitize@url \@url }%
\providecommand \@url [1]{\endgroup\@href {#1}{\urlprefix }}%
\providecommand \urlprefix  [0]{URL }%
\providecommand \Eprint [0]{\href }%
\providecommand \doibase [0]{http://dx.doi.org/}%
\providecommand \selectlanguage [0]{\@gobble}%
\providecommand \bibinfo  [0]{\@secondoftwo}%
\providecommand \bibfield  [0]{\@secondoftwo}%
\providecommand \translation [1]{[#1]}%
\providecommand \BibitemOpen [0]{}%
\providecommand \bibitemStop [0]{}%
\providecommand \bibitemNoStop [0]{.\EOS\space}%
\providecommand \EOS [0]{\spacefactor3000\relax}%
\providecommand \BibitemShut  [1]{\csname bibitem#1\endcsname}%
\let\auto@bib@innerbib\@empty
\bibitem [{\citenamefont {Kasevich}\ and\ \citenamefont
  {Chu}(1991)}]{Kasevich1991}%
  \BibitemOpen
  \bibfield  {author} {\bibinfo {author} {\bibfnamefont {M.}~\bibnamefont
  {Kasevich}}\ and\ \bibinfo {author} {\bibfnamefont {S.}~\bibnamefont {Chu}},\
  }\bibfield  {title} {\enquote {\bibinfo {title} {Atomic interferometry using
  stimulated raman transitions},}\ }\href {\doibase 10.1103/PhysRevLett.67.181}
  {\bibfield  {journal} {\bibinfo  {journal} {Phys. Rev. Lett.}\ }\textbf
  {\bibinfo {volume} {67}},\ \bibinfo {pages} {181--184} (\bibinfo {year}
  {1991})}\BibitemShut {NoStop}%
\bibitem [{\citenamefont {Peters}, \citenamefont {Chung},\ and\ \citenamefont
  {Chu}(2001)}]{Peters1998}%
  \BibitemOpen
  \bibfield  {author} {\bibinfo {author} {\bibfnamefont {A.}~\bibnamefont
  {Peters}}, \bibinfo {author} {\bibfnamefont {K.~Y.}\ \bibnamefont {Chung}}, \
  and\ \bibinfo {author} {\bibfnamefont {S.}~\bibnamefont {Chu}},\ }\bibfield
  {title} {\enquote {\bibinfo {title} {High-precision gravity measurements
  using atom interferometry},}\ }\href@noop {} {\bibfield  {journal} {\bibinfo
  {journal} {Metrologia}\ }\textbf {\bibinfo {volume} {38}},\ \bibinfo {pages}
  {25--61} (\bibinfo {year} {2001})}\BibitemShut {NoStop}%
\bibitem [{\citenamefont {Peters}\ \emph {et~al.}(1999)\citenamefont {Peters},
  \citenamefont {Achim}, \citenamefont {Chung}, \citenamefont {Yeow},
  \citenamefont {Chu},\ and\ \citenamefont {Steven}}]{Peters1999}%
  \BibitemOpen
  \bibfield  {author} {\bibinfo {author} {\bibnamefont {Peters}}, \bibinfo
  {author} {\bibnamefont {Achim}}, \bibinfo {author} {\bibnamefont {Chung}},
  \bibinfo {author} {\bibfnamefont {K.}~\bibnamefont {Yeow}}, \bibinfo {author}
  {\bibnamefont {Chu}}, \ and\ \bibinfo {author} {\bibnamefont {Steven}},\
  }\bibfield  {title} {\enquote {\bibinfo {title} {Measurement of gravitational
  acceleration by dropping atoms},}\ }\href {\doibase 10.1038/23655} {\bibfield
   {journal} {\bibinfo  {journal} {Nature}\ }\textbf {\bibinfo {volume}
  {400}},\ \bibinfo {pages} {849--} (\bibinfo {year} {1999})}\BibitemShut
  {NoStop}%
\bibitem [{\citenamefont {Lan}\ \emph {et~al.}(2012)\citenamefont {Lan},
  \citenamefont {Kuan}, \citenamefont {Estey}, \citenamefont {Haslinger},\ and\
  \citenamefont {M{\ifmmode\ddot{u}\else\"{u}\fi}ller}}]{Lan2012}%
  \BibitemOpen
  \bibfield  {author} {\bibinfo {author} {\bibfnamefont {S.-Y.}\ \bibnamefont
  {Lan}}, \bibinfo {author} {\bibfnamefont {P.-C.}\ \bibnamefont {Kuan}},
  \bibinfo {author} {\bibfnamefont {B.}~\bibnamefont {Estey}}, \bibinfo
  {author} {\bibfnamefont {P.}~\bibnamefont {Haslinger}}, \ and\ \bibinfo
  {author} {\bibfnamefont {H.}~\bibnamefont
  {M{\ifmmode\ddot{u}\else\"{u}\fi}ller}},\ }\bibfield  {title} {\enquote
  {\bibinfo {title} {{Influence of the Coriolis Force in Atom
  Interferometry}},}\ }\href {\doibase 10.1103/PhysRevLett.108.090402}
  {\bibfield  {journal} {\bibinfo  {journal} {Phys. Rev. Lett.}\ }\textbf
  {\bibinfo {volume} {108}},\ \bibinfo {pages} {090402} (\bibinfo {year}
  {2012})}\BibitemShut {NoStop}%
\bibitem [{\citenamefont {Barrett}\ \emph {et~al.}(2016)\citenamefont
  {Barrett}, \citenamefont {Antoni-Micollier}, \citenamefont {Chichet},
  \citenamefont {Battelier}, \citenamefont {L{\'{e}}v{\`{e}}que}, \citenamefont
  {Landragin},\ and\ \citenamefont {Bouyer}}]{Barrett2016}%
  \BibitemOpen
  \bibfield  {author} {\bibinfo {author} {\bibfnamefont {B.}~\bibnamefont
  {Barrett}}, \bibinfo {author} {\bibfnamefont {L.}~\bibnamefont
  {Antoni-Micollier}}, \bibinfo {author} {\bibfnamefont {L.}~\bibnamefont
  {Chichet}}, \bibinfo {author} {\bibfnamefont {B.}~\bibnamefont {Battelier}},
  \bibinfo {author} {\bibfnamefont {T.}~\bibnamefont {L{\'{e}}v{\`{e}}que}},
  \bibinfo {author} {\bibfnamefont {A.}~\bibnamefont {Landragin}}, \ and\
  \bibinfo {author} {\bibfnamefont {P.}~\bibnamefont {Bouyer}},\ }\bibfield
  {title} {\enquote {\bibinfo {title} {{Dual matter-wave inertial sensors in
  weightlessness}},}\ }\href {\doibase 10.1038/ncomms13786} {\bibfield
  {journal} {\bibinfo  {journal} {Nat. Commun.}\ }\textbf {\bibinfo {volume}
  {7}},\ \bibinfo {pages} {1--9} (\bibinfo {year} {2016})}\BibitemShut
  {NoStop}%
\bibitem [{\citenamefont {Wu}\ \emph {et~al.}(2019)\citenamefont {Wu},
  \citenamefont {Pagel}, \citenamefont {Malek}, \citenamefont {Nguyen},
  \citenamefont {Zi}, \citenamefont {Scheirer},\ and\ \citenamefont
  {M{\"{u}}ller}}]{Wu2019}%
  \BibitemOpen
  \bibfield  {author} {\bibinfo {author} {\bibfnamefont {X.}~\bibnamefont
  {Wu}}, \bibinfo {author} {\bibfnamefont {Z.}~\bibnamefont {Pagel}}, \bibinfo
  {author} {\bibfnamefont {B.~S.}\ \bibnamefont {Malek}}, \bibinfo {author}
  {\bibfnamefont {T.~H.}\ \bibnamefont {Nguyen}}, \bibinfo {author}
  {\bibfnamefont {F.}~\bibnamefont {Zi}}, \bibinfo {author} {\bibfnamefont
  {D.~S.}\ \bibnamefont {Scheirer}}, \ and\ \bibinfo {author} {\bibfnamefont
  {H.}~\bibnamefont {M{\"{u}}ller}},\ }\bibfield  {title} {\enquote {\bibinfo
  {title} {{Gravity surveys using a mobile atom interferometer}},}\ }\href
  {\doibase 10.1126/sciadv.aax0800} {\bibfield  {journal} {\bibinfo  {journal}
  {Sci. Adv.}\ }\textbf {\bibinfo {volume} {5}},\ \bibinfo {pages} {eaax0800}
  (\bibinfo {year} {2019})}\BibitemShut {NoStop}%
\bibitem [{\citenamefont {Templier}\ \emph {et~al.}(2022)\citenamefont
  {Templier}, \citenamefont {Cheiney}, \citenamefont {de~Castanet},
  \citenamefont {Gouraud}, \citenamefont {Porte}, \citenamefont {Napolitano},
  \citenamefont {Bouyer}, \citenamefont {Battelier},\ and\ \citenamefont
  {Barrett}}]{Templier2022}%
  \BibitemOpen
  \bibfield  {author} {\bibinfo {author} {\bibfnamefont {S.}~\bibnamefont
  {Templier}}, \bibinfo {author} {\bibfnamefont {P.}~\bibnamefont {Cheiney}},
  \bibinfo {author} {\bibfnamefont {Q.~d.}\ \bibnamefont {de~Castanet}},
  \bibinfo {author} {\bibfnamefont {B.}~\bibnamefont {Gouraud}}, \bibinfo
  {author} {\bibfnamefont {H.}~\bibnamefont {Porte}}, \bibinfo {author}
  {\bibfnamefont {F.}~\bibnamefont {Napolitano}}, \bibinfo {author}
  {\bibfnamefont {P.}~\bibnamefont {Bouyer}}, \bibinfo {author} {\bibfnamefont
  {B.}~\bibnamefont {Battelier}}, \ and\ \bibinfo {author} {\bibfnamefont
  {B.}~\bibnamefont {Barrett}},\ }\bibfield  {title} {\enquote {\bibinfo
  {title} {{Tracking the vector acceleration with a hybrid quantum
  accelerometer triad}},}\ }\href {\doibase 10.1126/sciadv.add3854} {\bibfield
  {journal} {\bibinfo  {journal} {Sci. Adv.}\ }\textbf {\bibinfo {volume}
  {8}},\ \bibinfo {pages} {eadd3854} (\bibinfo {year} {2022})}\BibitemShut
  {NoStop}%
\bibitem [{\citenamefont {Stray}\ \emph {et~al.}(2022)\citenamefont {Stray},
  \citenamefont {Lamb}, \citenamefont {Kaushik}, \citenamefont {Vovrosh},
  \citenamefont {Rodgers}, \citenamefont {Winch}, \citenamefont {Hayati},
  \citenamefont {Boddice}, \citenamefont {Stabrawa}, \citenamefont {Niggebaum},
  \citenamefont {Langlois}, \citenamefont {Lien}, \citenamefont {Lellouch},
  \citenamefont {Roshanmanesh}, \citenamefont {Ridley}, \citenamefont
  {de~Villiers}, \citenamefont {Brown}, \citenamefont {Cross}, \citenamefont
  {Tuckwell}, \citenamefont {Faramarzi}, \citenamefont {Metje}, \citenamefont
  {Bongs},\ and\ \citenamefont {Holynski}}]{Stray2022}%
  \BibitemOpen
  \bibfield  {author} {\bibinfo {author} {\bibfnamefont {B.}~\bibnamefont
  {Stray}}, \bibinfo {author} {\bibfnamefont {A.}~\bibnamefont {Lamb}},
  \bibinfo {author} {\bibfnamefont {A.}~\bibnamefont {Kaushik}}, \bibinfo
  {author} {\bibfnamefont {J.}~\bibnamefont {Vovrosh}}, \bibinfo {author}
  {\bibfnamefont {A.}~\bibnamefont {Rodgers}}, \bibinfo {author} {\bibfnamefont
  {J.}~\bibnamefont {Winch}}, \bibinfo {author} {\bibfnamefont
  {F.}~\bibnamefont {Hayati}}, \bibinfo {author} {\bibfnamefont
  {D.}~\bibnamefont {Boddice}}, \bibinfo {author} {\bibfnamefont
  {A.}~\bibnamefont {Stabrawa}}, \bibinfo {author} {\bibfnamefont
  {A.}~\bibnamefont {Niggebaum}}, \bibinfo {author} {\bibfnamefont
  {M.}~\bibnamefont {Langlois}}, \bibinfo {author} {\bibfnamefont {Y.-H.}\
  \bibnamefont {Lien}}, \bibinfo {author} {\bibfnamefont {S.}~\bibnamefont
  {Lellouch}}, \bibinfo {author} {\bibfnamefont {S.}~\bibnamefont
  {Roshanmanesh}}, \bibinfo {author} {\bibfnamefont {K.}~\bibnamefont
  {Ridley}}, \bibinfo {author} {\bibfnamefont {G.}~\bibnamefont {de~Villiers}},
  \bibinfo {author} {\bibfnamefont {G.}~\bibnamefont {Brown}}, \bibinfo
  {author} {\bibfnamefont {T.}~\bibnamefont {Cross}}, \bibinfo {author}
  {\bibfnamefont {G.}~\bibnamefont {Tuckwell}}, \bibinfo {author}
  {\bibfnamefont {A.}~\bibnamefont {Faramarzi}}, \bibinfo {author}
  {\bibfnamefont {N.}~\bibnamefont {Metje}}, \bibinfo {author} {\bibfnamefont
  {K.}~\bibnamefont {Bongs}}, \ and\ \bibinfo {author} {\bibfnamefont
  {M.}~\bibnamefont {Holynski}},\ }\bibfield  {title} {\enquote {\bibinfo
  {title} {{Quantum sensing for gravity cartography}},}\ }\href {\doibase
  10.1038/s41586-021-04315-3} {\bibfield  {journal} {\bibinfo  {journal}
  {Nature}\ }\textbf {\bibinfo {volume} {602}},\ \bibinfo {pages} {590--594}
  (\bibinfo {year} {2022})}\BibitemShut {NoStop}%
\bibitem [{\citenamefont {Rosi}\ \emph {et~al.}(2014)\citenamefont {Rosi},
  \citenamefont {Sorrentino}, \citenamefont {Cacciapuoti}, \citenamefont
  {Prevedelli},\ and\ \citenamefont {Tino}}]{Rosi2014}%
  \BibitemOpen
  \bibfield  {author} {\bibinfo {author} {\bibfnamefont {G.}~\bibnamefont
  {Rosi}}, \bibinfo {author} {\bibfnamefont {F.}~\bibnamefont {Sorrentino}},
  \bibinfo {author} {\bibfnamefont {L.}~\bibnamefont {Cacciapuoti}}, \bibinfo
  {author} {\bibfnamefont {M.}~\bibnamefont {Prevedelli}}, \ and\ \bibinfo
  {author} {\bibfnamefont {G.~M.}\ \bibnamefont {Tino}},\ }\bibfield  {title}
  {\enquote {\bibinfo {title} {Precision measurement of the newtonian
  gravitational constant using cold atoms},}\ }\href {\doibase
  10.1038/nature13433} {\bibfield  {journal} {\bibinfo  {journal} {Nature}\
  }\textbf {\bibinfo {volume} {510}},\ \bibinfo {pages} {518--21} (\bibinfo
  {year} {2014})}\BibitemShut {NoStop}%
\bibitem [{\citenamefont {Parker}\ \emph {et~al.}(2018)\citenamefont {Parker},
  \citenamefont {Yu}, \citenamefont {Zhong}, \citenamefont {Estey},\ and\
  \citenamefont {Müller}}]{Parker2018}%
  \BibitemOpen
  \bibfield  {author} {\bibinfo {author} {\bibfnamefont {R.}~\bibnamefont
  {Parker}}, \bibinfo {author} {\bibfnamefont {C.}~\bibnamefont {Yu}}, \bibinfo
  {author} {\bibfnamefont {W.}~\bibnamefont {Zhong}}, \bibinfo {author}
  {\bibfnamefont {B.}~\bibnamefont {Estey}}, \ and\ \bibinfo {author}
  {\bibfnamefont {H.}~\bibnamefont {Müller}},\ }\bibfield  {title} {\enquote
  {\bibinfo {title} {Measurement of the fine-structure constant as a test of
  the standard model},}\ }\href {\doibase 10.1126/science.aap7706} {\bibfield
  {journal} {\bibinfo  {journal} {Science}\ }\textbf {\bibinfo {volume}
  {360}},\ \bibinfo {pages} {191--195} (\bibinfo {year} {2018})}\BibitemShut
  {NoStop}%
\bibitem [{\citenamefont {Morel}\ \emph {et~al.}(2020)\citenamefont {Morel},
  \citenamefont {Yao}, \citenamefont {Clad{\'{e}}},\ and\ \citenamefont
  {Guellati-Kh{\'{e}}lifa}}]{Morel2020b}%
  \BibitemOpen
  \bibfield  {author} {\bibinfo {author} {\bibfnamefont {L.}~\bibnamefont
  {Morel}}, \bibinfo {author} {\bibfnamefont {Z.}~\bibnamefont {Yao}}, \bibinfo
  {author} {\bibfnamefont {P.}~\bibnamefont {Clad{\'{e}}}}, \ and\ \bibinfo
  {author} {\bibfnamefont {S.}~\bibnamefont {Guellati-Kh{\'{e}}lifa}},\
  }\bibfield  {title} {\enquote {\bibinfo {title} {{Determination of the
  fine-structure constant with an accuracy of 81 parts per trillion}},}\ }\href
  {\doibase 10.1038/s41586-020-2964-7} {\bibfield  {journal} {\bibinfo
  {journal} {Nature}\ }\textbf {\bibinfo {volume} {588}},\ \bibinfo {pages}
  {61--65} (\bibinfo {year} {2020})}\BibitemShut {NoStop}%
\bibitem [{\citenamefont {Dimopoulos}\ \emph
  {et~al.}(2008{\natexlab{a}})\citenamefont {Dimopoulos}, \citenamefont
  {Graham}, \citenamefont {Hogan}, \citenamefont {Kasevich},\ and\
  \citenamefont {Rajendran}}]{Dimopoulos2008b}%
  \BibitemOpen
  \bibfield  {author} {\bibinfo {author} {\bibfnamefont {S.}~\bibnamefont
  {Dimopoulos}}, \bibinfo {author} {\bibfnamefont {P.~W.}\ \bibnamefont
  {Graham}}, \bibinfo {author} {\bibfnamefont {J.~M.}\ \bibnamefont {Hogan}},
  \bibinfo {author} {\bibfnamefont {M.~A.}\ \bibnamefont {Kasevich}}, \ and\
  \bibinfo {author} {\bibfnamefont {S.}~\bibnamefont {Rajendran}},\ }\bibfield
  {title} {\enquote {\bibinfo {title} {{Atomic gravitational wave
  interferometric sensor}},}\ }\href {\doibase 10.1103/PhysRevD.78.122002}
  {\bibfield  {journal} {\bibinfo  {journal} {Phys. Rev. D}\ }\textbf {\bibinfo
  {volume} {78}},\ \bibinfo {pages} {122002} (\bibinfo {year}
  {2008}{\natexlab{a}})}\BibitemShut {NoStop}%
\bibitem [{\citenamefont {Dimopoulos}\ \emph {et~al.}(2009)\citenamefont
  {Dimopoulos}, \citenamefont {Graham}, \citenamefont {Hogan}, \citenamefont
  {Kasevich},\ and\ \citenamefont {Rajendran}}]{Dimopoulos2009}%
  \BibitemOpen
  \bibfield  {author} {\bibinfo {author} {\bibfnamefont {S.}~\bibnamefont
  {Dimopoulos}}, \bibinfo {author} {\bibfnamefont {P.~W.}\ \bibnamefont
  {Graham}}, \bibinfo {author} {\bibfnamefont {J.~M.}\ \bibnamefont {Hogan}},
  \bibinfo {author} {\bibfnamefont {M.~A.}\ \bibnamefont {Kasevich}}, \ and\
  \bibinfo {author} {\bibfnamefont {S.}~\bibnamefont {Rajendran}},\ }\bibfield
  {title} {\enquote {\bibinfo {title} {{Gravitational wave detection with atom
  interferometry}},}\ }\href {\doibase 10.1016/j.physletb.2009.06.011}
  {\bibfield  {journal} {\bibinfo  {journal} {Phys. Lett. B}\ }\textbf
  {\bibinfo {volume} {678}},\ \bibinfo {pages} {37--40} (\bibinfo {year}
  {2009})}\BibitemShut {NoStop}%
\bibitem [{\citenamefont {Graham}\ \emph {et~al.}(2016)\citenamefont {Graham},
  \citenamefont {Hogan}, \citenamefont {Kasevich},\ and\ \citenamefont
  {Rajendran}}]{Graham2016}%
  \BibitemOpen
  \bibfield  {author} {\bibinfo {author} {\bibfnamefont {P.~W.}\ \bibnamefont
  {Graham}}, \bibinfo {author} {\bibfnamefont {J.~M.}\ \bibnamefont {Hogan}},
  \bibinfo {author} {\bibfnamefont {M.~A.}\ \bibnamefont {Kasevich}}, \ and\
  \bibinfo {author} {\bibfnamefont {S.}~\bibnamefont {Rajendran}},\ }\bibfield
  {title} {\enquote {\bibinfo {title} {Resonant mode for gravitational wave
  detectors based on atom interferometry},}\ }\href {\doibase
  10.1103/PhysRevD.94.104022} {\bibfield  {journal} {\bibinfo  {journal} {Phys.
  Rev. D}\ }\textbf {\bibinfo {volume} {94}},\ \bibinfo {pages} {104022}
  (\bibinfo {year} {2016})}\BibitemShut {NoStop}%
\bibitem [{\citenamefont {Geraci}\ and\ \citenamefont
  {Derevianko}(2016)}]{Geraci2016}%
  \BibitemOpen
  \bibfield  {author} {\bibinfo {author} {\bibfnamefont {A.~A.}\ \bibnamefont
  {Geraci}}\ and\ \bibinfo {author} {\bibfnamefont {A.}~\bibnamefont
  {Derevianko}},\ }\bibfield  {title} {\enquote {\bibinfo {title} {{Sensitivity
  of Atom Interferometry to Ultralight Scalar Field Dark Matter}},}\ }\href
  {\doibase 10.1103/PhysRevLett.117.261301} {\bibfield  {journal} {\bibinfo
  {journal} {Phys. Rev. Lett.}\ }\textbf {\bibinfo {volume} {117}},\ \bibinfo
  {pages} {261301} (\bibinfo {year} {2016})}\BibitemShut {NoStop}%
\bibitem [{\citenamefont {Arvanitaki}\ \emph {et~al.}(2018)\citenamefont
  {Arvanitaki}, \citenamefont {Graham}, \citenamefont {Hogan}, \citenamefont
  {Rajendran},\ and\ \citenamefont {Van~Tilburg}}]{Arvanitaki2018}%
  \BibitemOpen
  \bibfield  {author} {\bibinfo {author} {\bibfnamefont {A.}~\bibnamefont
  {Arvanitaki}}, \bibinfo {author} {\bibfnamefont {P.~W.}\ \bibnamefont
  {Graham}}, \bibinfo {author} {\bibfnamefont {J.~M.}\ \bibnamefont {Hogan}},
  \bibinfo {author} {\bibfnamefont {S.}~\bibnamefont {Rajendran}}, \ and\
  \bibinfo {author} {\bibfnamefont {K.}~\bibnamefont {Van~Tilburg}},\
  }\bibfield  {title} {\enquote {\bibinfo {title} {{Search for light scalar
  dark matter with atomic gravitational wave detectors}},}\ }\href {\doibase
  10.1103/PhysRevD.97.075020} {\bibfield  {journal} {\bibinfo  {journal} {Phys.
  Rev. D}\ }\textbf {\bibinfo {volume} {97}},\ \bibinfo {pages} {075020}
  (\bibinfo {year} {2018})}\BibitemShut {NoStop}%
\bibitem [{\citenamefont {Badurina}\ \emph {et~al.}(2023)\citenamefont
  {Badurina}, \citenamefont {Gibson}, \citenamefont {McCabe},\ and\
  \citenamefont {Mitchell}}]{Badurina2023a}%
  \BibitemOpen
  \bibfield  {author} {\bibinfo {author} {\bibfnamefont {L.}~\bibnamefont
  {Badurina}}, \bibinfo {author} {\bibfnamefont {V.}~\bibnamefont {Gibson}},
  \bibinfo {author} {\bibfnamefont {C.}~\bibnamefont {McCabe}}, \ and\ \bibinfo
  {author} {\bibfnamefont {J.}~\bibnamefont {Mitchell}},\ }\bibfield  {title}
  {\enquote {\bibinfo {title} {{Ultralight dark matter searches at the sub-Hz
  frontier with atom multigradiometry}},}\ }\href {\doibase
  10.1103/PhysRevD.107.055002} {\bibfield  {journal} {\bibinfo  {journal}
  {Phys. Rev. D}\ }\textbf {\bibinfo {volume} {107}},\ \bibinfo {pages}
  {055002} (\bibinfo {year} {2023})}\BibitemShut {NoStop}%
\bibitem [{\citenamefont {Abe}\ \emph {et~al.}(2021)\citenamefont {Abe},
  \citenamefont {Adamson}, \citenamefont {Borcean}, \citenamefont {Bortoletto},
  \citenamefont {Bridges}, \citenamefont {Carman}, \citenamefont
  {Chattopadhyay}, \citenamefont {Coleman}, \citenamefont {Curfman},
  \citenamefont {DeRose}, \citenamefont {Deshpande}, \citenamefont
  {Dimopoulos}, \citenamefont {Foot}, \citenamefont {Frisch}, \citenamefont
  {Garber}, \citenamefont {Geer}, \citenamefont {Gibson}, \citenamefont
  {Glick}, \citenamefont {Graham}, \citenamefont {Hahn}, \citenamefont
  {Harnik}, \citenamefont {Hawkins}, \citenamefont {Hindley}, \citenamefont
  {Hogan}, \citenamefont {Jiang}, \citenamefont {Kasevich}, \citenamefont
  {Kellett}, \citenamefont {Kiburg}, \citenamefont {Kovachy}, \citenamefont
  {Lykken}, \citenamefont {March-Russell}, \citenamefont {Mitchell},
  \citenamefont {Murphy}, \citenamefont {Nantel}, \citenamefont {Nobrega},
  \citenamefont {Plunkett}, \citenamefont {Rajendran}, \citenamefont {Rudolph},
  \citenamefont {Sachdeva}, \citenamefont {Safdari}, \citenamefont {Santucci},
  \citenamefont {Schwartzman}, \citenamefont {Shipsey}, \citenamefont {Swan},
  \citenamefont {Valerio}, \citenamefont {Vasonis}, \citenamefont {Wang},\ and\
  \citenamefont {Wilkason}}]{Abe2021}%
  \BibitemOpen
  \bibfield  {author} {\bibinfo {author} {\bibfnamefont {M.}~\bibnamefont
  {Abe}}, \bibinfo {author} {\bibfnamefont {P.}~\bibnamefont {Adamson}},
  \bibinfo {author} {\bibfnamefont {M.}~\bibnamefont {Borcean}}, \bibinfo
  {author} {\bibfnamefont {D.}~\bibnamefont {Bortoletto}}, \bibinfo {author}
  {\bibfnamefont {K.}~\bibnamefont {Bridges}}, \bibinfo {author} {\bibfnamefont
  {S.~P.}\ \bibnamefont {Carman}}, \bibinfo {author} {\bibfnamefont
  {S.}~\bibnamefont {Chattopadhyay}}, \bibinfo {author} {\bibfnamefont
  {J.}~\bibnamefont {Coleman}}, \bibinfo {author} {\bibfnamefont {N.~M.}\
  \bibnamefont {Curfman}}, \bibinfo {author} {\bibfnamefont {K.}~\bibnamefont
  {DeRose}}, \bibinfo {author} {\bibfnamefont {T.}~\bibnamefont {Deshpande}},
  \bibinfo {author} {\bibfnamefont {S.}~\bibnamefont {Dimopoulos}}, \bibinfo
  {author} {\bibfnamefont {C.~J.}\ \bibnamefont {Foot}}, \bibinfo {author}
  {\bibfnamefont {J.~C.}\ \bibnamefont {Frisch}}, \bibinfo {author}
  {\bibfnamefont {B.~E.}\ \bibnamefont {Garber}}, \bibinfo {author}
  {\bibfnamefont {S.}~\bibnamefont {Geer}}, \bibinfo {author} {\bibfnamefont
  {V.}~\bibnamefont {Gibson}}, \bibinfo {author} {\bibfnamefont
  {J.}~\bibnamefont {Glick}}, \bibinfo {author} {\bibfnamefont {P.~W.}\
  \bibnamefont {Graham}}, \bibinfo {author} {\bibfnamefont {S.~R.}\
  \bibnamefont {Hahn}}, \bibinfo {author} {\bibfnamefont {R.}~\bibnamefont
  {Harnik}}, \bibinfo {author} {\bibfnamefont {L.}~\bibnamefont {Hawkins}},
  \bibinfo {author} {\bibfnamefont {S.}~\bibnamefont {Hindley}}, \bibinfo
  {author} {\bibfnamefont {J.~M.}\ \bibnamefont {Hogan}}, \bibinfo {author}
  {\bibfnamefont {Y.}~\bibnamefont {Jiang}}, \bibinfo {author} {\bibfnamefont
  {M.~A.}\ \bibnamefont {Kasevich}}, \bibinfo {author} {\bibfnamefont {R.~J.}\
  \bibnamefont {Kellett}}, \bibinfo {author} {\bibfnamefont {M.}~\bibnamefont
  {Kiburg}}, \bibinfo {author} {\bibfnamefont {T.}~\bibnamefont {Kovachy}},
  \bibinfo {author} {\bibfnamefont {J.~D.}\ \bibnamefont {Lykken}}, \bibinfo
  {author} {\bibfnamefont {J.}~\bibnamefont {March-Russell}}, \bibinfo {author}
  {\bibfnamefont {J.}~\bibnamefont {Mitchell}}, \bibinfo {author}
  {\bibfnamefont {M.}~\bibnamefont {Murphy}}, \bibinfo {author} {\bibfnamefont
  {M.}~\bibnamefont {Nantel}}, \bibinfo {author} {\bibfnamefont {L.~E.}\
  \bibnamefont {Nobrega}}, \bibinfo {author} {\bibfnamefont {R.~K.}\
  \bibnamefont {Plunkett}}, \bibinfo {author} {\bibfnamefont {S.}~\bibnamefont
  {Rajendran}}, \bibinfo {author} {\bibfnamefont {J.}~\bibnamefont {Rudolph}},
  \bibinfo {author} {\bibfnamefont {N.}~\bibnamefont {Sachdeva}}, \bibinfo
  {author} {\bibfnamefont {M.}~\bibnamefont {Safdari}}, \bibinfo {author}
  {\bibfnamefont {J.~K.}\ \bibnamefont {Santucci}}, \bibinfo {author}
  {\bibfnamefont {A.~G.}\ \bibnamefont {Schwartzman}}, \bibinfo {author}
  {\bibfnamefont {I.}~\bibnamefont {Shipsey}}, \bibinfo {author} {\bibfnamefont
  {H.}~\bibnamefont {Swan}}, \bibinfo {author} {\bibfnamefont {L.~R.}\
  \bibnamefont {Valerio}}, \bibinfo {author} {\bibfnamefont {A.}~\bibnamefont
  {Vasonis}}, \bibinfo {author} {\bibfnamefont {Y.}~\bibnamefont {Wang}}, \
  and\ \bibinfo {author} {\bibfnamefont {T.}~\bibnamefont {Wilkason}},\
  }\bibfield  {title} {\enquote {\bibinfo {title} {Matter-wave atomic
  gradiometer interferometric sensor (magis-100)},}\ }\href {\doibase
  10.1088/2058-9565/abf719} {\bibfield  {journal} {\bibinfo  {journal} {Quantum
  Sci. Technol.}\ }\textbf {\bibinfo {volume} {6}},\ \bibinfo {pages} {044003}
  (\bibinfo {year} {2021})}\BibitemShut {NoStop}%
\bibitem [{\citenamefont {Bertoldi}\ \emph {et~al.}(2021)\citenamefont
  {Bertoldi}, \citenamefont {Bongs}, \citenamefont {Bouyer}, \citenamefont
  {Buchmueller}, \citenamefont {Canuel}, \citenamefont {Caramete},
  \citenamefont {Chiofalo}, \citenamefont {Coleman}, \citenamefont {De~Roeck},
  \citenamefont {Ellis}, \citenamefont {Graham}, \citenamefont {Haehnelt},
  \citenamefont {Hees}, \citenamefont {Hogan}, \citenamefont {von Klitzing},
  \citenamefont {Krutzik}, \citenamefont {Lewicki}, \citenamefont {McCabe},
  \citenamefont {Peters}, \citenamefont {Rasel}, \citenamefont {Roura},
  \citenamefont {Sabulsky}, \citenamefont {Schiller}, \citenamefont {Schubert},
  \citenamefont {Signorini}, \citenamefont {Sorrentino}, \citenamefont {Singh},
  \citenamefont {Tino}, \citenamefont {Vaskonen},\ and\ \citenamefont
  {Zhan}}]{Bertoldi2021}%
  \BibitemOpen
  \bibfield  {author} {\bibinfo {author} {\bibfnamefont {A.}~\bibnamefont
  {Bertoldi}}, \bibinfo {author} {\bibfnamefont {K.}~\bibnamefont {Bongs}},
  \bibinfo {author} {\bibfnamefont {P.}~\bibnamefont {Bouyer}}, \bibinfo
  {author} {\bibfnamefont {O.}~\bibnamefont {Buchmueller}}, \bibinfo {author}
  {\bibfnamefont {B.}~\bibnamefont {Canuel}}, \bibinfo {author} {\bibfnamefont
  {L.-I.}\ \bibnamefont {Caramete}}, \bibinfo {author} {\bibfnamefont {M.~L.}\
  \bibnamefont {Chiofalo}}, \bibinfo {author} {\bibfnamefont {J.}~\bibnamefont
  {Coleman}}, \bibinfo {author} {\bibfnamefont {A.}~\bibnamefont {De~Roeck}},
  \bibinfo {author} {\bibfnamefont {J.}~\bibnamefont {Ellis}}, \bibinfo
  {author} {\bibfnamefont {P.~W.}\ \bibnamefont {Graham}}, \bibinfo {author}
  {\bibfnamefont {M.~G.}\ \bibnamefont {Haehnelt}}, \bibinfo {author}
  {\bibfnamefont {A.}~\bibnamefont {Hees}}, \bibinfo {author} {\bibfnamefont
  {J.}~\bibnamefont {Hogan}}, \bibinfo {author} {\bibfnamefont
  {W.}~\bibnamefont {von Klitzing}}, \bibinfo {author} {\bibfnamefont
  {M.}~\bibnamefont {Krutzik}}, \bibinfo {author} {\bibfnamefont
  {M.}~\bibnamefont {Lewicki}}, \bibinfo {author} {\bibfnamefont
  {C.}~\bibnamefont {McCabe}}, \bibinfo {author} {\bibfnamefont
  {A.}~\bibnamefont {Peters}}, \bibinfo {author} {\bibfnamefont {E.~M.}\
  \bibnamefont {Rasel}}, \bibinfo {author} {\bibfnamefont {A.}~\bibnamefont
  {Roura}}, \bibinfo {author} {\bibfnamefont {D.}~\bibnamefont {Sabulsky}},
  \bibinfo {author} {\bibfnamefont {S.}~\bibnamefont {Schiller}}, \bibinfo
  {author} {\bibfnamefont {C.}~\bibnamefont {Schubert}}, \bibinfo {author}
  {\bibfnamefont {C.}~\bibnamefont {Signorini}}, \bibinfo {author}
  {\bibfnamefont {F.}~\bibnamefont {Sorrentino}}, \bibinfo {author}
  {\bibfnamefont {Y.}~\bibnamefont {Singh}}, \bibinfo {author} {\bibfnamefont
  {G.~M.}\ \bibnamefont {Tino}}, \bibinfo {author} {\bibfnamefont
  {V.}~\bibnamefont {Vaskonen}}, \ and\ \bibinfo {author} {\bibfnamefont
  {M.-S.}\ \bibnamefont {Zhan}},\ }\bibfield  {title} {\enquote {\bibinfo
  {title} {{AEDGE: Atomic experiment for dark matter and gravity exploration in
  space}},}\ }\href {\doibase 10.1007/s10686-021-09701-3} {\bibfield  {journal}
  {\bibinfo  {journal} {Exp. Astron.}\ }\textbf {\bibinfo {volume} {51}},\
  \bibinfo {pages} {1417--1426} (\bibinfo {year} {2021})}\BibitemShut {NoStop}%
\bibitem [{\citenamefont {Badurina}, \citenamefont {Blas},\ and\ \citenamefont
  {McCabe}(2022)}]{Badurina2021}%
  \BibitemOpen
  \bibfield  {author} {\bibinfo {author} {\bibfnamefont {L.}~\bibnamefont
  {Badurina}}, \bibinfo {author} {\bibfnamefont {D.}~\bibnamefont {Blas}}, \
  and\ \bibinfo {author} {\bibfnamefont {C.}~\bibnamefont {McCabe}},\
  }\bibfield  {title} {\enquote {\bibinfo {title} {{Refined ultralight scalar
  dark matter searches with compact atom gradiometers}},}\ }\href {\doibase
  10.1103/PhysRevD.105.023006} {\bibfield  {journal} {\bibinfo  {journal}
  {Phys. Rev. D}\ }\textbf {\bibinfo {volume} {105}},\ \bibinfo {pages}
  {023006} (\bibinfo {year} {2022})}\BibitemShut {NoStop}%
\bibitem [{\citenamefont {Safronova}\ \emph {et~al.}(2018)\citenamefont
  {Safronova}, \citenamefont {Budker}, \citenamefont {DeMille}, \citenamefont
  {Kimball}, \citenamefont {Derevianko},\ and\ \citenamefont
  {Clark}}]{Safronova2018}%
  \BibitemOpen
  \bibfield  {author} {\bibinfo {author} {\bibfnamefont {M.~S.}\ \bibnamefont
  {Safronova}}, \bibinfo {author} {\bibfnamefont {D.}~\bibnamefont {Budker}},
  \bibinfo {author} {\bibfnamefont {D.}~\bibnamefont {DeMille}}, \bibinfo
  {author} {\bibfnamefont {D.~F.~J.}\ \bibnamefont {Kimball}}, \bibinfo
  {author} {\bibfnamefont {A.}~\bibnamefont {Derevianko}}, \ and\ \bibinfo
  {author} {\bibfnamefont {C.~W.}\ \bibnamefont {Clark}},\ }\bibfield  {title}
  {\enquote {\bibinfo {title} {Search for new physics with atoms and
  molecules},}\ }\href {\doibase 10.1103/RevModPhys.90.025008} {\bibfield
  {journal} {\bibinfo  {journal} {Rev. Mod. Phys.}\ }\textbf {\bibinfo {volume}
  {90}},\ \bibinfo {pages} {025008} (\bibinfo {year} {2018})}\BibitemShut
  {NoStop}%
\bibitem [{\citenamefont {Dimopoulos}\ \emph {et~al.}(2007)\citenamefont
  {Dimopoulos}, \citenamefont {Graham}, \citenamefont {Hogan},\ and\
  \citenamefont {Kasevich}}]{Dimopoulos2007}%
  \BibitemOpen
  \bibfield  {author} {\bibinfo {author} {\bibfnamefont {S.}~\bibnamefont
  {Dimopoulos}}, \bibinfo {author} {\bibfnamefont {P.~W.}\ \bibnamefont
  {Graham}}, \bibinfo {author} {\bibfnamefont {J.~M.}\ \bibnamefont {Hogan}}, \
  and\ \bibinfo {author} {\bibfnamefont {M.~A.}\ \bibnamefont {Kasevich}},\
  }\bibfield  {title} {\enquote {\bibinfo {title} {Testing general relativity
  with atom interferometry},}\ }\href {\doibase 10.1103/PhysRevLett.98.111102}
  {\bibfield  {journal} {\bibinfo  {journal} {Phys. Rev. Lett.}\ }\textbf
  {\bibinfo {volume} {98}},\ \bibinfo {pages} {111102} (\bibinfo {year}
  {2007})}\BibitemShut {NoStop}%
\bibitem [{\citenamefont {Dimopoulos}\ \emph
  {et~al.}(2008{\natexlab{b}})\citenamefont {Dimopoulos}, \citenamefont
  {Graham}, \citenamefont {Hogan},\ and\ \citenamefont
  {Kasevich}}]{Dimopoulos2008}%
  \BibitemOpen
  \bibfield  {author} {\bibinfo {author} {\bibfnamefont {S.}~\bibnamefont
  {Dimopoulos}}, \bibinfo {author} {\bibfnamefont {P.~W.}\ \bibnamefont
  {Graham}}, \bibinfo {author} {\bibfnamefont {J.~M.}\ \bibnamefont {Hogan}}, \
  and\ \bibinfo {author} {\bibfnamefont {M.~A.}\ \bibnamefont {Kasevich}},\
  }\bibfield  {title} {\enquote {\bibinfo {title} {General relativistic effects
  in atom interferometry},}\ }\href {\doibase 10.1103/PhysRevD.78.042003}
  {\bibfield  {journal} {\bibinfo  {journal} {Phys. Rev. D}\ }\textbf {\bibinfo
  {volume} {78}},\ \bibinfo {pages} {042003} (\bibinfo {year}
  {2008}{\natexlab{b}})}\BibitemShut {NoStop}%
\bibitem [{\citenamefont {Asenbaum}\ \emph {et~al.}(2020)\citenamefont
  {Asenbaum}, \citenamefont {Overstreet}, \citenamefont {Kim}, \citenamefont
  {Curti},\ and\ \citenamefont {Kasevich}}]{Asenbaum2020}%
  \BibitemOpen
  \bibfield  {author} {\bibinfo {author} {\bibfnamefont {P.}~\bibnamefont
  {Asenbaum}}, \bibinfo {author} {\bibfnamefont {C.}~\bibnamefont
  {Overstreet}}, \bibinfo {author} {\bibfnamefont {M.}~\bibnamefont {Kim}},
  \bibinfo {author} {\bibfnamefont {J.}~\bibnamefont {Curti}}, \ and\ \bibinfo
  {author} {\bibfnamefont {M.~A.}\ \bibnamefont {Kasevich}},\ }\bibfield
  {title} {\enquote {\bibinfo {title} {Atom-interferometric test of the
  equivalence principle at the ${10}^{\ensuremath{-}12}$ level},}\ }\href
  {\doibase 10.1103/PhysRevLett.125.191101} {\bibfield  {journal} {\bibinfo
  {journal} {Phys. Rev. Lett.}\ }\textbf {\bibinfo {volume} {125}},\ \bibinfo
  {pages} {191101} (\bibinfo {year} {2020})}\BibitemShut {NoStop}%
\bibitem [{\citenamefont {Badurina}\ \emph {et~al.}(2020)\citenamefont
  {Badurina}, \citenamefont {Bentine}, \citenamefont {Blas}, \citenamefont
  {Bongs}, \citenamefont {Bortoletto}, \citenamefont {Bowcock}, \citenamefont
  {Bridges}, \citenamefont {Bowden}, \citenamefont {Buchmueller}, \citenamefont
  {Burrage}, \citenamefont {Coleman}, \citenamefont {Elertas}, \citenamefont
  {Ellis}, \citenamefont {Foot}, \citenamefont {Gibson}, \citenamefont
  {Haehnelt}, \citenamefont {Harte}, \citenamefont {Hedges}, \citenamefont
  {Hobson},\ and\ \citenamefont {Wilmut}}]{Badurina2020}%
  \BibitemOpen
  \bibfield  {author} {\bibinfo {author} {\bibfnamefont {L.}~\bibnamefont
  {Badurina}}, \bibinfo {author} {\bibfnamefont {E.}~\bibnamefont {Bentine}},
  \bibinfo {author} {\bibfnamefont {D.}~\bibnamefont {Blas}}, \bibinfo {author}
  {\bibfnamefont {K.}~\bibnamefont {Bongs}}, \bibinfo {author} {\bibfnamefont
  {D.}~\bibnamefont {Bortoletto}}, \bibinfo {author} {\bibfnamefont
  {T.}~\bibnamefont {Bowcock}}, \bibinfo {author} {\bibfnamefont
  {K.}~\bibnamefont {Bridges}}, \bibinfo {author} {\bibfnamefont
  {W.}~\bibnamefont {Bowden}}, \bibinfo {author} {\bibfnamefont
  {O.}~\bibnamefont {Buchmueller}}, \bibinfo {author} {\bibfnamefont
  {C.}~\bibnamefont {Burrage}}, \bibinfo {author} {\bibfnamefont
  {J.}~\bibnamefont {Coleman}}, \bibinfo {author} {\bibfnamefont
  {G.}~\bibnamefont {Elertas}}, \bibinfo {author} {\bibfnamefont
  {J.}~\bibnamefont {Ellis}}, \bibinfo {author} {\bibfnamefont
  {C.}~\bibnamefont {Foot}}, \bibinfo {author} {\bibfnamefont {V.}~\bibnamefont
  {Gibson}}, \bibinfo {author} {\bibfnamefont {M.~G.}\ \bibnamefont
  {Haehnelt}}, \bibinfo {author} {\bibfnamefont {T.}~\bibnamefont {Harte}},
  \bibinfo {author} {\bibfnamefont {S.}~\bibnamefont {Hedges}}, \bibinfo
  {author} {\bibfnamefont {R.}~\bibnamefont {Hobson}}, \ and\ \bibinfo {author}
  {\bibfnamefont {I.}~\bibnamefont {Wilmut}},\ }\bibfield  {title} {\enquote
  {\bibinfo {title} {Aion: an atom interferometer observatory and network},}\
  }\href {\doibase 10.1088/1475-7516/2020/05/011} {\bibfield  {journal}
  {\bibinfo  {journal} {J. Cosmol. Astropart. Phys.}\ }\textbf {\bibinfo
  {volume} {2020}},\ \bibinfo {pages} {011--011} (\bibinfo {year}
  {2020})}\BibitemShut {NoStop}%
\bibitem [{\citenamefont {Canuel}\ \emph {et~al.}(2018)\citenamefont {Canuel},
  \citenamefont {Bertoldi}, \citenamefont {Amand}, \citenamefont {Pozzo
  Di~Borgo}, \citenamefont {Chantrait}, \citenamefont {Danquigny},
  \citenamefont {Fang}, \citenamefont {Freise}, \citenamefont {Geiger},
  \citenamefont {Gillot}, \citenamefont {Henry}, \citenamefont {Hinderer},
  \citenamefont {Holleville}, \citenamefont {Junca}, \citenamefont {Lefevre},
  \citenamefont {Merzougui}, \citenamefont {Mielec}, \citenamefont {Monfret},
  \citenamefont {Pelisson}, \citenamefont {Prevedelli}, \citenamefont
  {Reynaud}, \citenamefont {Riou}, \citenamefont {Rogister}, \citenamefont
  {Rosat}, \citenamefont {Cormier}, \citenamefont {Landragin}, \citenamefont
  {Chaibi}, \citenamefont {Gaffet},\ and\ \citenamefont {Bouyer}}]{Canuel2018}%
  \BibitemOpen
  \bibfield  {author} {\bibinfo {author} {\bibfnamefont {B.}~\bibnamefont
  {Canuel}}, \bibinfo {author} {\bibfnamefont {A.}~\bibnamefont {Bertoldi}},
  \bibinfo {author} {\bibfnamefont {L.}~\bibnamefont {Amand}}, \bibinfo
  {author} {\bibfnamefont {E.}~\bibnamefont {Pozzo Di~Borgo}}, \bibinfo
  {author} {\bibfnamefont {T.}~\bibnamefont {Chantrait}}, \bibinfo {author}
  {\bibfnamefont {C.}~\bibnamefont {Danquigny}}, \bibinfo {author}
  {\bibfnamefont {B.}~\bibnamefont {Fang}}, \bibinfo {author} {\bibfnamefont
  {A.}~\bibnamefont {Freise}}, \bibinfo {author} {\bibfnamefont
  {R.}~\bibnamefont {Geiger}}, \bibinfo {author} {\bibfnamefont
  {J.}~\bibnamefont {Gillot}}, \bibinfo {author} {\bibfnamefont
  {S.}~\bibnamefont {Henry}}, \bibinfo {author} {\bibfnamefont
  {J.}~\bibnamefont {Hinderer}}, \bibinfo {author} {\bibfnamefont
  {D.}~\bibnamefont {Holleville}}, \bibinfo {author} {\bibfnamefont
  {J.}~\bibnamefont {Junca}}, \bibinfo {author} {\bibfnamefont
  {G.}~\bibnamefont {Lefevre}}, \bibinfo {author} {\bibfnamefont
  {M.}~\bibnamefont {Merzougui}}, \bibinfo {author} {\bibfnamefont
  {N.}~\bibnamefont {Mielec}}, \bibinfo {author} {\bibfnamefont
  {T.}~\bibnamefont {Monfret}}, \bibinfo {author} {\bibfnamefont
  {S.}~\bibnamefont {Pelisson}}, \bibinfo {author} {\bibfnamefont
  {M.}~\bibnamefont {Prevedelli}}, \bibinfo {author} {\bibfnamefont
  {S.}~\bibnamefont {Reynaud}}, \bibinfo {author} {\bibfnamefont
  {I.}~\bibnamefont {Riou}}, \bibinfo {author} {\bibfnamefont {Y.}~\bibnamefont
  {Rogister}}, \bibinfo {author} {\bibfnamefont {S.}~\bibnamefont {Rosat}},
  \bibinfo {author} {\bibfnamefont {E.}~\bibnamefont {Cormier}}, \bibinfo
  {author} {\bibfnamefont {A.}~\bibnamefont {Landragin}}, \bibinfo {author}
  {\bibfnamefont {W.}~\bibnamefont {Chaibi}}, \bibinfo {author} {\bibfnamefont
  {S.}~\bibnamefont {Gaffet}}, \ and\ \bibinfo {author} {\bibfnamefont
  {P.}~\bibnamefont {Bouyer}},\ }\bibfield  {title} {\enquote {\bibinfo {title}
  {{Exploring gravity with the MIGA large scale atom interferometer}},}\ }\href
  {\doibase 10.1038/s41598-018-32165-z} {\bibfield  {journal} {\bibinfo
  {journal} {{Sci. Rep.}}\ }\textbf {\bibinfo {volume} {8}},\ \bibinfo {pages}
  {14064} (\bibinfo {year} {2018})}\BibitemShut {NoStop}%
\bibitem [{\citenamefont {Canuel}\ \emph {et~al.}(2020)\citenamefont {Canuel},
  \citenamefont {Abend}, \citenamefont {Amaro-Seoane}, \citenamefont
  {Badaracco}, \citenamefont {Beaufils}, \citenamefont {Bertoldi},
  \citenamefont {Bongs}, \citenamefont {Bouyer}, \citenamefont {Braxmaier},
  \citenamefont {Chaibi} \emph {et~al.}}]{Canuel2020}%
  \BibitemOpen
  \bibfield  {author} {\bibinfo {author} {\bibfnamefont {B.}~\bibnamefont
  {Canuel}}, \bibinfo {author} {\bibfnamefont {S.}~\bibnamefont {Abend}},
  \bibinfo {author} {\bibfnamefont {P.}~\bibnamefont {Amaro-Seoane}}, \bibinfo
  {author} {\bibfnamefont {F.}~\bibnamefont {Badaracco}}, \bibinfo {author}
  {\bibfnamefont {Q.}~\bibnamefont {Beaufils}}, \bibinfo {author}
  {\bibfnamefont {A.}~\bibnamefont {Bertoldi}}, \bibinfo {author}
  {\bibfnamefont {K.}~\bibnamefont {Bongs}}, \bibinfo {author} {\bibfnamefont
  {P.}~\bibnamefont {Bouyer}}, \bibinfo {author} {\bibfnamefont
  {C.}~\bibnamefont {Braxmaier}}, \bibinfo {author} {\bibfnamefont
  {W.}~\bibnamefont {Chaibi}},  \emph {et~al.},\ }\bibfield  {title} {\enquote
  {\bibinfo {title} {Elgar—a european laboratory for gravitation and
  atom-interferometric research},}\ }\href {\doibase 10.1088/1361-6382/aba80e}
  {\bibfield  {journal} {\bibinfo  {journal} {Class. Quantum Gravity}\ }\textbf
  {\bibinfo {volume} {37}},\ \bibinfo {pages} {225017} (\bibinfo {year}
  {2020})}\BibitemShut {NoStop}%
\bibitem [{\citenamefont {Zhan}\ \emph {et~al.}(2019)\citenamefont {Zhan},
  \citenamefont {Wang}, \citenamefont {Ni}, \citenamefont {Gao}, \citenamefont
  {Wang}, \citenamefont {He}, \citenamefont {Li}, \citenamefont {Zhou},
  \citenamefont {Chen}, \citenamefont {Zhong}, \citenamefont {Tang},
  \citenamefont {Yao}, \citenamefont {Zhu}, \citenamefont {Xiong},
  \citenamefont {Lu}, \citenamefont {Yu}, \citenamefont {Cheng}, \citenamefont
  {Liu}, \citenamefont {Liang}, \citenamefont {Xu}, \citenamefont {He},
  \citenamefont {Ke}, \citenamefont {Tan},\ and\ \citenamefont
  {Luo}}]{Zhan2019}%
  \BibitemOpen
  \bibfield  {author} {\bibinfo {author} {\bibfnamefont {M.-S.}\ \bibnamefont
  {Zhan}}, \bibinfo {author} {\bibfnamefont {J.}~\bibnamefont {Wang}}, \bibinfo
  {author} {\bibfnamefont {W.-T.}\ \bibnamefont {Ni}}, \bibinfo {author}
  {\bibfnamefont {D.-F.}\ \bibnamefont {Gao}}, \bibinfo {author} {\bibfnamefont
  {G.}~\bibnamefont {Wang}}, \bibinfo {author} {\bibfnamefont {L.-X.}\
  \bibnamefont {He}}, \bibinfo {author} {\bibfnamefont {R.-B.}\ \bibnamefont
  {Li}}, \bibinfo {author} {\bibfnamefont {L.}~\bibnamefont {Zhou}}, \bibinfo
  {author} {\bibfnamefont {X.}~\bibnamefont {Chen}}, \bibinfo {author}
  {\bibfnamefont {J.-Q.}\ \bibnamefont {Zhong}}, \bibinfo {author}
  {\bibfnamefont {B.}~\bibnamefont {Tang}}, \bibinfo {author} {\bibfnamefont
  {Z.-W.}\ \bibnamefont {Yao}}, \bibinfo {author} {\bibfnamefont
  {L.}~\bibnamefont {Zhu}}, \bibinfo {author} {\bibfnamefont {Z.-Y.}\
  \bibnamefont {Xiong}}, \bibinfo {author} {\bibfnamefont {S.-B.}\ \bibnamefont
  {Lu}}, \bibinfo {author} {\bibfnamefont {G.-H.}\ \bibnamefont {Yu}}, \bibinfo
  {author} {\bibfnamefont {Q.-F.}\ \bibnamefont {Cheng}}, \bibinfo {author}
  {\bibfnamefont {M.}~\bibnamefont {Liu}}, \bibinfo {author} {\bibfnamefont
  {Y.-R.}\ \bibnamefont {Liang}}, \bibinfo {author} {\bibfnamefont
  {P.}~\bibnamefont {Xu}}, \bibinfo {author} {\bibfnamefont {X.-D.}\
  \bibnamefont {He}}, \bibinfo {author} {\bibfnamefont {M.}~\bibnamefont {Ke}},
  \bibinfo {author} {\bibfnamefont {Z.}~\bibnamefont {Tan}}, \ and\ \bibinfo
  {author} {\bibfnamefont {J.}~\bibnamefont {Luo}},\ }\bibfield  {title}
  {\enquote {\bibinfo {title} {{ZAIGA}: Zhaoshan long-baseline atom
  interferometer gravitation antenna},}\ }\href {\doibase
  10.1142/s0218271819400054} {\bibfield  {journal} {\bibinfo  {journal} {Int.
  J. Mod. Phys. D}\ }\textbf {\bibinfo {volume} {29}},\ \bibinfo {pages}
  {1940005} (\bibinfo {year} {2019})}\BibitemShut {NoStop}%
\bibitem [{\citenamefont {Di~Pumpo}\ \emph {et~al.}(2023)\citenamefont
  {Di~Pumpo}, \citenamefont {Friedrich}, \citenamefont {Ufrecht},\ and\
  \citenamefont {Giese}}]{DiPumpo2023a}%
  \BibitemOpen
  \bibfield  {author} {\bibinfo {author} {\bibfnamefont {F.}~\bibnamefont
  {Di~Pumpo}}, \bibinfo {author} {\bibfnamefont {A.}~\bibnamefont {Friedrich}},
  \bibinfo {author} {\bibfnamefont {C.}~\bibnamefont {Ufrecht}}, \ and\
  \bibinfo {author} {\bibfnamefont {E.}~\bibnamefont {Giese}},\ }\bibfield
  {title} {\enquote {\bibinfo {title} {Universality-of-clock-rates test using
  atom interferometry with ${T}^{3}$ scaling},}\ }\href {\doibase
  10.1103/PhysRevD.107.064007} {\bibfield  {journal} {\bibinfo  {journal}
  {Phys. Rev. D}\ }\textbf {\bibinfo {volume} {107}},\ \bibinfo {pages}
  {064007} (\bibinfo {year} {2023})}\BibitemShut {NoStop}%
\bibitem [{\citenamefont {Will}(2014)}]{Will2014}%
  \BibitemOpen
  \bibfield  {author} {\bibinfo {author} {\bibfnamefont {C.~M.}\ \bibnamefont
  {Will}},\ }\bibfield  {title} {\enquote {\bibinfo {title} {{The confrontation
  between general relativity and experiment}},}\ }\href {\doibase
  10.12942/lrr-2014-4} {\bibfield  {journal} {\bibinfo  {journal} {Living Rev.
  Relativ.}\ }\textbf {\bibinfo {volume} {17}},\ \bibinfo {pages} {4} (\bibinfo
  {year} {2014})}\BibitemShut {NoStop}%
\bibitem [{\citenamefont {Di~Casola}, \citenamefont {Liberati},\ and\
  \citenamefont {Sonego}(2015)}]{DiCasola2015}%
  \BibitemOpen
  \bibfield  {author} {\bibinfo {author} {\bibfnamefont {E.}~\bibnamefont
  {Di~Casola}}, \bibinfo {author} {\bibfnamefont {S.}~\bibnamefont {Liberati}},
  \ and\ \bibinfo {author} {\bibfnamefont {S.}~\bibnamefont {Sonego}},\
  }\bibfield  {title} {\enquote {\bibinfo {title} {Nonequivalence of
  equivalence principles},}\ }\href {\doibase 10.1119/1.4895342} {\bibfield
  {journal} {\bibinfo  {journal} {Am. J. Phys.}\ }\textbf {\bibinfo {volume}
  {83}},\ \bibinfo {pages} {39--46} (\bibinfo {year} {2015})}\BibitemShut
  {NoStop}%
\bibitem [{\citenamefont {Vessot}\ \emph {et~al.}(1980)\citenamefont {Vessot},
  \citenamefont {Levine}, \citenamefont {Mattison}, \citenamefont {Blomberg},
  \citenamefont {Hoffman}, \citenamefont {Nystrom}, \citenamefont {Farrel},
  \citenamefont {Decher}, \citenamefont {Eby}, \citenamefont {Baugher},
  \citenamefont {Watts}, \citenamefont {Teuber},\ and\ \citenamefont
  {Wills}}]{Vessot1980}%
  \BibitemOpen
  \bibfield  {author} {\bibinfo {author} {\bibfnamefont {R.~F.~C.}\
  \bibnamefont {Vessot}}, \bibinfo {author} {\bibfnamefont {M.~W.}\
  \bibnamefont {Levine}}, \bibinfo {author} {\bibfnamefont {E.~M.}\
  \bibnamefont {Mattison}}, \bibinfo {author} {\bibfnamefont {E.~L.}\
  \bibnamefont {Blomberg}}, \bibinfo {author} {\bibfnamefont {T.~E.}\
  \bibnamefont {Hoffman}}, \bibinfo {author} {\bibfnamefont {G.~U.}\
  \bibnamefont {Nystrom}}, \bibinfo {author} {\bibfnamefont {B.~F.}\
  \bibnamefont {Farrel}}, \bibinfo {author} {\bibfnamefont {R.}~\bibnamefont
  {Decher}}, \bibinfo {author} {\bibfnamefont {P.~B.}\ \bibnamefont {Eby}},
  \bibinfo {author} {\bibfnamefont {C.~R.}\ \bibnamefont {Baugher}}, \bibinfo
  {author} {\bibfnamefont {J.~W.}\ \bibnamefont {Watts}}, \bibinfo {author}
  {\bibfnamefont {D.~L.}\ \bibnamefont {Teuber}}, \ and\ \bibinfo {author}
  {\bibfnamefont {F.~D.}\ \bibnamefont {Wills}},\ }\bibfield  {title} {\enquote
  {\bibinfo {title} {Test of relativistic gravitation with a space-borne
  hydrogen maser},}\ }\href {\doibase 10.1103/PhysRevLett.45.2081} {\bibfield
  {journal} {\bibinfo  {journal} {Phys. Rev. Lett.}\ }\textbf {\bibinfo
  {volume} {45}},\ \bibinfo {pages} {2081--2084} (\bibinfo {year}
  {1980})}\BibitemShut {NoStop}%
\bibitem [{\citenamefont {Chou}\ \emph {et~al.}(2010)\citenamefont {Chou},
  \citenamefont {Hume}, \citenamefont {Rosenband},\ and\ \citenamefont
  {Wineland}}]{Chou2010}%
  \BibitemOpen
  \bibfield  {author} {\bibinfo {author} {\bibfnamefont {C.~W.}\ \bibnamefont
  {Chou}}, \bibinfo {author} {\bibfnamefont {D.~B.}\ \bibnamefont {Hume}},
  \bibinfo {author} {\bibfnamefont {T.}~\bibnamefont {Rosenband}}, \ and\
  \bibinfo {author} {\bibfnamefont {D.~J.}\ \bibnamefont {Wineland}},\
  }\bibfield  {title} {\enquote {\bibinfo {title} {Optical clocks and
  relativity},}\ }\href {\doibase 10.1126/science.1192720} {\bibfield
  {journal} {\bibinfo  {journal} {Science}\ }\textbf {\bibinfo {volume}
  {329}},\ \bibinfo {pages} {1630--1633} (\bibinfo {year} {2010})}\BibitemShut
  {NoStop}%
\bibitem [{\citenamefont {Herrmann}\ \emph {et~al.}(2018)\citenamefont
  {Herrmann}, \citenamefont {Finke}, \citenamefont {L\"ulf}, \citenamefont
  {Kichakova}, \citenamefont {Puetzfeld}, \citenamefont {Knickmann},
  \citenamefont {List}, \citenamefont {Rievers}, \citenamefont {Giorgi},
  \citenamefont {G\"unther}, \citenamefont {Dittus}, \citenamefont
  {Prieto-Cerdeira}, \citenamefont {Dilssner}, \citenamefont {Gonzalez},
  \citenamefont {Sch\"onemann}, \citenamefont {Ventura-Traveset},\ and\
  \citenamefont {L\"ammerzahl}}]{Herrmann2018}%
  \BibitemOpen
  \bibfield  {author} {\bibinfo {author} {\bibfnamefont {S.}~\bibnamefont
  {Herrmann}}, \bibinfo {author} {\bibfnamefont {F.}~\bibnamefont {Finke}},
  \bibinfo {author} {\bibfnamefont {M.}~\bibnamefont {L\"ulf}}, \bibinfo
  {author} {\bibfnamefont {O.}~\bibnamefont {Kichakova}}, \bibinfo {author}
  {\bibfnamefont {D.}~\bibnamefont {Puetzfeld}}, \bibinfo {author}
  {\bibfnamefont {D.}~\bibnamefont {Knickmann}}, \bibinfo {author}
  {\bibfnamefont {M.}~\bibnamefont {List}}, \bibinfo {author} {\bibfnamefont
  {B.}~\bibnamefont {Rievers}}, \bibinfo {author} {\bibfnamefont
  {G.}~\bibnamefont {Giorgi}}, \bibinfo {author} {\bibfnamefont
  {C.}~\bibnamefont {G\"unther}}, \bibinfo {author} {\bibfnamefont
  {H.}~\bibnamefont {Dittus}}, \bibinfo {author} {\bibfnamefont
  {R.}~\bibnamefont {Prieto-Cerdeira}}, \bibinfo {author} {\bibfnamefont
  {F.}~\bibnamefont {Dilssner}}, \bibinfo {author} {\bibfnamefont
  {F.}~\bibnamefont {Gonzalez}}, \bibinfo {author} {\bibfnamefont
  {E.}~\bibnamefont {Sch\"onemann}}, \bibinfo {author} {\bibfnamefont
  {J.}~\bibnamefont {Ventura-Traveset}}, \ and\ \bibinfo {author}
  {\bibfnamefont {C.}~\bibnamefont {L\"ammerzahl}},\ }\bibfield  {title}
  {\enquote {\bibinfo {title} {Test of the gravitational redshift with
  {G}alileo satellites in an eccentric orbit},}\ }\href {\doibase
  10.1103/PhysRevLett.121.231102} {\bibfield  {journal} {\bibinfo  {journal}
  {Phys. Rev. Lett.}\ }\textbf {\bibinfo {volume} {121}},\ \bibinfo {pages}
  {231102} (\bibinfo {year} {2018})}\BibitemShut {NoStop}%
\bibitem [{\citenamefont {Delva}\ \emph {et~al.}(2019)\citenamefont {Delva},
  \citenamefont {Puchades}, \citenamefont {Sch{\"o}nemann}, \citenamefont
  {Dilssner}, \citenamefont {Courde}, \citenamefont {Bertone}, \citenamefont
  {Gonzalez}, \citenamefont {Hees}, \citenamefont {Poncin-Lafitte},
  \citenamefont {Meynadier}, \citenamefont {Prieto-Cerdeira}, \citenamefont
  {Sohet}, \citenamefont {Ventura-Traveset},\ and\ \citenamefont
  {Wolf}}]{Delva2019}%
  \BibitemOpen
  \bibfield  {author} {\bibinfo {author} {\bibfnamefont {P.}~\bibnamefont
  {Delva}}, \bibinfo {author} {\bibfnamefont {N.}~\bibnamefont {Puchades}},
  \bibinfo {author} {\bibfnamefont {E.}~\bibnamefont {Sch{\"o}nemann}},
  \bibinfo {author} {\bibfnamefont {F.}~\bibnamefont {Dilssner}}, \bibinfo
  {author} {\bibfnamefont {C.}~\bibnamefont {Courde}}, \bibinfo {author}
  {\bibfnamefont {S.}~\bibnamefont {Bertone}}, \bibinfo {author} {\bibfnamefont
  {F.}~\bibnamefont {Gonzalez}}, \bibinfo {author} {\bibfnamefont
  {A.}~\bibnamefont {Hees}}, \bibinfo {author} {\bibfnamefont {C.~L.}\
  \bibnamefont {Poncin-Lafitte}}, \bibinfo {author} {\bibfnamefont
  {F.}~\bibnamefont {Meynadier}}, \bibinfo {author} {\bibfnamefont
  {R.}~\bibnamefont {Prieto-Cerdeira}}, \bibinfo {author} {\bibfnamefont
  {B.}~\bibnamefont {Sohet}}, \bibinfo {author} {\bibfnamefont
  {J.}~\bibnamefont {Ventura-Traveset}}, \ and\ \bibinfo {author}
  {\bibfnamefont {P.}~\bibnamefont {Wolf}},\ }\bibfield  {title} {\enquote
  {\bibinfo {title} {A new test of gravitational redshift using {G}alileo
  satellites: The {GREAT} experiment},}\ }\href
  {http://www.sciencedirect.com/science/article/pii/S1631070519300271}
  {\bibfield  {journal} {\bibinfo  {journal} {C. R. Acad. Sci.}\ }\textbf
  {\bibinfo {volume} {20}},\ \bibinfo {pages} {176--182} (\bibinfo {year}
  {2019})}\BibitemShut {NoStop}%
\bibitem [{\citenamefont {Takamoto}\ \emph {et~al.}(2020)\citenamefont
  {Takamoto}, \citenamefont {Ushijima}, \citenamefont {Ohmae}, \citenamefont
  {Yahagi}, \citenamefont {Kokado}, \citenamefont {Shinkai},\ and\
  \citenamefont {Katori}}]{Takamoto2020}%
  \BibitemOpen
  \bibfield  {author} {\bibinfo {author} {\bibfnamefont {M.}~\bibnamefont
  {Takamoto}}, \bibinfo {author} {\bibfnamefont {I.}~\bibnamefont {Ushijima}},
  \bibinfo {author} {\bibfnamefont {N.}~\bibnamefont {Ohmae}}, \bibinfo
  {author} {\bibfnamefont {T.}~\bibnamefont {Yahagi}}, \bibinfo {author}
  {\bibfnamefont {K.}~\bibnamefont {Kokado}}, \bibinfo {author} {\bibfnamefont
  {H.}~\bibnamefont {Shinkai}}, \ and\ \bibinfo {author} {\bibfnamefont
  {H.}~\bibnamefont {Katori}},\ }\bibfield  {title} {\enquote {\bibinfo {title}
  {Test of general relativity by a pair of transportable optical lattice
  clocks},}\ }\href {\doibase 10.1038/s41566-020-0619-8} {\bibfield  {journal}
  {\bibinfo  {journal} {Nat. Photonics}\ }\textbf {\bibinfo {volume} {14}},\
  \bibinfo {pages} {411} (\bibinfo {year} {2020})}\BibitemShut {NoStop}%
\bibitem [{\citenamefont {Bothwell}\ \emph {et~al.}(2022)\citenamefont
  {Bothwell}, \citenamefont {Kennedy}, \citenamefont {Aeppli}, \citenamefont
  {Kedar}, \citenamefont {Robinson}, \citenamefont {Oelker}, \citenamefont
  {Staron},\ and\ \citenamefont {Ye}}]{Bothwell2022}%
  \BibitemOpen
  \bibfield  {author} {\bibinfo {author} {\bibfnamefont {T.}~\bibnamefont
  {Bothwell}}, \bibinfo {author} {\bibfnamefont {C.~J.}\ \bibnamefont
  {Kennedy}}, \bibinfo {author} {\bibfnamefont {A.}~\bibnamefont {Aeppli}},
  \bibinfo {author} {\bibfnamefont {D.}~\bibnamefont {Kedar}}, \bibinfo
  {author} {\bibfnamefont {J.~M.}\ \bibnamefont {Robinson}}, \bibinfo {author}
  {\bibfnamefont {E.}~\bibnamefont {Oelker}}, \bibinfo {author} {\bibfnamefont
  {A.}~\bibnamefont {Staron}}, \ and\ \bibinfo {author} {\bibfnamefont
  {J.}~\bibnamefont {Ye}},\ }\bibfield  {title} {\enquote {\bibinfo {title}
  {Resolving the gravitational redshift across a millimetre-scale atomic
  sample},}\ }\href {\doibase 10.1038/s41586-021-04349-7} {\bibfield  {journal}
  {\bibinfo  {journal} {Nature}\ }\textbf {\bibinfo {volume} {602}},\ \bibinfo
  {pages} {420--424} (\bibinfo {year} {2022})}\BibitemShut {NoStop}%
\bibitem [{\citenamefont {Brewer}\ \emph {et~al.}(2019)\citenamefont {Brewer},
  \citenamefont {Chen}, \citenamefont {Hankin}, \citenamefont {Clements},
  \citenamefont {Chou}, \citenamefont {Wineland}, \citenamefont {Hume},\ and\
  \citenamefont {Leibrandt}}]{Brewer2019}%
  \BibitemOpen
  \bibfield  {author} {\bibinfo {author} {\bibfnamefont {S.~M.}\ \bibnamefont
  {Brewer}}, \bibinfo {author} {\bibfnamefont {J.-S.}\ \bibnamefont {Chen}},
  \bibinfo {author} {\bibfnamefont {A.~M.}\ \bibnamefont {Hankin}}, \bibinfo
  {author} {\bibfnamefont {E.~R.}\ \bibnamefont {Clements}}, \bibinfo {author}
  {\bibfnamefont {C.~W.}\ \bibnamefont {Chou}}, \bibinfo {author}
  {\bibfnamefont {D.~J.}\ \bibnamefont {Wineland}}, \bibinfo {author}
  {\bibfnamefont {D.~B.}\ \bibnamefont {Hume}}, \ and\ \bibinfo {author}
  {\bibfnamefont {D.~R.}\ \bibnamefont {Leibrandt}},\ }\bibfield  {title}
  {\enquote {\bibinfo {title} {${}^{27}${A}l$^{+ }$ {Q}uantum-logic clock with
  a systematic uncertainty below $10^{-18}$},}\ }\href {\doibase
  10.1103/PhysRevLett.123.033201} {\bibfield  {journal} {\bibinfo  {journal}
  {Phys. Rev. Lett.}\ }\textbf {\bibinfo {volume} {123}},\ \bibinfo {pages}
  {033201} (\bibinfo {year} {2019})}\BibitemShut {NoStop}%
\bibitem [{\citenamefont {Oelker}\ \emph {et~al.}(2019)\citenamefont {Oelker},
  \citenamefont {Hutson}, \citenamefont {Kennedy}, \citenamefont {Sonderhouse},
  \citenamefont {Bothwell}, \citenamefont {Goban}, \citenamefont {Kedar},
  \citenamefont {Sanner}, \citenamefont {Robinson}, \citenamefont {Marti},
  \citenamefont {Matei}, \citenamefont {Legero}, \citenamefont {Giunta},
  \citenamefont {Holzwarth}, \citenamefont {Riehle}, \citenamefont {Sterr},\
  and\ \citenamefont {Ye}}]{Oelker2019}%
  \BibitemOpen
  \bibfield  {author} {\bibinfo {author} {\bibfnamefont {E.}~\bibnamefont
  {Oelker}}, \bibinfo {author} {\bibfnamefont {R.~B.}\ \bibnamefont {Hutson}},
  \bibinfo {author} {\bibfnamefont {C.~J.}\ \bibnamefont {Kennedy}}, \bibinfo
  {author} {\bibfnamefont {L.}~\bibnamefont {Sonderhouse}}, \bibinfo {author}
  {\bibfnamefont {T.}~\bibnamefont {Bothwell}}, \bibinfo {author}
  {\bibfnamefont {A.}~\bibnamefont {Goban}}, \bibinfo {author} {\bibfnamefont
  {D.}~\bibnamefont {Kedar}}, \bibinfo {author} {\bibfnamefont
  {C.}~\bibnamefont {Sanner}}, \bibinfo {author} {\bibfnamefont {J.~M.}\
  \bibnamefont {Robinson}}, \bibinfo {author} {\bibfnamefont {G.~E.}\
  \bibnamefont {Marti}}, \bibinfo {author} {\bibfnamefont {D.~G.}\ \bibnamefont
  {Matei}}, \bibinfo {author} {\bibfnamefont {T.}~\bibnamefont {Legero}},
  \bibinfo {author} {\bibfnamefont {M.}~\bibnamefont {Giunta}}, \bibinfo
  {author} {\bibfnamefont {R.}~\bibnamefont {Holzwarth}}, \bibinfo {author}
  {\bibfnamefont {F.}~\bibnamefont {Riehle}}, \bibinfo {author} {\bibfnamefont
  {U.}~\bibnamefont {Sterr}}, \ and\ \bibinfo {author} {\bibfnamefont
  {J.}~\bibnamefont {Ye}},\ }\bibfield  {title} {\enquote {\bibinfo {title}
  {Demonstration of 4.8 \texttimes{} 10${}^{-17}$ stability at 1 s for two
  independent optical clocks},}\ }\href
  {https://www.nature.com/articles/s41566-019-0493-4} {\bibfield  {journal}
  {\bibinfo  {journal} {Nat. Photonics}\ }\textbf {\bibinfo {volume} {13}},\
  \bibinfo {pages} {714--719} (\bibinfo {year} {2019})}\BibitemShut {NoStop}%
\bibitem [{\citenamefont {Madjarov}\ \emph {et~al.}(2019)\citenamefont
  {Madjarov}, \citenamefont {Cooper}, \citenamefont {Shaw}, \citenamefont
  {Covey}, \citenamefont {Schkolnik}, \citenamefont {Yoon}, \citenamefont
  {Williams},\ and\ \citenamefont {Endres}}]{Madjarov2019}%
  \BibitemOpen
  \bibfield  {author} {\bibinfo {author} {\bibfnamefont {I.~S.}\ \bibnamefont
  {Madjarov}}, \bibinfo {author} {\bibfnamefont {A.}~\bibnamefont {Cooper}},
  \bibinfo {author} {\bibfnamefont {A.~L.}\ \bibnamefont {Shaw}}, \bibinfo
  {author} {\bibfnamefont {J.~P.}\ \bibnamefont {Covey}}, \bibinfo {author}
  {\bibfnamefont {V.}~\bibnamefont {Schkolnik}}, \bibinfo {author}
  {\bibfnamefont {T.~H.}\ \bibnamefont {Yoon}}, \bibinfo {author}
  {\bibfnamefont {J.~R.}\ \bibnamefont {Williams}}, \ and\ \bibinfo {author}
  {\bibfnamefont {M.}~\bibnamefont {Endres}},\ }\bibfield  {title} {\enquote
  {\bibinfo {title} {An atomic-array optical clock with single-atom readout},}\
  }\href {\doibase 10.1103/PhysRevX.9.041052} {\bibfield  {journal} {\bibinfo
  {journal} {Phys. Rev. X}\ }\textbf {\bibinfo {volume} {9}},\ \bibinfo {pages}
  {041052} (\bibinfo {year} {2019})}\BibitemShut {NoStop}%
\bibitem [{\citenamefont {Touboul}\ \emph {et~al.}(2019)\citenamefont
  {Touboul}, \citenamefont {M{\'e}tris}, \citenamefont {Rodrigues},
  \citenamefont {Andr{\'e}}, \citenamefont {Baghi}, \citenamefont {Berg{\'e}},
  \citenamefont {Boulanger}, \citenamefont {Bremer}, \citenamefont {Chhun},
  \citenamefont {Christophe}, \citenamefont {Cipolla}, \citenamefont {Damour},
  \citenamefont {Danto}, \citenamefont {Dittus}, \citenamefont {Fayet},
  \citenamefont {Foulon}, \citenamefont {Guidotti}, \citenamefont {Hardy},
  \citenamefont {Huynh}, \citenamefont {L{\"a}mmerzahl}, \citenamefont {Lebat},
  \citenamefont {Liorzou}, \citenamefont {List}, \citenamefont {Panet},
  \citenamefont {Pires}, \citenamefont {Pouilloux}, \citenamefont {Prieur},
  \citenamefont {Reynaud}, \citenamefont {Rievers}, \citenamefont {Robert},
  \citenamefont {Selig}, \citenamefont {Serron}, \citenamefont {Sumner},\ and\
  \citenamefont {Visser}}]{Touboul2019}%
  \BibitemOpen
  \bibfield  {author} {\bibinfo {author} {\bibfnamefont {P.}~\bibnamefont
  {Touboul}}, \bibinfo {author} {\bibfnamefont {G.}~\bibnamefont {M{\'e}tris}},
  \bibinfo {author} {\bibfnamefont {M.}~\bibnamefont {Rodrigues}}, \bibinfo
  {author} {\bibfnamefont {Y.}~\bibnamefont {Andr{\'e}}}, \bibinfo {author}
  {\bibfnamefont {Q.}~\bibnamefont {Baghi}}, \bibinfo {author} {\bibfnamefont
  {J.}~\bibnamefont {Berg{\'e}}}, \bibinfo {author} {\bibfnamefont
  {D.}~\bibnamefont {Boulanger}}, \bibinfo {author} {\bibfnamefont
  {S.}~\bibnamefont {Bremer}}, \bibinfo {author} {\bibfnamefont
  {R.}~\bibnamefont {Chhun}}, \bibinfo {author} {\bibfnamefont
  {B.}~\bibnamefont {Christophe}}, \bibinfo {author} {\bibfnamefont
  {V.}~\bibnamefont {Cipolla}}, \bibinfo {author} {\bibfnamefont
  {T.}~\bibnamefont {Damour}}, \bibinfo {author} {\bibfnamefont
  {P.}~\bibnamefont {Danto}}, \bibinfo {author} {\bibfnamefont
  {H.}~\bibnamefont {Dittus}}, \bibinfo {author} {\bibfnamefont
  {P.}~\bibnamefont {Fayet}}, \bibinfo {author} {\bibfnamefont
  {B.}~\bibnamefont {Foulon}}, \bibinfo {author} {\bibfnamefont {P.-Y.}\
  \bibnamefont {Guidotti}}, \bibinfo {author} {\bibfnamefont {E.}~\bibnamefont
  {Hardy}}, \bibinfo {author} {\bibfnamefont {P.-A.}\ \bibnamefont {Huynh}},
  \bibinfo {author} {\bibfnamefont {C.}~\bibnamefont {L{\"a}mmerzahl}},
  \bibinfo {author} {\bibfnamefont {V.}~\bibnamefont {Lebat}}, \bibinfo
  {author} {\bibfnamefont {F.}~\bibnamefont {Liorzou}}, \bibinfo {author}
  {\bibfnamefont {M.}~\bibnamefont {List}}, \bibinfo {author} {\bibfnamefont
  {I.}~\bibnamefont {Panet}}, \bibinfo {author} {\bibfnamefont
  {S.}~\bibnamefont {Pires}}, \bibinfo {author} {\bibfnamefont
  {B.}~\bibnamefont {Pouilloux}}, \bibinfo {author} {\bibfnamefont
  {P.}~\bibnamefont {Prieur}}, \bibinfo {author} {\bibfnamefont
  {S.}~\bibnamefont {Reynaud}}, \bibinfo {author} {\bibfnamefont
  {B.}~\bibnamefont {Rievers}}, \bibinfo {author} {\bibfnamefont
  {A.}~\bibnamefont {Robert}}, \bibinfo {author} {\bibfnamefont
  {H.}~\bibnamefont {Selig}}, \bibinfo {author} {\bibfnamefont
  {L.}~\bibnamefont {Serron}}, \bibinfo {author} {\bibfnamefont
  {T.}~\bibnamefont {Sumner}}, \ and\ \bibinfo {author} {\bibfnamefont
  {P.}~\bibnamefont {Visser}},\ }\bibfield  {title} {\enquote {\bibinfo {title}
  {{Space test of the equivalence principle: first results of the MICROSCOPE
  mission}},}\ }\href {\doibase 10.1088/1361-6382/ab4707} {\bibfield  {journal}
  {\bibinfo  {journal} {Class. Quantum Gravity}\ }\textbf {\bibinfo {volume}
  {36}},\ \bibinfo {eid} {225006} (\bibinfo {year} {2019})}\BibitemShut
  {NoStop}%
\bibitem [{\citenamefont {Schlippert}\ \emph {et~al.}(2014)\citenamefont
  {Schlippert}, \citenamefont {Hartwig}, \citenamefont {Albers}, \citenamefont
  {Richardson}, \citenamefont {Schubert}, \citenamefont {Roura}, \citenamefont
  {Schleich}, \citenamefont {Ertmer},\ and\ \citenamefont
  {Rasel}}]{Schlippert2014}%
  \BibitemOpen
  \bibfield  {author} {\bibinfo {author} {\bibfnamefont {D.}~\bibnamefont
  {Schlippert}}, \bibinfo {author} {\bibfnamefont {J.}~\bibnamefont {Hartwig}},
  \bibinfo {author} {\bibfnamefont {H.}~\bibnamefont {Albers}}, \bibinfo
  {author} {\bibfnamefont {L.~L.}\ \bibnamefont {Richardson}}, \bibinfo
  {author} {\bibfnamefont {C.}~\bibnamefont {Schubert}}, \bibinfo {author}
  {\bibfnamefont {A.}~\bibnamefont {Roura}}, \bibinfo {author} {\bibfnamefont
  {W.~P.}\ \bibnamefont {Schleich}}, \bibinfo {author} {\bibfnamefont
  {W.}~\bibnamefont {Ertmer}}, \ and\ \bibinfo {author} {\bibfnamefont {E.~M.}\
  \bibnamefont {Rasel}},\ }\bibfield  {title} {\enquote {\bibinfo {title}
  {Quantum test of the universality of free fall},}\ }\href {\doibase
  10.1103/PhysRevLett.112.203002} {\bibfield  {journal} {\bibinfo  {journal}
  {Phys. Rev. Lett.}\ }\textbf {\bibinfo {volume} {112}},\ \bibinfo {pages}
  {203002} (\bibinfo {year} {2014})}\BibitemShut {NoStop}%
\bibitem [{\citenamefont {Archibald}\ \emph {et~al.}(2018)\citenamefont
  {Archibald}, \citenamefont {Gusinskaia}, \citenamefont {Hessels},
  \citenamefont {Deller}, \citenamefont {Kaplan}, \citenamefont {Lorimer},
  \citenamefont {Lynch}, \citenamefont {Ransom},\ and\ \citenamefont
  {Stairs}}]{Archibald2018}%
  \BibitemOpen
  \bibfield  {author} {\bibinfo {author} {\bibfnamefont {A.~M.}\ \bibnamefont
  {Archibald}}, \bibinfo {author} {\bibfnamefont {N.~V.}\ \bibnamefont
  {Gusinskaia}}, \bibinfo {author} {\bibfnamefont {J.~W.~T.}\ \bibnamefont
  {Hessels}}, \bibinfo {author} {\bibfnamefont {A.~T.}\ \bibnamefont {Deller}},
  \bibinfo {author} {\bibfnamefont {D.~L.}\ \bibnamefont {Kaplan}}, \bibinfo
  {author} {\bibfnamefont {D.~R.}\ \bibnamefont {Lorimer}}, \bibinfo {author}
  {\bibfnamefont {R.~S.}\ \bibnamefont {Lynch}}, \bibinfo {author}
  {\bibfnamefont {S.~M.}\ \bibnamefont {Ransom}}, \ and\ \bibinfo {author}
  {\bibfnamefont {I.~H.}\ \bibnamefont {Stairs}},\ }\bibfield  {title}
  {\enquote {\bibinfo {title} {{Universality of free fall from the orbital
  motion of a pulsar in a stellar triple system}},}\ }\href {\doibase
  10.1038/s41586-018-0265-1} {\bibfield  {journal} {\bibinfo  {journal}
  {Nature}\ }\textbf {\bibinfo {volume} {559}},\ \bibinfo {pages} {73--76}
  (\bibinfo {year} {2018})}\BibitemShut {NoStop}%
\bibitem [{\citenamefont {Zheng}\ \emph {et~al.}(2023)\citenamefont {Zheng},
  \citenamefont {Dolde}, \citenamefont {Cambria}, \citenamefont {Lim},\ and\
  \citenamefont {Kolkowitz}}]{Zheng2023}%
  \BibitemOpen
  \bibfield  {author} {\bibinfo {author} {\bibfnamefont {X.}~\bibnamefont
  {Zheng}}, \bibinfo {author} {\bibfnamefont {J.}~\bibnamefont {Dolde}},
  \bibinfo {author} {\bibfnamefont {M.}~\bibnamefont {Cambria}}, \bibinfo
  {author} {\bibfnamefont {H.}~\bibnamefont {Lim}}, \ and\ \bibinfo {author}
  {\bibfnamefont {S.}~\bibnamefont {Kolkowitz}},\ }\bibfield  {title} {\enquote
  {\bibinfo {title} {A lab-based test of the gravitational redshift with a
  miniature clock network},}\ }\href
  {https://www.nature.com/articles/s41467-023-40629-8} {\bibfield  {journal}
  {\bibinfo  {journal} {Nat. Commun.}\ }\textbf {\bibinfo {volume} {14}},\
  \bibinfo {pages} {4886} (\bibinfo {year} {2023})}\BibitemShut {NoStop}%
\bibitem [{\citenamefont {Rosi}\ \emph {et~al.}(2017)\citenamefont {Rosi},
  \citenamefont {D'Amico}, \citenamefont {Cacciapuoti}, \citenamefont
  {Sorrentino}, \citenamefont {Prevedelli}, \citenamefont {Zych}, \citenamefont
  {Brukner},\ and\ \citenamefont {Tino}}]{Rosi2017}%
  \BibitemOpen
  \bibfield  {author} {\bibinfo {author} {\bibfnamefont {G.}~\bibnamefont
  {Rosi}}, \bibinfo {author} {\bibfnamefont {G.}~\bibnamefont {D'Amico}},
  \bibinfo {author} {\bibfnamefont {L.}~\bibnamefont {Cacciapuoti}}, \bibinfo
  {author} {\bibfnamefont {F.}~\bibnamefont {Sorrentino}}, \bibinfo {author}
  {\bibfnamefont {M.}~\bibnamefont {Prevedelli}}, \bibinfo {author}
  {\bibfnamefont {M.}~\bibnamefont {Zych}}, \bibinfo {author} {\bibfnamefont
  {{\v{C}}.}~\bibnamefont {Brukner}}, \ and\ \bibinfo {author} {\bibfnamefont
  {G.~M.}\ \bibnamefont {Tino}},\ }\bibfield  {title} {\enquote {\bibinfo
  {title} {Quantum test of the equivalence principle for atoms in coherent
  superposition of internal energy states},}\ }\href
  {http://www.nature.com/articles/ncomms15529} {\bibfield  {journal} {\bibinfo
  {journal} {Nat. Commun.}\ }\textbf {\bibinfo {volume} {8}},\ \bibinfo {pages}
  {15529} (\bibinfo {year} {2017})}\BibitemShut {NoStop}%
\bibitem [{\citenamefont {Zhou}\ \emph {et~al.}(2021)\citenamefont {Zhou},
  \citenamefont {He}, \citenamefont {Yan}, \citenamefont {Chen}, \citenamefont
  {Gao}, \citenamefont {Duan}, \citenamefont {Ji}, \citenamefont {Xu},
  \citenamefont {Tang}, \citenamefont {Zhou}, \citenamefont {Barthwal},
  \citenamefont {Wang}, \citenamefont {Hou}, \citenamefont {Xiong},
  \citenamefont {Zhang}, \citenamefont {Liu}, \citenamefont {Ni}, \citenamefont
  {Wang},\ and\ \citenamefont {Zhan}}]{Zhou2021Aug}%
  \BibitemOpen
  \bibfield  {author} {\bibinfo {author} {\bibfnamefont {L.}~\bibnamefont
  {Zhou}}, \bibinfo {author} {\bibfnamefont {C.}~\bibnamefont {He}}, \bibinfo
  {author} {\bibfnamefont {S.-T.}\ \bibnamefont {Yan}}, \bibinfo {author}
  {\bibfnamefont {X.}~\bibnamefont {Chen}}, \bibinfo {author} {\bibfnamefont
  {D.-F.}\ \bibnamefont {Gao}}, \bibinfo {author} {\bibfnamefont {W.-T.}\
  \bibnamefont {Duan}}, \bibinfo {author} {\bibfnamefont {Y.-H.}\ \bibnamefont
  {Ji}}, \bibinfo {author} {\bibfnamefont {R.-D.}\ \bibnamefont {Xu}}, \bibinfo
  {author} {\bibfnamefont {B.}~\bibnamefont {Tang}}, \bibinfo {author}
  {\bibfnamefont {C.}~\bibnamefont {Zhou}}, \bibinfo {author} {\bibfnamefont
  {S.}~\bibnamefont {Barthwal}}, \bibinfo {author} {\bibfnamefont
  {Q.}~\bibnamefont {Wang}}, \bibinfo {author} {\bibfnamefont {Z.}~\bibnamefont
  {Hou}}, \bibinfo {author} {\bibfnamefont {Z.-Y.}\ \bibnamefont {Xiong}},
  \bibinfo {author} {\bibfnamefont {Y.-Z.}\ \bibnamefont {Zhang}}, \bibinfo
  {author} {\bibfnamefont {M.}~\bibnamefont {Liu}}, \bibinfo {author}
  {\bibfnamefont {W.-T.}\ \bibnamefont {Ni}}, \bibinfo {author} {\bibfnamefont
  {J.}~\bibnamefont {Wang}}, \ and\ \bibinfo {author} {\bibfnamefont {M.-S.}\
  \bibnamefont {Zhan}},\ }\bibfield  {title} {\enquote {\bibinfo {title}
  {{Joint mass-and-energy test of the equivalence principle at the
  ${10}^{\ensuremath{-}10}$ level using atoms with specified mass and internal
  energy}},}\ }\href {\doibase 10.1103/PhysRevA.104.022822} {\bibfield
  {journal} {\bibinfo  {journal} {Phys. Rev. A}\ }\textbf {\bibinfo {volume}
  {104}},\ \bibinfo {pages} {022822} (\bibinfo {year} {2021})}\BibitemShut
  {NoStop}%
\bibitem [{\citenamefont {Loriani}\ \emph {et~al.}(2019)\citenamefont
  {Loriani}, \citenamefont {Friedrich}, \citenamefont {Ufrecht}, \citenamefont
  {Di~Pumpo}, \citenamefont {Kleinert}, \citenamefont {Abend}, \citenamefont
  {Gaaloul}, \citenamefont {Meiners}, \citenamefont {Schubert}, \citenamefont
  {Tell}, \citenamefont {Wodey}, \citenamefont {Zych}, \citenamefont {Ertmer},
  \citenamefont {Roura}, \citenamefont {Schlippert}, \citenamefont {Schleich},
  \citenamefont {Rasel},\ and\ \citenamefont {Giese}}]{Loriani2019}%
  \BibitemOpen
  \bibfield  {author} {\bibinfo {author} {\bibfnamefont {S.}~\bibnamefont
  {Loriani}}, \bibinfo {author} {\bibfnamefont {A.}~\bibnamefont {Friedrich}},
  \bibinfo {author} {\bibfnamefont {C.}~\bibnamefont {Ufrecht}}, \bibinfo
  {author} {\bibfnamefont {F.}~\bibnamefont {Di~Pumpo}}, \bibinfo {author}
  {\bibfnamefont {S.}~\bibnamefont {Kleinert}}, \bibinfo {author}
  {\bibfnamefont {S.}~\bibnamefont {Abend}}, \bibinfo {author} {\bibfnamefont
  {N.}~\bibnamefont {Gaaloul}}, \bibinfo {author} {\bibfnamefont
  {C.}~\bibnamefont {Meiners}}, \bibinfo {author} {\bibfnamefont
  {C.}~\bibnamefont {Schubert}}, \bibinfo {author} {\bibfnamefont
  {D.}~\bibnamefont {Tell}}, \bibinfo {author} {\bibfnamefont
  {{\'{E}}.}~\bibnamefont {Wodey}}, \bibinfo {author} {\bibfnamefont
  {M.}~\bibnamefont {Zych}}, \bibinfo {author} {\bibfnamefont {W.}~\bibnamefont
  {Ertmer}}, \bibinfo {author} {\bibfnamefont {A.}~\bibnamefont {Roura}},
  \bibinfo {author} {\bibfnamefont {D.}~\bibnamefont {Schlippert}}, \bibinfo
  {author} {\bibfnamefont {W.~P.}\ \bibnamefont {Schleich}}, \bibinfo {author}
  {\bibfnamefont {E.~M.}\ \bibnamefont {Rasel}}, \ and\ \bibinfo {author}
  {\bibfnamefont {E.}~\bibnamefont {Giese}},\ }\bibfield  {title} {\enquote
  {\bibinfo {title} {Interference of clocks: {A} quantum twin paradox},}\
  }\href {\doibase 10.1126/sciadv.aax8966} {\bibfield  {journal} {\bibinfo
  {journal} {Sci. Adv.}\ }\textbf {\bibinfo {volume} {5}},\ \bibinfo {pages}
  {eaax8966} (\bibinfo {year} {2019})}\BibitemShut {NoStop}%
\bibitem [{\citenamefont {Giulini}(2012)}]{Giulini2012}%
  \BibitemOpen
  \bibfield  {author} {\bibinfo {author} {\bibfnamefont {D.}~\bibnamefont
  {Giulini}},\ }\enquote {\bibinfo {title} {Equivalence principle, quantum
  mechanics, and atom-interferometric tests},}\ in\ \href {\doibase
  10.1007/978-3-0348-0043-3_16} {\emph {\bibinfo {booktitle} {Quantum Field
  Theory and Gravity: {C}onceptual and Mathematical Advances in the Search for
  a Unified Framework}}},\ \bibinfo {editor} {edited by\ \bibinfo {editor}
  {\bibfnamefont {F.}~\bibnamefont {Finster}}, \bibinfo {editor} {\bibfnamefont
  {O.}~\bibnamefont {M{\"u}ller}}, \bibinfo {editor} {\bibfnamefont
  {M.}~\bibnamefont {Nardmann}}, \bibinfo {editor} {\bibfnamefont
  {J.}~\bibnamefont {Tolksdorf}}, \ and\ \bibinfo {editor} {\bibfnamefont
  {E.}~\bibnamefont {Zeidler}}}\ (\bibinfo  {publisher} {Springer Basel},\
  \bibinfo {address} {Basel},\ \bibinfo {year} {2012})\ pp.\ \bibinfo {pages}
  {345--370}\BibitemShut {NoStop}%
\bibitem [{\citenamefont {Roura}(2020)}]{Roura2020}%
  \BibitemOpen
  \bibfield  {author} {\bibinfo {author} {\bibfnamefont {A.}~\bibnamefont
  {Roura}},\ }\bibfield  {title} {\enquote {\bibinfo {title} {Gravitational
  redshift in quantum-clock interferometry},}\ }\href {\doibase
  10.1103/PhysRevX.10.021014} {\bibfield  {journal} {\bibinfo  {journal} {Phys.
  Rev. X}\ }\textbf {\bibinfo {volume} {10}},\ \bibinfo {pages} {021014}
  (\bibinfo {year} {2020})}\BibitemShut {NoStop}%
\bibitem [{\citenamefont {Ufrecht}\ and\ \citenamefont
  {Giese}(2020)}]{Ufrecht20202}%
  \BibitemOpen
  \bibfield  {author} {\bibinfo {author} {\bibfnamefont {C.}~\bibnamefont
  {Ufrecht}}\ and\ \bibinfo {author} {\bibfnamefont {E.}~\bibnamefont
  {Giese}},\ }\bibfield  {title} {\enquote {\bibinfo {title} {Perturbative
  operator approach to high-precision light-pulse atom interferometry},}\
  }\href {\doibase 10.1103/PhysRevA.101.053615} {\bibfield  {journal} {\bibinfo
   {journal} {Phys. Rev. A}\ }\textbf {\bibinfo {volume} {101}},\ \bibinfo
  {pages} {053615} (\bibinfo {year} {2020})}\BibitemShut {NoStop}%
\bibitem [{\citenamefont {Di~Pumpo}\ \emph {et~al.}(2021)\citenamefont
  {Di~Pumpo}, \citenamefont {Ufrecht}, \citenamefont {Friedrich}, \citenamefont
  {Giese}, \citenamefont {Schleich},\ and\ \citenamefont
  {Unruh}}]{DiPumpo2021}%
  \BibitemOpen
  \bibfield  {author} {\bibinfo {author} {\bibfnamefont {F.}~\bibnamefont
  {Di~Pumpo}}, \bibinfo {author} {\bibfnamefont {C.}~\bibnamefont {Ufrecht}},
  \bibinfo {author} {\bibfnamefont {A.}~\bibnamefont {Friedrich}}, \bibinfo
  {author} {\bibfnamefont {E.}~\bibnamefont {Giese}}, \bibinfo {author}
  {\bibfnamefont {W.~P.}\ \bibnamefont {Schleich}}, \ and\ \bibinfo {author}
  {\bibfnamefont {W.~G.}\ \bibnamefont {Unruh}},\ }\bibfield  {title} {\enquote
  {\bibinfo {title} {Gravitational redshift tests with atomic clocks and atom
  interferometers},}\ }\href {\doibase 10.1103/PRXQuantum.2.040333} {\bibfield
  {journal} {\bibinfo  {journal} {PRX Quantum}\ }\textbf {\bibinfo {volume}
  {2}},\ \bibinfo {pages} {040333} (\bibinfo {year} {2021})}\BibitemShut
  {NoStop}%
\bibitem [{\citenamefont {Ufrecht}\ \emph {et~al.}(2020)\citenamefont
  {Ufrecht}, \citenamefont {Di~Pumpo}, \citenamefont {Friedrich}, \citenamefont
  {Roura}, \citenamefont {Schubert}, \citenamefont {Schlippert}, \citenamefont
  {Rasel}, \citenamefont {Schleich},\ and\ \citenamefont
  {Giese}}]{Ufrecht2020}%
  \BibitemOpen
  \bibfield  {author} {\bibinfo {author} {\bibfnamefont {C.}~\bibnamefont
  {Ufrecht}}, \bibinfo {author} {\bibfnamefont {F.}~\bibnamefont {Di~Pumpo}},
  \bibinfo {author} {\bibfnamefont {A.}~\bibnamefont {Friedrich}}, \bibinfo
  {author} {\bibfnamefont {A.}~\bibnamefont {Roura}}, \bibinfo {author}
  {\bibfnamefont {C.}~\bibnamefont {Schubert}}, \bibinfo {author}
  {\bibfnamefont {D.}~\bibnamefont {Schlippert}}, \bibinfo {author}
  {\bibfnamefont {E.~M.}\ \bibnamefont {Rasel}}, \bibinfo {author}
  {\bibfnamefont {W.~P.}\ \bibnamefont {Schleich}}, \ and\ \bibinfo {author}
  {\bibfnamefont {E.}~\bibnamefont {Giese}},\ }\bibfield  {title} {\enquote
  {\bibinfo {title} {Atom-interferometric test of the universality of
  gravitational redshift and free fall},}\ }\href {\doibase
  10.1103/PhysRevResearch.2.043240} {\bibfield  {journal} {\bibinfo  {journal}
  {Phys. Rev. Res.}\ }\textbf {\bibinfo {volume} {2}},\ \bibinfo {pages}
  {043240} (\bibinfo {year} {2020})}\BibitemShut {NoStop}%
\bibitem [{\citenamefont {Zych}\ \emph {et~al.}(2011)\citenamefont {Zych},
  \citenamefont {Costa}, \citenamefont {Pikovski},\ and\ \citenamefont
  {Brukner}}]{Zych2011}%
  \BibitemOpen
  \bibfield  {author} {\bibinfo {author} {\bibfnamefont {M.}~\bibnamefont
  {Zych}}, \bibinfo {author} {\bibfnamefont {F.}~\bibnamefont {Costa}},
  \bibinfo {author} {\bibfnamefont {I.}~\bibnamefont {Pikovski}}, \ and\
  \bibinfo {author} {\bibfnamefont {{\v{C}}.}~\bibnamefont {Brukner}},\
  }\bibfield  {title} {\enquote {\bibinfo {title} {Quantum interferometric
  visibility as a witness of general relativistic proper time},}\ }\href
  {https://www.nature.com/articles/ncomms1498} {\bibfield  {journal} {\bibinfo
  {journal} {Nat. Commun.}\ }\textbf {\bibinfo {volume} {2}},\ \bibinfo {pages}
  {505} (\bibinfo {year} {2011})}\BibitemShut {NoStop}%
\bibitem [{\citenamefont {Sinha}\ and\ \citenamefont
  {Samuel}(2011)}]{Sinha2011}%
  \BibitemOpen
  \bibfield  {author} {\bibinfo {author} {\bibfnamefont {S.}~\bibnamefont
  {Sinha}}\ and\ \bibinfo {author} {\bibfnamefont {J.}~\bibnamefont {Samuel}},\
  }\bibfield  {title} {\enquote {\bibinfo {title} {Atom interferometry and the
  gravitational redshift},}\ }\href {\doibase 10.1088/0264-9381/28/14/145018}
  {\bibfield  {journal} {\bibinfo  {journal} {Class. Quantum Gravity}\ }\textbf
  {\bibinfo {volume} {28}},\ \bibinfo {pages} {145018} (\bibinfo {year}
  {2011})}\BibitemShut {NoStop}%
\bibitem [{\citenamefont {Grynberg}(1983)}]{Grynberg1983}%
  \BibitemOpen
  \bibfield  {author} {\bibinfo {author} {\bibfnamefont {G.}~\bibnamefont
  {Grynberg}},\ }\bibfield  {title} {\enquote {\bibinfo {title} {Remarks on
  e1-e2 and e1-m1 two-photon transitions},}\ }\href {\doibase
  10.1051/jphys:01983004406067900} {\bibfield  {journal} {\bibinfo  {journal}
  {J. Phys.}\ }\textbf {\bibinfo {volume} {44}},\ \bibinfo {pages} {679--682}
  (\bibinfo {year} {1983})}\BibitemShut {NoStop}%
\bibitem [{\citenamefont {Rahaman}, \citenamefont {Wright},\ and\ \citenamefont
  {Dutta}(2023)}]{Rahaman2023}%
  \BibitemOpen
  \bibfield  {author} {\bibinfo {author} {\bibfnamefont {B.}~\bibnamefont
  {Rahaman}}, \bibinfo {author} {\bibfnamefont {S.~C.}\ \bibnamefont {Wright}},
  \ and\ \bibinfo {author} {\bibfnamefont {S.}~\bibnamefont {Dutta}},\
  }\href@noop {} {\enquote {\bibinfo {title} {Observation of quantum
  interference of optical transition pathways in doppler-free two-photon
  spectroscopy and implications for precision measurements},}\ } (\bibinfo
  {year} {2023}),\ \Eprint {http://arxiv.org/abs/2308.12386} {arXiv:2308.12386
  [physics.atom-ph]} \BibitemShut {NoStop}%
\bibitem [{\citenamefont {Alden}, \citenamefont {Moore},\ and\ \citenamefont
  {Leanhardt}(2014)}]{Alden2014}%
  \BibitemOpen
  \bibfield  {author} {\bibinfo {author} {\bibfnamefont {E.~A.}\ \bibnamefont
  {Alden}}, \bibinfo {author} {\bibfnamefont {K.~R.}\ \bibnamefont {Moore}}, \
  and\ \bibinfo {author} {\bibfnamefont {A.~E.}\ \bibnamefont {Leanhardt}},\
  }\bibfield  {title} {\enquote {\bibinfo {title} {Two-photon {$E1$}-{$M1$}
  optical clock},}\ }\href {\doibase 10.1103/PhysRevA.90.012523} {\bibfield
  {journal} {\bibinfo  {journal} {Phys. Rev. A}\ }\textbf {\bibinfo {volume}
  {90}},\ \bibinfo {pages} {012523} (\bibinfo {year} {2014})}\BibitemShut
  {NoStop}%
\bibitem [{\citenamefont {Alden}(2014)}]{AldenDiss2014}%
  \BibitemOpen
  \bibfield  {author} {\bibinfo {author} {\bibfnamefont {E.}~\bibnamefont
  {Alden}},\ }\emph {\bibinfo {title} {A Two-Photon E1-M1 Optical Clock.}},\
  \href@noop {} {Ph.D. thesis},\ \bibinfo  {school} {The University of
  Michigan} (\bibinfo {year} {2014})\BibitemShut {NoStop}%
\bibitem [{\citenamefont {Lopp}\ and\ \citenamefont
  {Mart\'{\i}n-Mart\'{\i}nez}(2021)}]{Lopp2021}%
  \BibitemOpen
  \bibfield  {author} {\bibinfo {author} {\bibfnamefont {R.}~\bibnamefont
  {Lopp}}\ and\ \bibinfo {author} {\bibfnamefont {E.}~\bibnamefont
  {Mart\'{\i}n-Mart\'{\i}nez}},\ }\bibfield  {title} {\enquote {\bibinfo
  {title} {Quantum delocalization, gauge, and quantum optics: Light-matter
  interaction in relativistic quantum information},}\ }\href {\doibase
  10.1103/PhysRevA.103.013703} {\bibfield  {journal} {\bibinfo  {journal}
  {Phys. Rev. A}\ }\textbf {\bibinfo {volume} {103}},\ \bibinfo {pages}
  {013703} (\bibinfo {year} {2021})}\BibitemShut {NoStop}%
\bibitem [{\citenamefont {Gillot}\ \emph {et~al.}(2016)\citenamefont {Gillot},
  \citenamefont {Cheng}, \citenamefont {Merlet},\ and\ \citenamefont {Pereira
  Dos~Santos}}]{Gillot2016}%
  \BibitemOpen
  \bibfield  {author} {\bibinfo {author} {\bibfnamefont {P.}~\bibnamefont
  {Gillot}}, \bibinfo {author} {\bibfnamefont {B.}~\bibnamefont {Cheng}},
  \bibinfo {author} {\bibfnamefont {S.}~\bibnamefont {Merlet}}, \ and\ \bibinfo
  {author} {\bibfnamefont {F.}~\bibnamefont {Pereira Dos~Santos}},\ }\bibfield
  {title} {\enquote {\bibinfo {title} {Limits to the symmetry of a
  mach-zehnder-type atom interferometer},}\ }\href {\doibase
  10.1103/PhysRevA.93.013609} {\bibfield  {journal} {\bibinfo  {journal} {Phys.
  Rev. A}\ }\textbf {\bibinfo {volume} {93}},\ \bibinfo {pages} {013609}
  (\bibinfo {year} {2016})}\BibitemShut {NoStop}%
\bibitem [{\citenamefont {Weiss}, \citenamefont {Young},\ and\ \citenamefont
  {Chu}(1994)}]{Weiss1994}%
  \BibitemOpen
  \bibfield  {author} {\bibinfo {author} {\bibfnamefont {D.}~\bibnamefont
  {Weiss}}, \bibinfo {author} {\bibfnamefont {B.}~\bibnamefont {Young}}, \ and\
  \bibinfo {author} {\bibfnamefont {S.}~\bibnamefont {Chu}},\ }\bibfield
  {title} {\enquote {\bibinfo {title} {Precision measurement of
  $\hbar/m_{\text{cs}}$ based on photon recoil using laser-cooled atoms and
  atomic interferometry},}\ }\href {\doibase 10.1007/BF01081393} {\bibfield
  {journal} {\bibinfo  {journal} {Appl. Phys. B}\ }\textbf {\bibinfo {volume}
  {59}},\ \bibinfo {pages} {217--256} (\bibinfo {year} {1994})}\BibitemShut
  {NoStop}%
\bibitem [{\citenamefont {M{\"u}ller}, \citenamefont {Peters},\ and\
  \citenamefont {Chu}(2010{\natexlab{a}})}]{Mueller2010}%
  \BibitemOpen
  \bibfield  {author} {\bibinfo {author} {\bibfnamefont {H.}~\bibnamefont
  {M{\"u}ller}}, \bibinfo {author} {\bibfnamefont {A.}~\bibnamefont {Peters}},
  \ and\ \bibinfo {author} {\bibfnamefont {S.}~\bibnamefont {Chu}},\ }\bibfield
   {title} {\enquote {\bibinfo {title} {A precision measurement of the
  gravitational redshift by the interference of matter waves},}\ }\href
  {\doibase 10.1038/nature08776} {\bibfield  {journal} {\bibinfo  {journal}
  {Nature}\ }\textbf {\bibinfo {volume} {463}},\ \bibinfo {pages} {926--929}
  (\bibinfo {year} {2010}{\natexlab{a}})}\BibitemShut {NoStop}%
\bibitem [{\citenamefont {M{\"u}ller}, \citenamefont {Peters},\ and\
  \citenamefont {Chu}(2010{\natexlab{b}})}]{Mueller2010b}%
  \BibitemOpen
  \bibfield  {author} {\bibinfo {author} {\bibfnamefont {H.}~\bibnamefont
  {M{\"u}ller}}, \bibinfo {author} {\bibfnamefont {A.}~\bibnamefont {Peters}},
  \ and\ \bibinfo {author} {\bibfnamefont {S.}~\bibnamefont {Chu}},\ }\bibfield
   {title} {\enquote {\bibinfo {title} {M{\"u}ller, {P}eters \& {C}hu reply},}\
  }\href {https://www.nature.com/articles/nature09341} {\bibfield  {journal}
  {\bibinfo  {journal} {Nature}\ }\textbf {\bibinfo {volume} {467}},\ \bibinfo
  {pages} {E2} (\bibinfo {year} {2010}{\natexlab{b}})}\BibitemShut {NoStop}%
\bibitem [{\citenamefont {Wolf}\ \emph {et~al.}(2010)\citenamefont {Wolf},
  \citenamefont {Blanchet}, \citenamefont {Bord\'{e}}, \citenamefont {Reynaud},
  \citenamefont {Salomon},\ and\ \citenamefont {Cohen-Tannoudji}}]{Wolf2010}%
  \BibitemOpen
  \bibfield  {author} {\bibinfo {author} {\bibfnamefont {P.}~\bibnamefont
  {Wolf}}, \bibinfo {author} {\bibfnamefont {L.}~\bibnamefont {Blanchet}},
  \bibinfo {author} {\bibfnamefont {C.~J.}\ \bibnamefont {Bord\'{e}}}, \bibinfo
  {author} {\bibfnamefont {S.}~\bibnamefont {Reynaud}}, \bibinfo {author}
  {\bibfnamefont {C.}~\bibnamefont {Salomon}}, \ and\ \bibinfo {author}
  {\bibfnamefont {C.}~\bibnamefont {Cohen-Tannoudji}},\ }\bibfield  {title}
  {\enquote {\bibinfo {title} {Atom gravimeters and gravitational redshift},}\
  }\href {\doibase 10.1038/nature09340} {\bibfield  {journal} {\bibinfo
  {journal} {Nature}\ }\textbf {\bibinfo {volume} {467}},\ \bibinfo {pages}
  {E1} (\bibinfo {year} {2010})}\BibitemShut {NoStop}%
\bibitem [{\citenamefont {Wolf}\ \emph {et~al.}(2011)\citenamefont {Wolf},
  \citenamefont {Blanchet}, \citenamefont {Bord\'{e}}, \citenamefont {Reynaud},
  \citenamefont {Salomon},\ and\ \citenamefont {Cohen-Tannoudji}}]{Wolf2011}%
  \BibitemOpen
  \bibfield  {author} {\bibinfo {author} {\bibfnamefont {P.}~\bibnamefont
  {Wolf}}, \bibinfo {author} {\bibfnamefont {L.}~\bibnamefont {Blanchet}},
  \bibinfo {author} {\bibfnamefont {C.~J.}\ \bibnamefont {Bord\'{e}}}, \bibinfo
  {author} {\bibfnamefont {S.}~\bibnamefont {Reynaud}}, \bibinfo {author}
  {\bibfnamefont {C.}~\bibnamefont {Salomon}}, \ and\ \bibinfo {author}
  {\bibfnamefont {C.}~\bibnamefont {Cohen-Tannoudji}},\ }\bibfield  {title}
  {\enquote {\bibinfo {title} {Does an atom interferometer test the
  gravitational redshift at the {Compton} frequency?}}\ }\href {\doibase
  10.1088/0264-9381/28/14/145017} {\bibfield  {journal} {\bibinfo  {journal}
  {Class. Quantum Gravity}\ }\textbf {\bibinfo {volume} {28}},\ \bibinfo
  {pages} {145017} (\bibinfo {year} {2011})}\BibitemShut {NoStop}%
\bibitem [{\citenamefont {Wolf}\ \emph {et~al.}(2012)\citenamefont {Wolf},
  \citenamefont {Blanchet}, \citenamefont {Bord\'{e}}, \citenamefont {Reynaud},
  \citenamefont {Salomon},\ and\ \citenamefont {Cohen-Tannoudji}}]{Wolf2012}%
  \BibitemOpen
  \bibfield  {author} {\bibinfo {author} {\bibfnamefont {P.}~\bibnamefont
  {Wolf}}, \bibinfo {author} {\bibfnamefont {L.}~\bibnamefont {Blanchet}},
  \bibinfo {author} {\bibfnamefont {C.~J.}\ \bibnamefont {Bord\'{e}}}, \bibinfo
  {author} {\bibfnamefont {S.}~\bibnamefont {Reynaud}}, \bibinfo {author}
  {\bibfnamefont {C.}~\bibnamefont {Salomon}}, \ and\ \bibinfo {author}
  {\bibfnamefont {C.}~\bibnamefont {Cohen-Tannoudji}},\ }\bibfield  {title}
  {\enquote {\bibinfo {title} {Reply to comment on: `{Does} an atom
  interferometer test the gravitational redshift at the {Compton}
  frequency?'},}\ }\href {\doibase 10.1088/0264-9381/29/4/048002} {\bibfield
  {journal} {\bibinfo  {journal} {Class. Quantum Gravity}\ }\textbf {\bibinfo
  {volume} {29}},\ \bibinfo {pages} {048002} (\bibinfo {year}
  {2012})}\BibitemShut {NoStop}%
\bibitem [{\citenamefont {Schleich}, \citenamefont {Greenberger},\ and\
  \citenamefont {Rasel}(2013)}]{Schleich2013}%
  \BibitemOpen
  \bibfield  {author} {\bibinfo {author} {\bibfnamefont {W.~P.}\ \bibnamefont
  {Schleich}}, \bibinfo {author} {\bibfnamefont {D.~M.}\ \bibnamefont
  {Greenberger}}, \ and\ \bibinfo {author} {\bibfnamefont {E.~M.}\ \bibnamefont
  {Rasel}},\ }\bibfield  {title} {\enquote {\bibinfo {title} {Redshift
  controversy in atom interferometry: {Representation} dependence of the origin
  of phase shift},}\ }\href {\doibase 10.1103/PhysRevLett.110.010401}
  {\bibfield  {journal} {\bibinfo  {journal} {Phys. Rev. Lett.}\ }\textbf
  {\bibinfo {volume} {110}},\ \bibinfo {pages} {010401} (\bibinfo {year}
  {2013})}\BibitemShut {NoStop}%
\bibitem [{\citenamefont {Pikovski}\ \emph {et~al.}(2017)\citenamefont
  {Pikovski}, \citenamefont {Zych}, \citenamefont {Costa},\ and\ \citenamefont
  {Brukner}}]{Pikovski2017}%
  \BibitemOpen
  \bibfield  {author} {\bibinfo {author} {\bibfnamefont {I.}~\bibnamefont
  {Pikovski}}, \bibinfo {author} {\bibfnamefont {M.}~\bibnamefont {Zych}},
  \bibinfo {author} {\bibfnamefont {F.}~\bibnamefont {Costa}}, \ and\ \bibinfo
  {author} {\bibfnamefont {{\v{C}}.}~\bibnamefont {Brukner}},\ }\bibfield
  {title} {\enquote {\bibinfo {title} {Time dilation in quantum systems and
  decoherence},}\ }\href {\doibase 10.1088/1367-2630/aa5d92} {\bibfield
  {journal} {\bibinfo  {journal} {New J. Phys.}\ }\textbf {\bibinfo {volume}
  {19}},\ \bibinfo {pages} {025011} (\bibinfo {year} {2017})}\BibitemShut
  {NoStop}%
\bibitem [{\citenamefont {Sebastian}(1981)}]{Sebastian1981}%
  \BibitemOpen
  \bibfield  {author} {\bibinfo {author} {\bibfnamefont {K.~J.}\ \bibnamefont
  {Sebastian}},\ }\bibfield  {title} {\enquote {\bibinfo {title} {{Interaction
  of a composite system with the quantized radiation field in an approximately
  relativistic theory}},}\ }\href {\doibase 10.1103/PhysRevA.23.2810}
  {\bibfield  {journal} {\bibinfo  {journal} {Phys. Rev. A}\ }\textbf {\bibinfo
  {volume} {23}},\ \bibinfo {pages} {2810--2825} (\bibinfo {year}
  {1981})}\BibitemShut {NoStop}%
\bibitem [{\citenamefont {Sebastian}(1986)}]{Sebastian1986}%
  \BibitemOpen
  \bibfield  {author} {\bibinfo {author} {\bibfnamefont {K.~J.}\ \bibnamefont
  {Sebastian}},\ }\bibfield  {title} {\enquote {\bibinfo {title} {{Relativistic
  invariance of the two-photon transition amplitude of a composite system}},}\
  }\href {\doibase 10.1103/PhysRevA.34.3839} {\bibfield  {journal} {\bibinfo
  {journal} {Phys. Rev. A}\ }\textbf {\bibinfo {volume} {34}},\ \bibinfo
  {pages} {3839--3846} (\bibinfo {year} {1986})}\BibitemShut {NoStop}%
\bibitem [{\citenamefont {Osborn}(1968)}]{Osborn1968}%
  \BibitemOpen
  \bibfield  {author} {\bibinfo {author} {\bibfnamefont {H.}~\bibnamefont
  {Osborn}},\ }\bibfield  {title} {\enquote {\bibinfo {title} {Relativistic
  center-of-mass variables for two-particle systems with spin},}\ }\href
  {\doibase 10.1103/PhysRev.176.1514} {\bibfield  {journal} {\bibinfo
  {journal} {Phys. Rev.}\ }\textbf {\bibinfo {volume} {176}},\ \bibinfo {pages}
  {1514--1522} (\bibinfo {year} {1968})}\BibitemShut {NoStop}%
\bibitem [{\citenamefont {Close}\ and\ \citenamefont
  {Osborn}(1970)}]{Close1970}%
  \BibitemOpen
  \bibfield  {author} {\bibinfo {author} {\bibfnamefont {F.~E.}\ \bibnamefont
  {Close}}\ and\ \bibinfo {author} {\bibfnamefont {H.}~\bibnamefont {Osborn}},\
  }\bibfield  {title} {\enquote {\bibinfo {title} {{Relativistic Center-of-Mass
  Motion and the Electromagnetic Interaction of Systems of Charged
  Particles}},}\ }\href {\doibase 10.1103/PhysRevD.2.2127} {\bibfield
  {journal} {\bibinfo  {journal} {Phys. Rev. D}\ }\textbf {\bibinfo {volume}
  {2}},\ \bibinfo {pages} {2127--2140} (\bibinfo {year} {1970})}\BibitemShut
  {NoStop}%
\bibitem [{\citenamefont {Krajcik}\ and\ \citenamefont
  {Foldy}(1974)}]{Krajcik1974}%
  \BibitemOpen
  \bibfield  {author} {\bibinfo {author} {\bibfnamefont {R.~A.}\ \bibnamefont
  {Krajcik}}\ and\ \bibinfo {author} {\bibfnamefont {L.~L.}\ \bibnamefont
  {Foldy}},\ }\bibfield  {title} {\enquote {\bibinfo {title} {{Relativistic
  center-of-mass variables for composite systems with arbitrary internal
  interactions}},}\ }\href {\doibase 10.1103/PhysRevD.10.1777} {\bibfield
  {journal} {\bibinfo  {journal} {Phys. Rev. D}\ }\textbf {\bibinfo {volume}
  {10}},\ \bibinfo {pages} {1777--1795} (\bibinfo {year} {1974})}\BibitemShut
  {NoStop}%
\bibitem [{\citenamefont {Sonnleitner}\ and\ \citenamefont
  {Barnett}(2018)}]{Sonnleitner2018}%
  \BibitemOpen
  \bibfield  {author} {\bibinfo {author} {\bibfnamefont {M.}~\bibnamefont
  {Sonnleitner}}\ and\ \bibinfo {author} {\bibfnamefont {S.~M.}\ \bibnamefont
  {Barnett}},\ }\bibfield  {title} {\enquote {\bibinfo {title} {Mass-energy and
  anomalous friction in quantum optics},}\ }\href {\doibase
  10.1103/PhysRevA.98.042106} {\bibfield  {journal} {\bibinfo  {journal} {Phys.
  Rev. A}\ }\textbf {\bibinfo {volume} {98}},\ \bibinfo {pages} {042106}
  (\bibinfo {year} {2018})}\BibitemShut {NoStop}%
\bibitem [{\citenamefont {Schwartz}\ and\ \citenamefont
  {Giulini}(2019)}]{Schwartz2019}%
  \BibitemOpen
  \bibfield  {author} {\bibinfo {author} {\bibfnamefont {P.~K.}\ \bibnamefont
  {Schwartz}}\ and\ \bibinfo {author} {\bibfnamefont {D.}~\bibnamefont
  {Giulini}},\ }\bibfield  {title} {\enquote {\bibinfo {title}
  {Post-{N}ewtonian {H}amiltonian description of an atom in a weak
  gravitational field},}\ }\href {\doibase 10.1103/PhysRevA.100.052116}
  {\bibfield  {journal} {\bibinfo  {journal} {Phys. Rev. A}\ }\textbf {\bibinfo
  {volume} {100}},\ \bibinfo {pages} {052116} (\bibinfo {year}
  {2019})}\BibitemShut {NoStop}%
\bibitem [{\citenamefont {Schwartz}(2020)}]{Schwartz2020b}%
  \BibitemOpen
  \bibfield  {author} {\bibinfo {author} {\bibfnamefont {P.~K.}\ \bibnamefont
  {Schwartz}},\ }\emph {\bibinfo {title} {{Post-Newtonian Description of
  Quantum Systems in Gravitational Fields}}},\ \href {\doibase 10.15488/10085}
  {Ph.D. thesis},\ \bibinfo  {school} {{Gottfried Willhelm Leibniz
  Universit{\"{a}}t Hannover}} (\bibinfo {year} {2020})\BibitemShut {NoStop}%
\bibitem [{\citenamefont {A{\ss}mann}, \citenamefont {Giese},\ and\
  \citenamefont {Di~Pumpo}(2023)}]{Assmann2023}%
  \BibitemOpen
  \bibfield  {author} {\bibinfo {author} {\bibfnamefont {T.}~\bibnamefont
  {A{\ss}mann}}, \bibinfo {author} {\bibfnamefont {E.}~\bibnamefont {Giese}}, \
  and\ \bibinfo {author} {\bibfnamefont {F.}~\bibnamefont {Di~Pumpo}},\
  }\bibfield  {title} {\enquote {\bibinfo {title} {{Quantum field theory for
  multipolar composite bosons with mass defect and relativistic
  corrections}},}\ }\href {\doibase 10.48550/arXiv.2307.06110} {\bibfield
  {journal} {\bibinfo  {journal} {arXiv}\ } (\bibinfo {year} {2023}),\
  10.48550/arXiv.2307.06110},\ \Eprint {http://arxiv.org/abs/2307.06110}
  {2307.06110} \BibitemShut {NoStop}%
\bibitem [{\citenamefont {Perche}\ and\ \citenamefont
  {Neuser}(2021)}]{Perche2021}%
  \BibitemOpen
  \bibfield  {author} {\bibinfo {author} {\bibfnamefont {T.~R.}\ \bibnamefont
  {Perche}}\ and\ \bibinfo {author} {\bibfnamefont {J.}~\bibnamefont
  {Neuser}},\ }\bibfield  {title} {\enquote {\bibinfo {title} {{A wavefunction
  description for a localized quantum particle in curved spacetimes}},}\ }\href
  {\doibase 10.1088/1361-6382/ac103d} {\bibfield  {journal} {\bibinfo
  {journal} {Classical Quantum Gravity}\ }\textbf {\bibinfo {volume} {38}},\
  \bibinfo {pages} {175002} (\bibinfo {year} {2021})}\BibitemShut {NoStop}%
\bibitem [{\citenamefont {Perche}(2022)}]{Perche2022}%
  \BibitemOpen
  \bibfield  {author} {\bibinfo {author} {\bibfnamefont {T.~R.}\ \bibnamefont
  {Perche}},\ }\bibfield  {title} {\enquote {\bibinfo {title} {{Localized
  nonrelativistic quantum systems in curved spacetimes: A general
  characterization of particle detector models}},}\ }\href {\doibase
  10.1103/PhysRevD.106.025018} {\bibfield  {journal} {\bibinfo  {journal}
  {Phys. Rev. D}\ }\textbf {\bibinfo {volume} {106}},\ \bibinfo {pages}
  {025018} (\bibinfo {year} {2022})}\BibitemShut {NoStop}%
\bibitem [{\citenamefont {Ito}(2021)}]{Ito2021}%
  \BibitemOpen
  \bibfield  {author} {\bibinfo {author} {\bibfnamefont {A.}~\bibnamefont
  {Ito}},\ }\bibfield  {title} {\enquote {\bibinfo {title} {{Inertial and
  gravitational effects on a geonium atom}},}\ }\href {\doibase
  10.1088/1361-6382/ac1be9} {\bibfield  {journal} {\bibinfo  {journal}
  {Classical Quantum Gravity}\ }\textbf {\bibinfo {volume} {38}},\ \bibinfo
  {pages} {195015} (\bibinfo {year} {2021})}\BibitemShut {NoStop}%
\bibitem [{\citenamefont {Alibabaei}, \citenamefont {Schwartz},\ and\
  \citenamefont {Giulini}(2023)}]{Alibabaei2023}%
  \BibitemOpen
  \bibfield  {author} {\bibinfo {author} {\bibfnamefont {A.}~\bibnamefont
  {Alibabaei}}, \bibinfo {author} {\bibfnamefont {P.~K.}\ \bibnamefont
  {Schwartz}}, \ and\ \bibinfo {author} {\bibfnamefont {D.}~\bibnamefont
  {Giulini}},\ }\bibfield  {title} {\enquote {\bibinfo {title} {{Geometric
  post-Newtonian description of massive spin-half particles in curved
  spacetime}},}\ }\href {\doibase 10.48550/arXiv.2307.04743} {\bibfield
  {journal} {\bibinfo  {journal} {arXiv}\ } (\bibinfo {year} {2023}),\
  10.48550/arXiv.2307.04743},\ \Eprint {http://arxiv.org/abs/2307.04743}
  {2307.04743} \BibitemShut {NoStop}%
\bibitem [{\citenamefont {Giulini}, \citenamefont {Gro{\ss}ardt},\ and\
  \citenamefont {Schwartz}(2022)}]{Giulini2022}%
  \BibitemOpen
  \bibfield  {author} {\bibinfo {author} {\bibfnamefont {D.}~\bibnamefont
  {Giulini}}, \bibinfo {author} {\bibfnamefont {A.}~\bibnamefont
  {Gro{\ss}ardt}}, \ and\ \bibinfo {author} {\bibfnamefont {P.~K.}\
  \bibnamefont {Schwartz}},\ }\bibfield  {title} {\enquote {\bibinfo {title}
  {{Coupling Quantum Matter and Gravity}},}\ }\href {\doibase
  10.48550/arXiv.2207.05029} {\bibfield  {journal} {\bibinfo  {journal}
  {arXiv}\ } (\bibinfo {year} {2022}),\ 10.48550/arXiv.2207.05029},\ \Eprint
  {http://arxiv.org/abs/2207.05029} {2207.05029} \BibitemShut {NoStop}%
\bibitem [{\citenamefont {Anastopoulos}\ and\ \citenamefont
  {Hu}(2018)}]{Anastopoulos2018}%
  \BibitemOpen
  \bibfield  {author} {\bibinfo {author} {\bibfnamefont {C.}~\bibnamefont
  {Anastopoulos}}\ and\ \bibinfo {author} {\bibfnamefont {B.~L.~H.}\
  \bibnamefont {Hu}},\ }\bibfield  {title} {\enquote {\bibinfo {title}
  {{Equivalence principle for quantum systems: Dephasing and phase shift of
  free-falling particles}},}\ }\href {\doibase 10.1088/1361-6382/aaa0e8}
  {\bibfield  {journal} {\bibinfo  {journal} {Class. Quantum Gravity}\ }\textbf
  {\bibinfo {volume} {35}},\ \bibinfo {pages} {035011} (\bibinfo {year}
  {2018})}\BibitemShut {NoStop}%
\bibitem [{\citenamefont {Mart{\'{i}}nez-Lahuerta}\ \emph
  {et~al.}(2022)\citenamefont {Mart{\'{i}}nez-Lahuerta}, \citenamefont
  {Eilers}, \citenamefont {Mehlst{\"{a}}ubler}, \citenamefont {Schmidt},\ and\
  \citenamefont {Hammerer}}]{MartinezLahuerta2022b}%
  \BibitemOpen
  \bibfield  {author} {\bibinfo {author} {\bibfnamefont {V.~J.}\ \bibnamefont
  {Mart{\'{i}}nez-Lahuerta}}, \bibinfo {author} {\bibfnamefont
  {S.}~\bibnamefont {Eilers}}, \bibinfo {author} {\bibfnamefont {T.~E.}\
  \bibnamefont {Mehlst{\"{a}}ubler}}, \bibinfo {author} {\bibfnamefont {P.~O.}\
  \bibnamefont {Schmidt}}, \ and\ \bibinfo {author} {\bibfnamefont
  {K.}~\bibnamefont {Hammerer}},\ }\bibfield  {title} {\enquote {\bibinfo
  {title} {{Ab initio quantum theory of mass defect and time dilation in
  trapped-ion optical clocks}},}\ }\href {\doibase 10.1103/PhysRevA.106.032803}
  {\bibfield  {journal} {\bibinfo  {journal} {Phys. Rev. A}\ }\textbf {\bibinfo
  {volume} {106}},\ \bibinfo {pages} {032803} (\bibinfo {year}
  {2022})}\BibitemShut {NoStop}%
\bibitem [{\citenamefont {Wood}\ and\ \citenamefont {Zych}(2022)}]{Wood2022}%
  \BibitemOpen
  \bibfield  {author} {\bibinfo {author} {\bibfnamefont {C.~E.}\ \bibnamefont
  {Wood}}\ and\ \bibinfo {author} {\bibfnamefont {M.}~\bibnamefont {Zych}},\
  }\bibfield  {title} {\enquote {\bibinfo {title} {Quantized mass-energy
  effects in an unruh-dewitt detector},}\ }\href {\doibase
  10.1103/PhysRevD.106.025012} {\bibfield  {journal} {\bibinfo  {journal}
  {Phys. Rev. D}\ }\textbf {\bibinfo {volume} {106}},\ \bibinfo {pages}
  {025012} (\bibinfo {year} {2022})}\BibitemShut {NoStop}%
\bibitem [{\citenamefont {Wood}\ and\ \citenamefont {Zych}(2021)}]{Wood2021}%
  \BibitemOpen
  \bibfield  {author} {\bibinfo {author} {\bibfnamefont {C.~E.}\ \bibnamefont
  {Wood}}\ and\ \bibinfo {author} {\bibfnamefont {M.}~\bibnamefont {Zych}},\
  }\bibfield  {title} {\enquote {\bibinfo {title} {Composite particles with
  minimum uncertainty in spacetime},}\ }\href {\doibase
  10.1103/PhysRevResearch.3.013049} {\bibfield  {journal} {\bibinfo  {journal}
  {Phys. Rev. Res.}\ }\textbf {\bibinfo {volume} {3}},\ \bibinfo {pages}
  {013049} (\bibinfo {year} {2021})}\BibitemShut {NoStop}%
\bibitem [{\citenamefont {Storey}\ and\ \citenamefont
  {Cohen-Tannoudji}(1994)}]{Storey1994}%
  \BibitemOpen
  \bibfield  {author} {\bibinfo {author} {\bibfnamefont {P.}~\bibnamefont
  {Storey}}\ and\ \bibinfo {author} {\bibfnamefont {C.}~\bibnamefont
  {Cohen-Tannoudji}},\ }\bibfield  {title} {\enquote {\bibinfo {title} {The
  {Feynman} path integral approach to atomic interferometry. {A} tutorial},}\
  }\href {\doibase 10.1051/jp2:1994103} {\bibfield  {journal} {\bibinfo
  {journal} {J. Phys. II France}\ }\textbf {\bibinfo {volume} {4}},\ \bibinfo
  {pages} {1999--2027} (\bibinfo {year} {1994})}\BibitemShut {NoStop}%
\bibitem [{\citenamefont {Kleinert}(2009)}]{Kleinert2009}%
  \BibitemOpen
  \bibfield  {author} {\bibinfo {author} {\bibfnamefont {H.}~\bibnamefont
  {Kleinert}},\ }\href {\doibase 10.1142/7305} {\emph {\bibinfo {title} {{Path
  Integrals in Quantum Mechanics, Statistics, Polymer Physics, and Financial
  Markets}}}}\ (\bibinfo  {publisher} {World Scientific Publishing Company},\
  \bibinfo {address} {Singapore},\ \bibinfo {year} {2009})\BibitemShut
  {NoStop}%
\bibitem [{\citenamefont {Kleinert}\ \emph {et~al.}(2015)\citenamefont
  {Kleinert}, \citenamefont {Kajari}, \citenamefont {Roura},\ and\
  \citenamefont {Schleich}}]{Kleinert2015}%
  \BibitemOpen
  \bibfield  {author} {\bibinfo {author} {\bibfnamefont {S.}~\bibnamefont
  {Kleinert}}, \bibinfo {author} {\bibfnamefont {E.}~\bibnamefont {Kajari}},
  \bibinfo {author} {\bibfnamefont {A.}~\bibnamefont {Roura}}, \ and\ \bibinfo
  {author} {\bibfnamefont {W.~P.}\ \bibnamefont {Schleich}},\ }\bibfield
  {title} {\enquote {\bibinfo {title} {Representation-free description of
  light-pulse atom interferometry including non-inertial effects},}\ }\href
  {http://www.sciencedirect.com/science/article/pii/S0370157315003968}
  {\bibfield  {journal} {\bibinfo  {journal} {Phys. Rep.}\ }\textbf {\bibinfo
  {volume} {605}},\ \bibinfo {pages} {1--50} (\bibinfo {year}
  {2015})}\BibitemShut {NoStop}%
\bibitem [{\citenamefont {Ufrecht}(2021)}]{Ufrecht2021}%
  \BibitemOpen
  \bibfield  {author} {\bibinfo {author} {\bibfnamefont {C.}~\bibnamefont
  {Ufrecht}},\ }\bibfield  {title} {\enquote {\bibinfo {title} {Generalized
  gravity-gradient mitigation scheme},}\ }\href {\doibase
  10.1103/PhysRevA.103.023305} {\bibfield  {journal} {\bibinfo  {journal}
  {Phys. Rev. A}\ }\textbf {\bibinfo {volume} {103}},\ \bibinfo {pages}
  {023305} (\bibinfo {year} {2021})}\BibitemShut {NoStop}%
\bibitem [{\citenamefont {Giese}\ \emph {et~al.}(2014)\citenamefont {Giese},
  \citenamefont {Zeller}, \citenamefont {Kleinert}, \citenamefont {Meister},
  \citenamefont {Tamma}, \citenamefont {Roura},\ and\ \citenamefont
  {Schleich}}]{Giese2014}%
  \BibitemOpen
  \bibfield  {author} {\bibinfo {author} {\bibfnamefont {E.}~\bibnamefont
  {Giese}}, \bibinfo {author} {\bibfnamefont {W.}~\bibnamefont {Zeller}},
  \bibinfo {author} {\bibfnamefont {S.}~\bibnamefont {Kleinert}}, \bibinfo
  {author} {\bibfnamefont {M.}~\bibnamefont {Meister}}, \bibinfo {author}
  {\bibfnamefont {V.}~\bibnamefont {Tamma}}, \bibinfo {author} {\bibfnamefont
  {A.}~\bibnamefont {Roura}}, \ and\ \bibinfo {author} {\bibfnamefont {W.~P.}\
  \bibnamefont {Schleich}},\ }\bibfield  {title} {\enquote {\bibinfo {title}
  {{The interface of gravity and quantum mechanics illuminated by Wigner phase
  space}},}\ }in\ \href {\doibase 10.3254/978-1-61499-448-0-171} {\emph
  {\bibinfo {booktitle} {{Atom Interferometry}}}}\ (\bibinfo  {publisher} {IOS
  Press},\ \bibinfo {year} {2014})\ pp.\ \bibinfo {pages}
  {171--236}\BibitemShut {NoStop}%
\bibitem [{\citenamefont {Roura}, \citenamefont {Zeller},\ and\ \citenamefont
  {Schleich}(2014)}]{Roura2014}%
  \BibitemOpen
  \bibfield  {author} {\bibinfo {author} {\bibfnamefont {A.}~\bibnamefont
  {Roura}}, \bibinfo {author} {\bibfnamefont {W.}~\bibnamefont {Zeller}}, \
  and\ \bibinfo {author} {\bibfnamefont {W.~P.}\ \bibnamefont {Schleich}},\
  }\bibfield  {title} {\enquote {\bibinfo {title} {Overcoming loss of contrast
  in atom interferometry due to gravity gradients},}\ }\href
  {http://stacks.iop.org/1367-2630/16/i=12/a=123012?key=crossref.dba3a4e048fecb35939f0a052e6f83ab}
  {\bibfield  {journal} {\bibinfo  {journal} {New J. Phys.}\ }\textbf {\bibinfo
  {volume} {16}},\ \bibinfo {pages} {123012} (\bibinfo {year}
  {2014})}\BibitemShut {NoStop}%
\bibitem [{\citenamefont {Zimmermann}(2021)}]{Zimmermann2021}%
  \BibitemOpen
  \bibfield  {author} {\bibinfo {author} {\bibfnamefont {M.}~\bibnamefont
  {Zimmermann}},\ }\bibfield  {title} {\enquote {\bibinfo {title}
  {{Interference of matter waves: branch-dependent dynamics, the Kennard phase,
  and T{${^3}$} Stern-Gerlach interferometry}},}\ }\href {\doibase
  10.18725/OPARU-39705} {\bibfield  {journal} {\bibinfo  {journal}
  {Universit{\"{a}}t Ulm}\ } (\bibinfo {year} {2021}),\
  10.18725/OPARU-39705}\BibitemShut {NoStop}%
\bibitem [{\citenamefont {Jenewein}\ \emph {et~al.}(2022)\citenamefont
  {Jenewein}, \citenamefont {Hartmann}, \citenamefont {Roura},\ and\
  \citenamefont {Giese}}]{Jenewein2022}%
  \BibitemOpen
  \bibfield  {author} {\bibinfo {author} {\bibfnamefont {J.}~\bibnamefont
  {Jenewein}}, \bibinfo {author} {\bibfnamefont {S.}~\bibnamefont {Hartmann}},
  \bibinfo {author} {\bibfnamefont {A.}~\bibnamefont {Roura}}, \ and\ \bibinfo
  {author} {\bibfnamefont {E.}~\bibnamefont {Giese}},\ }\bibfield  {title}
  {\enquote {\bibinfo {title} {{Bragg-diffraction-induced imperfections of the
  signal in retroreflective atom interferometers}},}\ }\href {\doibase
  10.1103/PhysRevA.105.063316} {\bibfield  {journal} {\bibinfo  {journal}
  {Phys. Rev. A}\ }\textbf {\bibinfo {volume} {105}},\ \bibinfo {pages}
  {063316} (\bibinfo {year} {2022})}\BibitemShut {NoStop}%
\bibitem [{\citenamefont {Kirsten-Siem{\ss}}\ \emph {et~al.}(2023)\citenamefont
  {Kirsten-Siem{\ss}}, \citenamefont {Fitzek}, \citenamefont {Schubert},
  \citenamefont {Rasel}, \citenamefont {Gaaloul},\ and\ \citenamefont
  {Hammerer}}]{Kirsten-Siemss2023}%
  \BibitemOpen
  \bibfield  {author} {\bibinfo {author} {\bibfnamefont {J.-N.}\ \bibnamefont
  {Kirsten-Siem{\ss}}}, \bibinfo {author} {\bibfnamefont {F.}~\bibnamefont
  {Fitzek}}, \bibinfo {author} {\bibfnamefont {C.}~\bibnamefont {Schubert}},
  \bibinfo {author} {\bibfnamefont {E.~M.}\ \bibnamefont {Rasel}}, \bibinfo
  {author} {\bibfnamefont {N.}~\bibnamefont {Gaaloul}}, \ and\ \bibinfo
  {author} {\bibfnamefont {K.}~\bibnamefont {Hammerer}},\ }\bibfield  {title}
  {\enquote {\bibinfo {title} {{Large-Momentum-Transfer Atom Interferometers
  with $\mathrm{\ensuremath{\mu}}\mathrm{rad}$-Accuracy Using Bragg
  Diffraction}},}\ }\href {\doibase 10.1103/PhysRevLett.131.033602} {\bibfield
  {journal} {\bibinfo  {journal} {Phys. Rev. Lett.}\ }\textbf {\bibinfo
  {volume} {131}},\ \bibinfo {pages} {033602} (\bibinfo {year}
  {2023})}\BibitemShut {NoStop}%
\bibitem [{\citenamefont {Sarkar}\ \emph {et~al.}(2022)\citenamefont {Sarkar},
  \citenamefont {Piccon}, \citenamefont {Merlet},\ and\ \citenamefont {dos
  Santos}}]{Sarkar2022}%
  \BibitemOpen
  \bibfield  {author} {\bibinfo {author} {\bibfnamefont {S.}~\bibnamefont
  {Sarkar}}, \bibinfo {author} {\bibfnamefont {R.}~\bibnamefont {Piccon}},
  \bibinfo {author} {\bibfnamefont {S.}~\bibnamefont {Merlet}}, \ and\ \bibinfo
  {author} {\bibfnamefont {F.~P.}\ \bibnamefont {dos Santos}},\ }\bibfield
  {title} {\enquote {\bibinfo {title} {Simple and robust architecture of a
  laser system for atom interferometry},}\ }\href {\doibase 10.1364/OE.447073}
  {\bibfield  {journal} {\bibinfo  {journal} {Opt. Express}\ }\textbf {\bibinfo
  {volume} {30}},\ \bibinfo {pages} {3358--3366} (\bibinfo {year}
  {2022})}\BibitemShut {NoStop}%
\bibitem [{\citenamefont {Zanon-Willette}\ \emph {et~al.}(2023)\citenamefont
  {Zanon-Willette}, \citenamefont {Impens}, \citenamefont {Arimondo},
  \citenamefont {Wilkowski}, \citenamefont {Taichenachev},\ and\ \citenamefont
  {Yudin}}]{ZanonWillette2023}%
  \BibitemOpen
  \bibfield  {author} {\bibinfo {author} {\bibfnamefont {T.}~\bibnamefont
  {Zanon-Willette}}, \bibinfo {author} {\bibfnamefont {F.}~\bibnamefont
  {Impens}}, \bibinfo {author} {\bibfnamefont {E.}~\bibnamefont {Arimondo}},
  \bibinfo {author} {\bibfnamefont {D.}~\bibnamefont {Wilkowski}}, \bibinfo
  {author} {\bibfnamefont {A.~V.}\ \bibnamefont {Taichenachev}}, \ and\
  \bibinfo {author} {\bibfnamefont {V.~I.}\ \bibnamefont {Yudin}},\ }\href@noop
  {} {\enquote {\bibinfo {title} {Robust quantum sensors with twisted-light
  fields induced optical transitions},}\ } (\bibinfo {year} {2023}),\ \Eprint
  {http://arxiv.org/abs/2306.17620} {arXiv:2306.17620 [physics.atom-ph]}
  \BibitemShut {NoStop}%
\bibitem [{\citenamefont {Marzlin}\ and\ \citenamefont
  {Audretsch}(1996)}]{Marzlin1996}%
  \BibitemOpen
  \bibfield  {author} {\bibinfo {author} {\bibfnamefont {K.-P.}\ \bibnamefont
  {Marzlin}}\ and\ \bibinfo {author} {\bibfnamefont {J.}~\bibnamefont
  {Audretsch}},\ }\bibfield  {title} {\enquote {\bibinfo {title} {``freely''
  falling two-level atom in a running laser wave},}\ }\href {\doibase
  10.1103/PhysRevA.53.1004} {\bibfield  {journal} {\bibinfo  {journal} {Phys.
  Rev. A}\ }\textbf {\bibinfo {volume} {53}},\ \bibinfo {pages} {1004--1013}
  (\bibinfo {year} {1996})}\BibitemShut {NoStop}%
\bibitem [{\citenamefont {Bott}, \citenamefont {Di~Pumpo},\ and\ \citenamefont
  {Giese}(2023)}]{Bott2023}%
  \BibitemOpen
  \bibfield  {author} {\bibinfo {author} {\bibfnamefont {A.}~\bibnamefont
  {Bott}}, \bibinfo {author} {\bibfnamefont {F.}~\bibnamefont {Di~Pumpo}}, \
  and\ \bibinfo {author} {\bibfnamefont {E.}~\bibnamefont {Giese}},\ }\bibfield
   {title} {\enquote {\bibinfo {title} {{Atomic diffraction from single-photon
  transitions in gravity and Standard-Model extensions}},}\ }\href {\doibase
  10.1116/5.0174258} {\bibfield  {journal} {\bibinfo  {journal} {AVS Quantum
  Science}\ }\textbf {\bibinfo {volume} {5}},\ \bibinfo {pages} {044402}
  (\bibinfo {year} {2023})}\BibitemShut {NoStop}%
\bibitem [{\citenamefont {Bloch}(1946)}]{Bloch1946}%
  \BibitemOpen
  \bibfield  {author} {\bibinfo {author} {\bibfnamefont {F.}~\bibnamefont
  {Bloch}},\ }\bibfield  {title} {\enquote {\bibinfo {title} {Nuclear
  induction},}\ }\href {\doibase 10.1103/PhysRev.70.460} {\bibfield  {journal}
  {\bibinfo  {journal} {Phys. Rev.}\ }\textbf {\bibinfo {volume} {70}},\
  \bibinfo {pages} {460--474} (\bibinfo {year} {1946})}\BibitemShut {NoStop}%
\bibitem [{\citenamefont {Sanz}, \citenamefont {Solano},\ and\ \citenamefont
  {Egusquiza}(2016)}]{Sanz2015}%
  \BibitemOpen
  \bibfield  {author} {\bibinfo {author} {\bibfnamefont {M.}~\bibnamefont
  {Sanz}}, \bibinfo {author} {\bibfnamefont {E.}~\bibnamefont {Solano}}, \ and\
  \bibinfo {author} {\bibfnamefont {{\'I}.~L.}\ \bibnamefont {Egusquiza}},\
  }\bibfield  {title} {\enquote {\bibinfo {title} {Beyond adiabatic
  elimination: Effective hamiltonians and singular perturbation},}\ }in\
  \href@noop {} {\emph {\bibinfo {booktitle} {Applications + Practical
  Conceptualization + Mathematics = fruitful Innovation}}},\ \bibinfo {editor}
  {edited by\ \bibinfo {editor} {\bibfnamefont {R.~S.}\ \bibnamefont
  {Anderssen}}, \bibinfo {editor} {\bibfnamefont {P.}~\bibnamefont
  {Broadbridge}}, \bibinfo {editor} {\bibfnamefont {Y.}~\bibnamefont
  {Fukumoto}}, \bibinfo {editor} {\bibfnamefont {K.}~\bibnamefont {Kajiwara}},
  \bibinfo {editor} {\bibfnamefont {T.}~\bibnamefont {Takagi}}, \bibinfo
  {editor} {\bibfnamefont {E.}~\bibnamefont {Verbitskiy}}, \ and\ \bibinfo
  {editor} {\bibfnamefont {M.}~\bibnamefont {Wakayama}}}\ (\bibinfo
  {publisher} {Springer Japan},\ \bibinfo {address} {Tokyo},\ \bibinfo {year}
  {2016})\ pp.\ \bibinfo {pages} {127--142}\BibitemShut {NoStop}%
\bibitem [{\citenamefont {Fewell}(2005)}]{Fewell2005}%
  \BibitemOpen
  \bibfield  {author} {\bibinfo {author} {\bibfnamefont {M.~P.}\ \bibnamefont
  {Fewell}},\ }\bibfield  {title} {\enquote {\bibinfo {title} {Adiabatic
  elimination, the rotating-wave approximation and two-photon transitions},}\
  }\href {\doibase 10.1016/j.optcom.2005.04.049} {\bibfield  {journal}
  {\bibinfo  {journal} {Opt. Commun.}\ }\textbf {\bibinfo {volume} {253}},\
  \bibinfo {pages} {125--137} (\bibinfo {year} {2005})}\BibitemShut {NoStop}%
\bibitem [{\citenamefont {Englert}(2014)}]{Englert2014}%
  \BibitemOpen
  \bibfield  {author} {\bibinfo {author} {\bibfnamefont {B.-G.}\ \bibnamefont
  {Englert}},\ }\href@noop {} {\emph {\bibinfo {title} {Semiclassical Theory of
  Atoms}}},\ Lecture Notes in Physics\ (\bibinfo  {publisher} {Springer Berlin
  Heidelberg},\ \bibinfo {year} {2014})\BibitemShut {NoStop}%
\bibitem [{\citenamefont {Garcia}\ \emph {et~al.}(2012)\citenamefont {Garcia},
  \citenamefont {Galindo}, \citenamefont {Alvarez-Gaume},\ and\ \citenamefont
  {Pascual}}]{GalindoPascualQM1}%
  \BibitemOpen
  \bibfield  {author} {\bibinfo {author} {\bibfnamefont {J.~D.}\ \bibnamefont
  {Garcia}}, \bibinfo {author} {\bibfnamefont {A.}~\bibnamefont {Galindo}},
  \bibinfo {author} {\bibfnamefont {L.}~\bibnamefont {Alvarez-Gaume}}, \ and\
  \bibinfo {author} {\bibfnamefont {P.}~\bibnamefont {Pascual}},\ }\href@noop
  {} {\emph {\bibinfo {title} {Quantum Mechanics I}}},\ Theoretical and
  Mathematical Physics\ (\bibinfo  {publisher} {Springer Berlin Heidelberg},\
  \bibinfo {year} {2012})\BibitemShut {NoStop}%
\bibitem [{\citenamefont {Alvarez-Gaume}, \citenamefont {Galindo},\ and\
  \citenamefont {Pascual}(2012)}]{GalindoPascualQM2}%
  \BibitemOpen
  \bibfield  {author} {\bibinfo {author} {\bibfnamefont {L.}~\bibnamefont
  {Alvarez-Gaume}}, \bibinfo {author} {\bibfnamefont {A.}~\bibnamefont
  {Galindo}}, \ and\ \bibinfo {author} {\bibfnamefont {P.}~\bibnamefont
  {Pascual}},\ }\href@noop {} {\emph {\bibinfo {title} {Quantum Mechanics
  II}}},\ Theoretical and Mathematical Physics\ (\bibinfo  {publisher}
  {Springer Berlin Heidelberg},\ \bibinfo {year} {2012})\BibitemShut {NoStop}%
\bibitem [{\citenamefont {Shore}(1990)}]{ShoreBook}%
  \BibitemOpen
  \bibfield  {author} {\bibinfo {author} {\bibfnamefont {B.~W.}\ \bibnamefont
  {Shore}},\ }\href@noop {} {\emph {\bibinfo {title} {The Theory of Coherent
  Atomic Excitation, Simple Atoms and Fields}}},\ The Theory of Coherent Atomic
  Excitation\ (\bibinfo  {publisher} {Wiley},\ \bibinfo {year}
  {1990})\BibitemShut {NoStop}%
\bibitem [{\citenamefont {Gardiner}\ and\ \citenamefont
  {Zoller}(2014)}]{GardinerZoller1}%
  \BibitemOpen
  \bibfield  {author} {\bibinfo {author} {\bibfnamefont {C.}~\bibnamefont
  {Gardiner}}\ and\ \bibinfo {author} {\bibfnamefont {P.}~\bibnamefont
  {Zoller}},\ }\href {\doibase 10.1142/p941} {\emph {\bibinfo {title} {The
  Quantum World of Ultra-Cold Atoms and Light Book I: Foundations of Quantum
  Optics}}}\ (\bibinfo  {publisher} {Imperial College Press},\ \bibinfo {year}
  {2014})\BibitemShut {NoStop}%
\bibitem [{\citenamefont {Gardiner}\ and\ \citenamefont
  {Zoller}(2015)}]{GardinerZoller2}%
  \BibitemOpen
  \bibfield  {author} {\bibinfo {author} {\bibfnamefont {C.}~\bibnamefont
  {Gardiner}}\ and\ \bibinfo {author} {\bibfnamefont {P.}~\bibnamefont
  {Zoller}},\ }\href {\doibase 10.1142/p983} {\emph {\bibinfo {title} {The
  Quantum World of Ultra-Cold Atoms and Light Book II: The Physics of
  Quantum-Optical Devices}}}\ (\bibinfo  {publisher} {Imperial College Press},\
  \bibinfo {year} {2015})\BibitemShut {NoStop}%
\bibitem [{\citenamefont {Dalibard}\ and\ \citenamefont
  {Cohen-Tannoudji}(1989)}]{DalibardCohen-Tannoudji1989}%
  \BibitemOpen
  \bibfield  {author} {\bibinfo {author} {\bibfnamefont {J.}~\bibnamefont
  {Dalibard}}\ and\ \bibinfo {author} {\bibfnamefont {C.}~\bibnamefont
  {Cohen-Tannoudji}},\ }\bibfield  {title} {\enquote {\bibinfo {title} {Laser
  cooling below the doppler limit by polarization gradients: simple theoretical
  models},}\ }\href {\doibase 10.1364/JOSAB.6.002023} {\bibfield  {journal}
  {\bibinfo  {journal} {J. Opt. Soc. Am. B}\ }\textbf {\bibinfo {volume} {6}},\
  \bibinfo {pages} {2023--2045} (\bibinfo {year} {1989})}\BibitemShut {NoStop}%
\bibitem [{\citenamefont {Bade}\ \emph {et~al.}(2018)\citenamefont {Bade},
  \citenamefont {Djadaojee}, \citenamefont {Andia}, \citenamefont {Clad\'e},\
  and\ \citenamefont {Guellati-Khelifa}}]{Bade2018}%
  \BibitemOpen
  \bibfield  {author} {\bibinfo {author} {\bibfnamefont {S.}~\bibnamefont
  {Bade}}, \bibinfo {author} {\bibfnamefont {L.}~\bibnamefont {Djadaojee}},
  \bibinfo {author} {\bibfnamefont {M.}~\bibnamefont {Andia}}, \bibinfo
  {author} {\bibfnamefont {P.}~\bibnamefont {Clad\'e}}, \ and\ \bibinfo
  {author} {\bibfnamefont {S.}~\bibnamefont {Guellati-Khelifa}},\ }\bibfield
  {title} {\enquote {\bibinfo {title} {Observation of extra photon recoil in a
  distorted optical field},}\ }\href {\doibase 10.1103/PhysRevLett.121.073603}
  {\bibfield  {journal} {\bibinfo  {journal} {Phys. Rev. Lett.}\ }\textbf
  {\bibinfo {volume} {121}},\ \bibinfo {pages} {073603} (\bibinfo {year}
  {2018})}\BibitemShut {NoStop}%
\bibitem [{\citenamefont {Loriani}\ \emph {et~al.}(2020)\citenamefont
  {Loriani}, \citenamefont {Schubert}, \citenamefont {Schlippert},
  \citenamefont {Ertmer}, \citenamefont {Pereira Dos~Santos}, \citenamefont
  {Rasel}, \citenamefont {Gaaloul},\ and\ \citenamefont {Wolf}}]{Loriani2020}%
  \BibitemOpen
  \bibfield  {author} {\bibinfo {author} {\bibfnamefont {S.}~\bibnamefont
  {Loriani}}, \bibinfo {author} {\bibfnamefont {C.}~\bibnamefont {Schubert}},
  \bibinfo {author} {\bibfnamefont {D.}~\bibnamefont {Schlippert}}, \bibinfo
  {author} {\bibfnamefont {W.}~\bibnamefont {Ertmer}}, \bibinfo {author}
  {\bibfnamefont {F.}~\bibnamefont {Pereira Dos~Santos}}, \bibinfo {author}
  {\bibfnamefont {E.~M.}\ \bibnamefont {Rasel}}, \bibinfo {author}
  {\bibfnamefont {N.}~\bibnamefont {Gaaloul}}, \ and\ \bibinfo {author}
  {\bibfnamefont {P.}~\bibnamefont {Wolf}},\ }\bibfield  {title} {\enquote
  {\bibinfo {title} {Resolution of the colocation problem in satellite quantum
  tests of the universality of free fall},}\ }\href {\doibase
  10.1103/PhysRevD.102.124043} {\bibfield  {journal} {\bibinfo  {journal}
  {Phys. Rev. D}\ }\textbf {\bibinfo {volume} {102}},\ \bibinfo {pages}
  {124043} (\bibinfo {year} {2020})}\BibitemShut {NoStop}%
\bibitem [{\citenamefont {Glick}\ \emph {et~al.}(2023)\citenamefont {Glick},
  \citenamefont {Chen}, \citenamefont {Deshpande}, \citenamefont {Wang},\ and\
  \citenamefont {Kovachy}}]{Glick2023}%
  \BibitemOpen
  \bibfield  {author} {\bibinfo {author} {\bibfnamefont {J.}~\bibnamefont
  {Glick}}, \bibinfo {author} {\bibfnamefont {Z.}~\bibnamefont {Chen}},
  \bibinfo {author} {\bibfnamefont {T.}~\bibnamefont {Deshpande}}, \bibinfo
  {author} {\bibfnamefont {Y.}~\bibnamefont {Wang}}, \ and\ \bibinfo {author}
  {\bibfnamefont {T.}~\bibnamefont {Kovachy}},\ }\href@noop {} {\enquote
  {\bibinfo {title} {Coriolis force compensation and laser beam delivery for
  100-meter baseline atom interferometry},}\ } (\bibinfo {year} {2023}),\
  \Eprint {http://arxiv.org/abs/2311.05714} {arXiv:2311.05714
  [physics.atom-ph]} \BibitemShut {NoStop}%
\bibitem [{\citenamefont {Meschede}(2015)}]{Meschede2015}%
  \BibitemOpen
  \bibfield  {author} {\bibinfo {author} {\bibfnamefont {D.}~\bibnamefont
  {Meschede}},\ }\href {\doibase 10.1007/978-3-8348-9288-1} {\emph {\bibinfo
  {title} {Optik, Licht und Laser}}},\ Teubner Studienb{\"u}cher Physik\
  (\bibinfo  {publisher} {Vieweg+Teubner Verlag},\ \bibinfo {year}
  {2015})\BibitemShut {NoStop}%
\bibitem [{Note1()}]{Note1}%
  \BibitemOpen
  \bibinfo {note} {Note, that the plane wave factor in $Z$-direction is already
  considered in the equations of the previous section.}\BibitemShut {Stop}%
\bibitem [{\citenamefont {Abend}\ \emph {et~al.}(2023)\citenamefont {Abend},
  \citenamefont {Allard}, \citenamefont {Alonso}, \citenamefont {Antoniadis},
  \citenamefont {Araujo}, \citenamefont {Arduini}, \citenamefont {Arnold},
  \citenamefont {Aßmann}, \citenamefont {Augst}, \citenamefont {Badurina}
  \emph {et~al.}}]{Abend2023}%
  \BibitemOpen
  \bibfield  {author} {\bibinfo {author} {\bibfnamefont {S.}~\bibnamefont
  {Abend}}, \bibinfo {author} {\bibfnamefont {B.}~\bibnamefont {Allard}},
  \bibinfo {author} {\bibfnamefont {I.}~\bibnamefont {Alonso}}, \bibinfo
  {author} {\bibfnamefont {J.}~\bibnamefont {Antoniadis}}, \bibinfo {author}
  {\bibfnamefont {H.}~\bibnamefont {Araujo}}, \bibinfo {author} {\bibfnamefont
  {G.}~\bibnamefont {Arduini}}, \bibinfo {author} {\bibfnamefont
  {A.}~\bibnamefont {Arnold}}, \bibinfo {author} {\bibfnamefont
  {T.}~\bibnamefont {Aßmann}}, \bibinfo {author} {\bibfnamefont
  {N.}~\bibnamefont {Augst}}, \bibinfo {author} {\bibfnamefont
  {L.}~\bibnamefont {Badurina}},  \emph {et~al.},\ }\href@noop {} {\enquote
  {\bibinfo {title} {Terrestrial very-long-baseline atom interferometry:
  Workshop summary},}\ } (\bibinfo {year} {2023}),\ \Eprint
  {http://arxiv.org/abs/2310.08183} {arXiv:2310.08183 [hep-ex]} \BibitemShut
  {NoStop}%
\bibitem [{\citenamefont {Beloy}\ \emph {et~al.}(2012)\citenamefont {Beloy},
  \citenamefont {Sherman}, \citenamefont {Lemke}, \citenamefont {Hinkley},
  \citenamefont {Oates},\ and\ \citenamefont {Ludlow}}]{Beloy2012}%
  \BibitemOpen
  \bibfield  {author} {\bibinfo {author} {\bibfnamefont {K.}~\bibnamefont
  {Beloy}}, \bibinfo {author} {\bibfnamefont {J.~A.}\ \bibnamefont {Sherman}},
  \bibinfo {author} {\bibfnamefont {N.~D.}\ \bibnamefont {Lemke}}, \bibinfo
  {author} {\bibfnamefont {N.}~\bibnamefont {Hinkley}}, \bibinfo {author}
  {\bibfnamefont {C.~W.}\ \bibnamefont {Oates}}, \ and\ \bibinfo {author}
  {\bibfnamefont {A.~D.}\ \bibnamefont {Ludlow}},\ }\bibfield  {title}
  {\enquote {\bibinfo {title} {Determination of the $5d6s$
  ${}^{3}\phantom{\rule{-0.16em}{0ex}}{D}_{1}$ state lifetime and
  blackbody-radiation clock shift in yb},}\ }\href {\doibase
  10.1103/PhysRevA.86.051404} {\bibfield  {journal} {\bibinfo  {journal} {Phys.
  Rev. A}\ }\textbf {\bibinfo {volume} {86}},\ \bibinfo {pages} {051404}
  (\bibinfo {year} {2012})}\BibitemShut {NoStop}%
\bibitem [{\citenamefont {Hong}\ \emph {et~al.}(2005)\citenamefont {Hong},
  \citenamefont {Cramer}, \citenamefont {Cook}, \citenamefont {Nagourney},\
  and\ \citenamefont {Fortson}}]{Hong2005}%
  \BibitemOpen
  \bibfield  {author} {\bibinfo {author} {\bibfnamefont {T.}~\bibnamefont
  {Hong}}, \bibinfo {author} {\bibfnamefont {C.}~\bibnamefont {Cramer}},
  \bibinfo {author} {\bibfnamefont {E.}~\bibnamefont {Cook}}, \bibinfo {author}
  {\bibfnamefont {W.}~\bibnamefont {Nagourney}}, \ and\ \bibinfo {author}
  {\bibfnamefont {E.~N.}\ \bibnamefont {Fortson}},\ }\bibfield  {title}
  {\enquote {\bibinfo {title} {Observation of the s01--p03 transition in atomic
  ytterbium for optical clocks and qubit arrays},}\ }\href {\doibase
  10.1364/OL.30.002644} {\bibfield  {journal} {\bibinfo  {journal} {Opt.
  Lett.}\ }\textbf {\bibinfo {volume} {30}},\ \bibinfo {pages} {2644--2646}
  (\bibinfo {year} {2005})}\BibitemShut {NoStop}%
\bibitem [{\citenamefont {Porsev}\ \emph {et~al.}(2001)\citenamefont {Porsev},
  \citenamefont {Kozlov}, \citenamefont {Rakhlina},\ and\ \citenamefont
  {Derevianko}}]{Porsev2001}%
  \BibitemOpen
  \bibfield  {author} {\bibinfo {author} {\bibfnamefont {S.~G.}\ \bibnamefont
  {Porsev}}, \bibinfo {author} {\bibfnamefont {M.~G.}\ \bibnamefont {Kozlov}},
  \bibinfo {author} {\bibfnamefont {Y.~G.}\ \bibnamefont {Rakhlina}}, \ and\
  \bibinfo {author} {\bibfnamefont {A.}~\bibnamefont {Derevianko}},\ }\bibfield
   {title} {\enquote {\bibinfo {title} {Many-body calculations of
  electric-dipole amplitudes for transitions between low-lying levels of mg,
  ca, and sr},}\ }\href {https://api.semanticscholar.org/CorpusID:7899140}
  {\bibfield  {journal} {\bibinfo  {journal} {Phys. Rev. A}\ }\textbf {\bibinfo
  {volume} {64}} (\bibinfo {year} {2001})}\BibitemShut {NoStop}%
\bibitem [{\citenamefont {Curtis}(2001)}]{Curtis2001}%
  \BibitemOpen
  \bibfield  {author} {\bibinfo {author} {\bibfnamefont {L.~J.}\ \bibnamefont
  {Curtis}},\ }\bibfield  {title} {\enquote {\bibinfo {title} {Atomic structure
  and lifetimes},}\ }\href
  {https://assets.cambridge.org/97805218/29397/sample/9780521829397ws.pdf}
  {\bibfield  {journal} {\bibinfo  {journal} {Cambridge University Press}\ }
  (\bibinfo {year} {2001})}\BibitemShut {NoStop}%
\bibitem [{\citenamefont {Ludlow}\ \emph {et~al.}(2006)\citenamefont {Ludlow},
  \citenamefont {Boyd}, \citenamefont {Zelevinsky}, \citenamefont {Foreman},
  \citenamefont {Blatt}, \citenamefont {Notcutt}, \citenamefont {Ido},\ and\
  \citenamefont {Ye}}]{Ludlow2006}%
  \BibitemOpen
  \bibfield  {author} {\bibinfo {author} {\bibfnamefont {A.~D.}\ \bibnamefont
  {Ludlow}}, \bibinfo {author} {\bibfnamefont {M.~M.}\ \bibnamefont {Boyd}},
  \bibinfo {author} {\bibfnamefont {T.}~\bibnamefont {Zelevinsky}}, \bibinfo
  {author} {\bibfnamefont {S.~M.}\ \bibnamefont {Foreman}}, \bibinfo {author}
  {\bibfnamefont {S.}~\bibnamefont {Blatt}}, \bibinfo {author} {\bibfnamefont
  {M.}~\bibnamefont {Notcutt}}, \bibinfo {author} {\bibfnamefont
  {T.}~\bibnamefont {Ido}}, \ and\ \bibinfo {author} {\bibfnamefont
  {J.}~\bibnamefont {Ye}},\ }\bibfield  {title} {\enquote {\bibinfo {title}
  {Systematic study of the $^{87}\mathrm{Sr}$ clock transition in an optical
  lattice},}\ }\href {\doibase 10.1103/PhysRevLett.96.033003} {\bibfield
  {journal} {\bibinfo  {journal} {Phys. Rev. Lett.}\ }\textbf {\bibinfo
  {volume} {96}},\ \bibinfo {pages} {033003} (\bibinfo {year}
  {2006})}\BibitemShut {NoStop}%
\bibitem [{\citenamefont {Wodey}\ \emph {et~al.}(2020)\citenamefont {Wodey},
  \citenamefont {Tell}, \citenamefont {Rasel}, \citenamefont {Schlippert},
  \citenamefont {Baur}, \citenamefont {Kissling}, \citenamefont {Kölliker},
  \citenamefont {Lorenz}, \citenamefont {Marrer}, \citenamefont {Schläpfer},
  \citenamefont {Widmer}, \citenamefont {Ufrecht}, \citenamefont {Stuiber},\
  and\ \citenamefont {Fierlinger}}]{Wodey2020}%
  \BibitemOpen
  \bibfield  {author} {\bibinfo {author} {\bibfnamefont {E.}~\bibnamefont
  {Wodey}}, \bibinfo {author} {\bibfnamefont {D.}~\bibnamefont {Tell}},
  \bibinfo {author} {\bibfnamefont {E.~M.}\ \bibnamefont {Rasel}}, \bibinfo
  {author} {\bibfnamefont {D.}~\bibnamefont {Schlippert}}, \bibinfo {author}
  {\bibfnamefont {R.}~\bibnamefont {Baur}}, \bibinfo {author} {\bibfnamefont
  {U.}~\bibnamefont {Kissling}}, \bibinfo {author} {\bibfnamefont
  {B.}~\bibnamefont {Kölliker}}, \bibinfo {author} {\bibfnamefont
  {M.}~\bibnamefont {Lorenz}}, \bibinfo {author} {\bibfnamefont
  {M.}~\bibnamefont {Marrer}}, \bibinfo {author} {\bibfnamefont
  {U.}~\bibnamefont {Schläpfer}}, \bibinfo {author} {\bibfnamefont
  {M.}~\bibnamefont {Widmer}}, \bibinfo {author} {\bibfnamefont
  {C.}~\bibnamefont {Ufrecht}}, \bibinfo {author} {\bibfnamefont
  {S.}~\bibnamefont {Stuiber}}, \ and\ \bibinfo {author} {\bibfnamefont
  {P.}~\bibnamefont {Fierlinger}},\ }\bibfield  {title} {\enquote {\bibinfo
  {title} {{A scalable high-performance magnetic shield for very long baseline
  atom interferometry}},}\ }\href {\doibase 10.1063/1.5141340} {\bibfield
  {journal} {\bibinfo  {journal} {Rev. Sci. Instrum.}\ }\textbf {\bibinfo
  {volume} {91}},\ \bibinfo {pages} {035117} (\bibinfo {year}
  {2020})}\BibitemShut {NoStop}%
\bibitem [{\citenamefont {Lezeik}\ \emph {et~al.}(2023)\citenamefont {Lezeik},
  \citenamefont {Tell}, \citenamefont {Zipfel}, \citenamefont {Gupta},
  \citenamefont {Wodey}, \citenamefont {Rasel}, \citenamefont {Schubert},\ and\
  \citenamefont {Schlippert}}]{Lezeik2023}%
  \BibitemOpen
  \bibfield  {author} {\bibinfo {author} {\bibfnamefont {A.}~\bibnamefont
  {Lezeik}}, \bibinfo {author} {\bibfnamefont {D.}~\bibnamefont {Tell}},
  \bibinfo {author} {\bibfnamefont {K.~H.}\ \bibnamefont {Zipfel}}, \bibinfo
  {author} {\bibfnamefont {V.}~\bibnamefont {Gupta}}, \bibinfo {author}
  {\bibfnamefont {E.}~\bibnamefont {Wodey}}, \bibinfo {author} {\bibfnamefont
  {E.~M.}\ \bibnamefont {Rasel}}, \bibinfo {author} {\bibfnamefont
  {C.}~\bibnamefont {Schubert}}, \ and\ \bibinfo {author} {\bibfnamefont
  {D.}~\bibnamefont {Schlippert}},\ }\bibfield  {title} {\enquote {\bibinfo
  {title} {{Understanding the Gravitational and Magnetic Environment of a Very
  Long Baseline Atom Interferometer}},}\ }in\ \href {\doibase
  10.1142/9789811275388_0014} {\emph {\bibinfo {booktitle} {{CPT and Lorentz
  Symmetry}}}}\ (\bibinfo  {publisher} {WORLD SCIENTIFIC},\ \bibinfo {address}
  {Singapore},\ \bibinfo {year} {2023})\ pp.\ \bibinfo {pages}
  {64--68}\BibitemShut {NoStop}%
\bibitem [{\citenamefont {Dickerson}\ \emph {et~al.}(2013)\citenamefont
  {Dickerson}, \citenamefont {Hogan}, \citenamefont {Sugarbaker}, \citenamefont
  {Johnson},\ and\ \citenamefont {Kasevich}}]{Dickerson2013}%
  \BibitemOpen
  \bibfield  {author} {\bibinfo {author} {\bibfnamefont {S.~M.}\ \bibnamefont
  {Dickerson}}, \bibinfo {author} {\bibfnamefont {J.~M.}\ \bibnamefont
  {Hogan}}, \bibinfo {author} {\bibfnamefont {A.}~\bibnamefont {Sugarbaker}},
  \bibinfo {author} {\bibfnamefont {D.~M.~S.}\ \bibnamefont {Johnson}}, \ and\
  \bibinfo {author} {\bibfnamefont {M.~A.}\ \bibnamefont {Kasevich}},\
  }\bibfield  {title} {\enquote {\bibinfo {title} {Multiaxis inertial sensing
  with long-time point source atom interferometry},}\ }\href {\doibase
  10.1103/PhysRevLett.111.083001} {\bibfield  {journal} {\bibinfo  {journal}
  {Phys. Rev. Lett.}\ }\textbf {\bibinfo {volume} {111}},\ \bibinfo {pages}
  {083001} (\bibinfo {year} {2013})}\BibitemShut {NoStop}%
\bibitem [{\citenamefont {Roura}(2017)}]{Roura2017}%
  \BibitemOpen
  \bibfield  {author} {\bibinfo {author} {\bibfnamefont {A.}~\bibnamefont
  {Roura}},\ }\bibfield  {title} {\enquote {\bibinfo {title} {Circumventing
  heisenberg's uncertainty principle in atom interferometry tests of the
  equivalence principle},}\ }\href {\doibase 10.1103/PhysRevLett.118.160401}
  {\bibfield  {journal} {\bibinfo  {journal} {Phys. Rev. Lett.}\ }\textbf
  {\bibinfo {volume} {118}},\ \bibinfo {pages} {160401} (\bibinfo {year}
  {2017})}\BibitemShut {NoStop}%
\bibitem [{\citenamefont {Overstreet}\ \emph {et~al.}(2018)\citenamefont
  {Overstreet}, \citenamefont {Asenbaum}, \citenamefont {Kovachy},
  \citenamefont {Notermans}, \citenamefont {Hogan},\ and\ \citenamefont
  {Kasevich}}]{Overstreet2018}%
  \BibitemOpen
  \bibfield  {author} {\bibinfo {author} {\bibfnamefont {C.}~\bibnamefont
  {Overstreet}}, \bibinfo {author} {\bibfnamefont {P.}~\bibnamefont
  {Asenbaum}}, \bibinfo {author} {\bibfnamefont {T.}~\bibnamefont {Kovachy}},
  \bibinfo {author} {\bibfnamefont {R.}~\bibnamefont {Notermans}}, \bibinfo
  {author} {\bibfnamefont {J.~M.}\ \bibnamefont {Hogan}}, \ and\ \bibinfo
  {author} {\bibfnamefont {M.~A.}\ \bibnamefont {Kasevich}},\ }\bibfield
  {title} {\enquote {\bibinfo {title} {Effective inertial frame in an atom
  interferometric test of the equivalence principle},}\ }\href {\doibase
  10.1103/PhysRevLett.120.183604} {\bibfield  {journal} {\bibinfo  {journal}
  {Phys. Rev. Lett.}\ }\textbf {\bibinfo {volume} {120}},\ \bibinfo {pages}
  {183604} (\bibinfo {year} {2018})}\BibitemShut {NoStop}%
\bibitem [{\citenamefont {Neumann}, \citenamefont {Gebbe},\ and\ \citenamefont
  {Walser}(2021)}]{Neumann2021}%
  \BibitemOpen
  \bibfield  {author} {\bibinfo {author} {\bibfnamefont {A.}~\bibnamefont
  {Neumann}}, \bibinfo {author} {\bibfnamefont {M.}~\bibnamefont {Gebbe}}, \
  and\ \bibinfo {author} {\bibfnamefont {R.}~\bibnamefont {Walser}},\
  }\bibfield  {title} {\enquote {\bibinfo {title} {Aberrations in
  (3+1)-dimensional bragg diffraction using pulsed laguerre-gaussian laser
  beams},}\ }\href {\doibase 10.1103/PhysRevA.103.043306} {\bibfield  {journal}
  {\bibinfo  {journal} {Phys. Rev. A}\ }\textbf {\bibinfo {volume} {103}},\
  \bibinfo {pages} {043306} (\bibinfo {year} {2021})}\BibitemShut {NoStop}%
\end{thebibliography}%

\end{document}